\documentclass[longauth]{aa}

\usepackage{graphicx}
\usepackage{txfonts}
\usepackage{caption}
\usepackage{multirow}
\usepackage{threeparttablex}
\usepackage{xcolor}

\makeatletter

\newcommand{\figref[1]}{Fig.~\ref{#1}}

\newcommand{\kms}{$\mathrm{km~s^{-1}}$}

\newcommand{\grp}{{G_\mathrm{RP}}}
\newcommand{\grvs}{{G_\mathrm{RVS}}}

\makeatother

\begin{document}

   \title{\textit{Gaia} Early Data Release 3: Updated radial velocities from \textit{Gaia} DR2}

  % \subtitle{}

   \author{
G. M. ~Seabroke  \inst{\ref{inst:0002}} 
\and C. ~Fabricius\inst{\ref{inst:0018}} 
\and    D. ~Teyssier\inst{\ref{inst:0017}} 
\and    P. ~Sartoretti\inst{\ref{inst:0001}} 
\and    D. ~Katz\inst{\ref{inst:0001}} 
\and      M. ~Cropper \inst{\ref{inst:0002}} 
\and            T. ~Antoja\inst{\ref{inst:0018}} 
\and       K. ~Benson\inst{\ref{inst:0002}}
\and      M.~Smith\inst{\ref{inst:0002}}
\and      C.        ~Dolding                   \inst{\ref{inst:0002}}
\and E.        ~Gosset                 \inst{\ref{inst:0007},\ref{inst:0009}}\relax
\and P.      ~Panuzzo                     \inst{\ref{inst:0001}} 
\and F.      ~Th\'{e}venin            \inst{\ref{inst:0012}}
\and C.        ~Allende Prieto          \inst{\ref{inst:0002},\ref{inst:tenerife1},\ref{inst:tenerife2}}\relax
\and R.      ~Blomme                    \inst{\ref{inst:0003}}
\and A.      ~Guerrier                 \inst{\ref{inst:0005}}
\and H.~Huckle                      \inst{\ref{inst:0002}}
\and A.      ~Jean-Antoine            \inst{\ref{inst:0005}}
\and R.       ~Haigron                       \inst{\ref{inst:0001}}\relax
\and O.~Marchal                     \inst{\ref{inst:0019},\ref{inst:0001}}
\and S.        ~Baker                     \inst{\ref{inst:0002}} 
\and Y.        ~Damerdji                \inst{\ref{inst:0006},\ref{inst:0007}}
\and M.        ~David                         \inst{\ref{inst:0008}}
\and Y.        ~Fr\'{e}mat                \inst{\ref{inst:0003}}\relax
\and K.        ~Jan{\ss}en                 \inst{\ref{inst:aip}}\relax
\and G.      ~Jasniewicz                \inst{\ref{inst:0004}}
\and A.        ~Lobel               \inst{\ref{inst:0003}}\relax
\and N.      ~Samaras                   \inst{\ref{inst:0003}}
 \and G.     ~Plum                      \inst{\ref{inst:0001}}
\and C.       ~Soubiran                 \inst{\ref{inst:0011}}
\and O.       ~Vanel                      \inst{\ref{inst:0001}}
\and T.       ~Zwitter                       \inst{\ref{inst:0016}}\relax 
\and M.        ~Ajaj                         \inst{\ref{inst:0001}} 
\and E. ~Caffau \inst{\ref{inst:0001}}
\and L.       ~Chemin                    \inst{\ref{inst:0014}}\relax
\and F.       ~Royer                       \inst{\ref{inst:0001}}  
\and N.~Brouillet  \inst{\ref{inst:0011}}
\and F.        ~Crifo                       \inst{\ref{inst:0001}}
\and L. P.     ~Guy                           \inst{\ref{inst:0010},\ref{inst:0020}}
\and N.~C.~Hambly                     \inst{\ref{inst:0015}}\relax
\and N.       ~Leclerc                        \inst{\ref{inst:0001}} \relax
\and A.~Mastrobuono-Battisti \inst{\ref{inst:0001},\ref{inst:lund}}  \relax
\and Y.       ~Viala                          \inst{\ref{inst:0001}}
   }  

\institute{
Mullard Space Science Laboratory, University College London, Holmbury St Mary, Dorking, Surrey, RH5 6NT, United Kingdom\relax                                                                           
\label{inst:0002}
\and Institut de Ci\`encies del Cosmos, Universitat de Barcelona (IEEC-UB), Mart\'i i Franqu\`es 1, 08028 Barcelona, Spain
\label{inst:0018}
\and Telespazio Vega UK Ltd for ESA/ESAC, Camino bajo del Castillo, s/n, Urbanizacion Villafranca del Castillo, Villanueva de la Ca\~nada, 28692 Madrid, Spain
\label{inst:0017}
  \and  GEPI, Observatoire de Paris, Universit\'{e} PSL, CNRS, 5 Place Jules Janssen, F-92190 Meudon, France\relax    
 \label{inst:0001}
  \and Universit\'{e} de Nice Sophia-antipolis, CNRS, Observatoire de la C\^{o}te d'Azur, Laboratoire Lagrange, BP 4229, F-06304 Nice, France\relax                                                       
 \label{inst:0012}
 \and Instituto de Astrof\'{\i}sica de Canarias, E-38205 La Laguna, Tenerife, Spain\relax                                                                                                                     
 \label{inst:tenerife1}
\and Universidad de La Laguna, Departamento de Astrof\'{\i}sica, E-38206 La Laguna, Tenerife, Spain\relax    
\label{inst:tenerife2} 
\and Royal Observatory of Belgium, Ringlaan 3, B-1180 Brussels, Belgium\relax         
\label{inst:0003}
\and CNES Centre Spatial de Toulouse, 18 avenue Edouard Belin, F-31401 Toulouse Cedex 9, France\relax                                                                                                          
\label{inst:0005}
\and Observatoire astronomique de Strasbourg, Universit\'{e} de Strasbourg, CNRS, 11 rue de l'Universit\'{e}, F-67000 Strasbourg, France
\label{inst:0019}
\and CRAAG - Centre de Recherche en Astronomie, Astrophysique et G\'{e}ophysique, Route de l'Observatoire, Bp 63 Bouzareah, DZ-16340, Alger, Alg\'{e}rie\relax                                                      
 \label{inst:0006}
 \and Institut d'Astrophysique et de G\'{e}ophysique, Universit\'{e} de Li\`{e}ge, 19c, All\'{e}e du 6 Ao\^{u}t, B-4000 Li\`{e}ge, Belgium\relax   
\label{inst:0007}
 \and Universiteit Antwerpen, Onderzoeksgroep Toegepaste Wiskunde, Middelheimlaan 1, B-2020 Antwerpen, Belgium\relax                                                                                            
 \label{inst:0008}
 \and F.R.S.-FNRS, Rue d'Egmont 5, B-1000 Brussels, Belgium\relax                                                                                                                                               
\label{inst:0009}
\and Leibniz Institute for Astrophysics Potsdam (AIP), An der Sternwarte 16, D-14482 Potsdam, Germany\relax                                                                                                    
\label{inst:aip}
\and Laboratoire Univers et Particules de Montpellier, Universit\'{e} Montpellier, CNRS, Place Eug\`{e}ne Bataillon, CC72, F-34095 Montpellier Cedex 05, France\relax                                               
 \label{inst:0004}
 \and Laboratoire d'astrophysique de Bordeaux, Universit\'{e} de Bordeaux, CNRS, B18N, all{\'e}e Geoffroy Saint-Hilaire, F-33615 Pessac, France\relax                                                           
\label{inst:0011}
 \and Centro de Astronom\'ia, Universidad de Antofagasta, Avda. U. de Antofagasta, 02800 Antofagasta, Chile
\label{inst:0014}
\and ATOS for CNES Centre Spatial de Toulouse, 18 avenue Edouard Belin, F-31401 Toulouse Cedex 9, France\relax                                                                                                      
 \label{inst:atos}
\and Thales Services for CNES Centre Spatial de Toulouse, 18 avenue Edouard Belin, F-31401 Toulouse Cedex 9, France\relax    
\label{inst:thales}
\and Faculty of Mathematics and Physics, University of Ljubljana, Jadranska ulica 19, SLO-1000 Ljubljana, Slovenia\relax                                                                                         
\label{inst:0016}
\and Department of Astronomy, University of Geneva, Chemin d'Ecogia 16, CH-1290 Versoix, Switzerland\relax                                                                                                   
\label{inst:0010}
\and Large Synoptic Survey Telescope, 950 N Cherry Avenue, Tucson, Arizona 85719, USA
\label{inst:0020}
\and Institute for Astronomy, University of Edinburgh, Royal Observatory, Blackford Hill, Edinburgh EH9 3HJ, United Kingdom\relax                                                                           
 \label{inst:0015}
\and Department of Astronomy and Theoretical Physics, Lund Observatory, Box 43, SE--221 00, Lund, Sweden
 \label{inst:lund}
 }

\date{Received ; accepted }

  \abstract
   %{} %leave it empty if necessary  
   {\textit{Gaia}'s Early Third Data Release (EDR3) does not contain new radial velocities because these will be published in \textit{Gaia}'s full third data release (DR3), expected in the first half of 2022.  To maximise the usefulness of EDR3, \textit{Gaia}'s second data release (DR2) sources (with radial velocities) are matched to EDR3 sources to allow their DR2 radial velocities to also be included in EDR3.  This presents two considerations: ($i$) \citet{boubert2019}, hereafter B19, published a list of 70\,365 sources with potentially contaminated DR2 radial velocities; and ($ii$) EDR3 is based on a new astrometric solution and a new source list, which means sources in DR2 may not be in EDR3. }
   {The two aims of this work are: ($i$) investigate the B19 list in order to improve the DR2 radial velocities being included in EDR3 and to avoid false-positive hyper-velocity candidates; and ($ii$) match the DR2 sources (with radial velocities) to EDR3 sources.}
   {The two methods of this work are: ($i$) unpublished, preliminary DR3 radial velocities of sources in the B19 list, and high-velocity stars not in the B19 list, are compared with their DR2 radial velocities to identify and remove contaminated DR2 radial velocities from EDR3; and ($ii$) proper motions and epoch position propagation is used to attempt to match all sources with radial velocities in DR2 to EDR3 sources.  The comparison of DR2 and DR3 radial velocities are used to resolve match ambiguities.}
   {EDR3 contains 7\,209\,831 sources with a DR2 radial velocity, which is 99.8\% of sources with a radial velocity in DR2 (7\,224\,631).  14\,800 radial velocities from DR2 are not propagated to any EDR3 sources because ($i$) 3871 from the B19 list are found to either not have a DR3 radial velocity or it differs significantly from its DR2 value, and five high-velocity stars not in the B19 list are confirmed to have contaminated radial velocities, in one case because of contamination from the non-overlapping RVS windows of a nearby, bright star; and ($ii$) 10\,924 DR2 sources could not be satisfactorily matched to any EDR3 sources so their DR2 radial velocities are also missing from EDR3.}
   {The reliability of radial velocities in EDR3 has improved compared to DR2 because the update removes a small fraction of erroneous radial velocities (0.05\% of DR2 radial velocities and 5.5\% of the B19 list).  Lessons learnt from EDR3 (e.g. bright star contamination) will improve the radial velocities in future {\it Gaia} data releases.  The main reason for radial velocities from DR2 not propagating to EDR3 is not related to DR2 radial velocity quality.  It is because the DR2 astrometry is based on one component of close binary pairs, while EDR3 astrometry is based on the other component, which prevents these sources being unambiguously matched.}


%\titlerunning{ Early Data Release 3: updated radial velocities from  DR2}
%\authorrunning{G. M. Seabroke {\it et al}.}
\maketitle
%-------------------------------------------------------------------

\section{Introduction}

The second \textit{Gaia} data release (DR2, \citealt{gaiabrown2018}) included line-of-sight radial velocities from \textit{Gaia}'s Radial Velocity Spectrometer (RVS, \citealt{cropper2018}) for the first time.  The radial velocities are for 7\,224\,631 stars brighter than $\grvs = 12$~mag \citep{sartoretti2018,katz2019}, based on 22 months of data.  

The third \textit{Gaia} data release (DR3) will include new radial velocities for stars brighter than $\grvs = 14$~mag,  based on 34 months of data,  expected in the first half of 2022\footnote{\url{https://www.cosmos.esa.int/web/gaia/release}}.  DR3 has been split into two releases in order to release data products as soon as they are ready. \textit{Gaia}'s Early Third Data Release (EDR3) includes improved astrometry and integrated photometry \citep{brown2021}.

EDR3 does not include the new radial velocities for two reasons.  Firstly, their derivation is based on an internal release of EDR3 astrometry and so there is a time lag between these Data Processing and Analysis Consortium (DPAC) products being ready.
Secondly, once validated by DPAC-Co-ordination Unit (CU) 6 ({\it Spectroscopic Processing}), the new radial velocities are further validated by CU4 ({\it Object Processing}) to derive binary-star orbital parameters.  The magnitude derived from RVS spectra ($\grvs$) and the RVS spectra themselves are also being validated by CU7 ({\it Variability Processing}) and CU8 ({\it Astrophysical Parameters}), deriving variability and astrophysical parameters, respectively.  CU9 ({\it Catalogue Access}) validates DPAC data products holistically before they are publicly released in DR3.

To maximise the usefulness of EDR3, radial velocities from DR2 have been transferred and improved in EDR3.  Various filters were applied to DR2 radial velocities at different stages prior to publication \citep{sartoretti2018,katz2019,arenou2018}.  It was decided not to revisit any DR2 filters when transferring and improving the DR2 radial velocities to EDR3 sources i.e. the starting point for this work are the 7\,224\,631 sources with radial velocities in DR2.

By design, the number of sources with radial velocities in EDR3 cannot exceed the number in DR2 but it can be fewer.  Firstly, in propagating radial velocities from DR2 sources to EDR3 sources, there is the opportunity to filter out DR2 radial velocities identified as potentially erroneous after DR2 was published.

Filtering is a
choice consistent with DPAC policy for the early data releases.  Adding trustworthy quality flags would take much more validation time so the
filtering is a compromise DPAC chooses. Users cannot undo DPAC filters but they are free to further
restrict the sample they select based on fields in the catalogue.

This work focuses on the list that was published\footnote{\url{https://arxiv.org/src/1901.10460v1/anc/}} by \citet{boubert2019}, hereafter B19.  They found 70\,365  DR2 sources that have potentially contaminated radial velocities (less than 1\% of the total number of sources in DR2 with radial velocities).  In addition, we examine high-velocity stars not in the B19 list because of their high scientific interest.  We do not revisit all the DR2 radial velocities because this will be done with the new radial velocities in DR3.

Secondly, the number of sources with radial velocities in EDR3 cannot exceed the number in DR2 but it can be fewer because the matching of DR2 sources to EDR3 sources is not always successful.  EDR3 is based on a new astrometric solution \citep{lindegren2021} and a new source list \citep{torra2021}, which means sources in DR2 may not be present in EDR3 owing to the sources merging in with others or splitting into new ones.   

In this paper, we describe the observations and pipelines used in this work (Sect. \ref{sec:data}). Unpublished, preliminary DR3 radial velocities were used to investigate the B19 list (Sect. \ref{sec:rv}) and high-velocity stars not in the B19 list (Sect. \ref{sec:hvs_dr3}).  High-velocity stars not in the B19 list and without a DR3 radial velocity are investigated collectively (Sect. \ref{sec:hvs}) and individually (Appendix \ref{sec:appendix_hvs}).  A bespoke cross-match between DR2 sources with radial velocities and EDR3 sources was required (Sect. \ref{Sect:Xmatch}) and source ambiguity was resolved (Sect. \ref{Sect:source}).  The final propagation of DR2 radial velocities to EDR3 is presented in Sect.~\ref{Sect:results} and Appendix \ref{sect:appendix2}, along with validation of the decision to remove some radial velocities from EDR3 by comparing these with literature radial velocities (Sect.~\ref{sec:lit}).  The results are discussed in Sect.~\ref{sec:discussion}, before concluding in Sect. \ref{sec:conc}.

\section{Observations}
\label{sec:data}

\subsection{RVS windows}
\label{sec:obs}

RVS is slitless because {\it Gaia} operates in time-delay integration (TDI) mode in which the RVS spectra scan over the focal plane at the same rate at which the CCD detectors are being read out.  In other words, RVS is effectively an integral field unit but it disperses spectra along one direction of the on-sky image on its CCDs.  This means that the images of the spectra can overlap or be `blended'.  All {\it Gaia} observations are read out as `windows' of pixels surrounding their data.  RVS windows are 10 pixels in the across-scan (AC) direction and in the along-scan (AL) direction, they are 1260 pixels (until June 2015) or 1296 pixels (after June 2015) \citep{cropper2018}.  The windows are centred on the spectra so when the images of spectra overlap, their windows are also overlapped (\figref[fig:window]).

\begin{figure*}%[t]
\centering
\includegraphics[width=\textwidth]{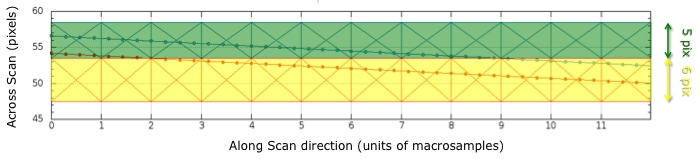}
\caption[]{An illustration of two RVS windows overlapping and how they are truncated at readout.  The vertical axis is an example pixel number in the spatial AC direction.  Prior to truncation, the nominal 10-AC pixel extent of the green and yellow windows are [48.5,58.5] and [47.5,57.5] pixels, respectively. Because they overlap each other in AC, the green and yellow shading delineates the AC truncation at readout.  The truncated green and yellow windows have 5 and 6 AC pixels, respectively.  The macrosamples are numbered along the bottom.  The top and bottom diagonal lines track the AC peak of the spectrum in the green and yellow windows, respectively.}
\label{fig:window}
\end{figure*}

The Video Processing Unit (VPU) uses the onboard Red Photometer (RP) spectra to calculate an onboard $\grvs$.\footnote{If the RP spectra are saturated, onboard $\grvs$ is calculated from the Astrometric Field (AF) $G$ magnitude instead.}  If this is brighter than 7 mag, RVS windows have every pixel read out to produce a two-dimensional (2D) window.  If it is fainter than 7 mag, RVS windows have their pixels summed on the CCD in the AC direction to produce one-dimensional (1D) spectra.  Because pixels can be selected only once at the CCD readout node, they are generally assigned to one window only.\footnote{The exceptions to this rule are rare cases of overlapping 2D windows and co-located 1D windows (Sect. \ref{sec:4658865791827681536}), when pixel values are duplicated in each window by the VPU.}  This leads to window truncation in the AC direction.  When there are two overlapping windows, the truncated window AC width is apportioned equally (to the nearest integer, \figref[fig:window]).  RVS windows always start or end in AL on multiples of 105 pixels (before June 2015) or 108 pixels (after June 2015), called macrosample boundaries so there are always 12 macrosamples in each RVS window \citep{cropper2018}.  The AC truncation is identical within a macrosample.  If windows overlap in AC and are aligned in AL, both windows can have rectangular geometries after truncation.  This is illustrated in \figref[fig:window].  \citet{cropper2018} fig. 7 illustrates non-rectangular truncation when windows are not aligned in AL.  

\subsection{CU6-DR2 pipeline}
\label{sec:dr2pipe}

The CU6-DR2 pipeline did not include the functionality to `deblend' overlapping windows.  Therefore, the original plan for DR2 was to filter out overlapping windows so the CU6-DR2 pipeline processed non-overlapping windows only.  However, approximately 40\% of the stars with $7 < \grvs < 9$ mag had overlapping windows.  This occurs because of spurious detections around and along the diffraction spikes of sources brighter than about 16 mag in the {\it Gaia}-SkyMapper CCDs \citep{fabricius2016}.  Spurious detection events decrease rapidly with the magnitude of the star, such that at $\grvs \approx 11$ mag, $\approx$5\% of the stars have overlapped windows \citep{sartoretti2018}.   If the spurious detection is brighter than the RVS limiting magnitude\footnote{Prior to June 2015, the RVS limiting magnitude was the onboard magnitude $\grvs = 16.2$ mag.  After June 2015, the RVS limiting magnitude was adapted to the level of the instantaneous straylight in each RVS CCD row, varying from onboard magnitude $\grvs = 15.3 - 16.2$ mag, following the straylight pattern \citep{cropper2018}.}, then it also gets a RVS window.  This spurious detection window is sufficiently close to the bright source's window that they are overlapping each other and so the bright source's flux is distributed in both windows.  They typically truncate each other to have a window AC size of 5 or 6 pixels, as illustrated in \figref[fig:window].  Normally, they are aligned in AL so both truncated windows are rectangular.    

Excluding 40\% of the stars with $7 < \grvs < 9$ mag with overlapping windows would degrade the RVS wavelength calibration \citep{sartoretti2018}.  Thus, it was decided that the CU6-DR2 pipeline should process all rectangular windows, regardless of truncation.  This allowed rectangular windows truncated by spurious detections to be processed, as well as the nominal untruncated, rectangular windows.  They were mostly single source and self-truncated i.e. not contaminated by another source.  Windows were cross-matched with the  working catalogue \citep{fabricius2016} and if found to be spurious were not processed by the CU6-DR2 pipeline.  Therefore, it was not possible at the end of the CU6-DR2 pipeline to verify that a brighter rectangularly truncated window, that had not been filtered out, had been truncated by a window generated by a spurious detection or truncated by a window containing a different source.  All the truncated windows looked at in an offline study were the result of spurious detections but this was not an exhaustive check.  

\subsection{DR2 5932173855446728064}
\label{sec:b19}

The availability of the largest ever number of stars with radial velocities in DR2 was a boon for searching for stars travelling so fast that they can escape from the Milky Way's gravitational potential: `hypervelocity stars'.  The premier hypervelocity star candidate in DR2 had the DR2 source ID 5932173855446728064 \citep{bromley2018,marchetti2019}, hereafter referred to as S1 (Table \ref{table:symbol}).  Its radial velocity of $-614.3 \pm 2.4$ \kms was sufficient on its own to class it as a hypervelocity star.  B19 obtained spectroscopic follow-up of  S1 and found a very different median radial velocity of $-56.5 \pm 5.3$ \kms.  

Using the Gaia Observation Forecasting Tool (GOST)\footnote{\url{https://gaia.esac.esa.int/gost/}}, B19 were able to construct the probable scan angles of the transits that contributed to  S1's radial velocity (B19 fig. 7).  They found all these scans also passed through another star with DR2 source ID 5932173855446724352, hereafter referred to as S2 (Table \ref{table:symbol}).  A disturbing source is more of a problem if all (accepted) observations have nearly the same position angle of the scan, which is the case here.  CU6 required sources to have two or more transits for a radial velocity to be published in DR2 \citep{sartoretti2018}.  Both sources have radial velocities in DR2 suggesting that at least two of their transits have rectangular windows.

The aforementioned 105 or 108 pixels in a RVS macrosample corresponds to approximately 6.2 or 6.4 arcseconds, respectively.  Two sources with angular separations in the AL direction smaller than these values will have RVS windows starting on the same macrosample boundary.  The angular separation between  S1 and  S2 is 4.284 arcsec.  This is a large fraction of the macrosample size so most transits will have their windows starting on different macrosample boundaries.  If the windows are overlapping, they will truncate each other in a non-rectangular way.  Only certain phasings of the positions of these sources relative to the macrosample boundaries will result in transits with their windows starting on the same macrosample boundaries.  If the windows are overlapping, they will truncate each other in a rectangular way i.e. the situation illustrated in \figref[fig:window].  CU6 compared the RVS window positions of transits of  S1 and  S2 to confirm B19's prediction that their windows were overlapping and truncating each other.  
	
Although RVS windows always start on a macrosample boundary, RVS spectra do not generally start at the same AL position because they have different sky positions.  The wavelength scale of each RVS spectrum is determined from the known position of the star from {\it Gaia} astrometry.  Pixels in the AL direction are 0.0589 arcsec long and the dispersion varies from 8.51 \kms pix$^{-1}$ at 847 nm to 8.58 \kms pix$^{-1}$ at 873 nm, which corresponds to 144.5 $-$ 145.7 \kms arcsec$^{-1}$ with a mean of 145.1 \kms arcsec$^{-1}$ \citep{cropper2018}.   B19 realised that the light from each star in a RVS window containing two blended RVS spectra can be offset by 145.1 \kms arcsec$^{-1}$.

The angular separation between  S1 and  S2 corresponds to a radial velocity offset of 619.0$-$624.2 \kms.   S2's DR2 radial velocity is $5.40 \pm 2.85$ \kms.  Subtracting the radial velocity offset from the measured radial velocity gives $-613.6$ to $-618.8$ \kms, which is consistent with  S1's DR2 radial velocity of $-614.3 \pm 2.5$ \kms. Therefore B19 demonstrated that S1's anomalous DR2 radial velocity can be fully explained if the spectra used to determine the radial velocity was dominated by contaminating flux from  S2 in each transit.  

B19 were not able to confirm that S2's flux was dominating S1's spectra.  $\grvs$ magnitudes in general and  S2's $\grp$ magnitude in particular were not published in DR2.  Both  S1 and  S2 had their $G$ magnitudes published in DR2: 13.8 and 13.4 mag, respectively, but these are sufficiently close that a colour difference could make either source the brighter one in $\grvs$.  

Comparing the RVS spectra of the two transits of S1 and  S2 would ideally be done in pixel space to see the alignment of the spectra on each CCD but this information was not persisted by either the DR2 or DR3 versions of the CU6 pipeline.  Transit spectra were persisted from the DR2 pipeline.  These are the three CCD spectra from the transit of the RVS focal plane resampled onto a uniform, barycentric wavelength grid between 846 and 870 nm with 0.025 nm~pixel$^{-1}$.

\begin{figure*}
\centering
\includegraphics[width=0.75\textwidth]{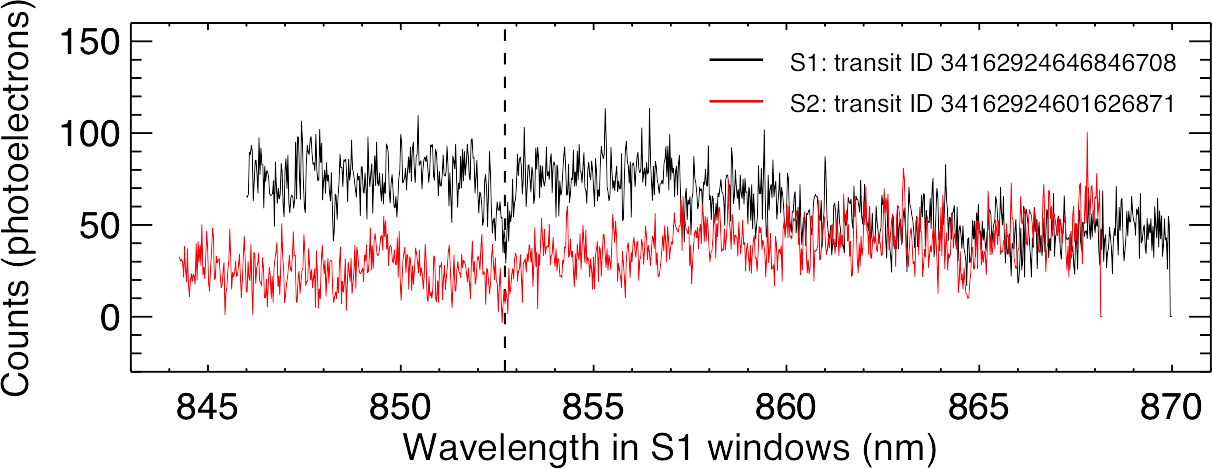}
\hspace{0.5cm}
\includegraphics[width=0.75\textwidth]{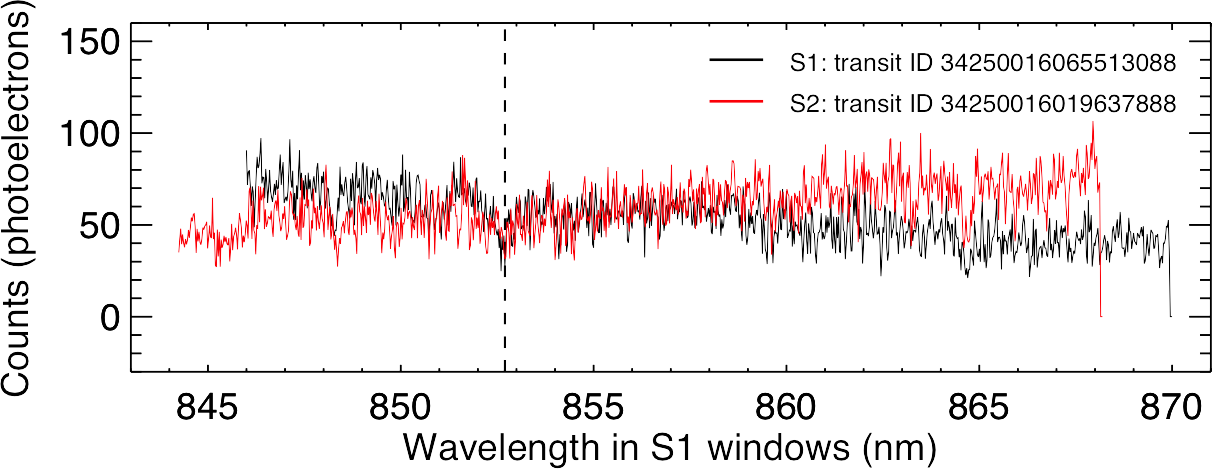}
\caption[]{RVS transit spectra of the two transits of S1 and S2.  The wavelength scale of the S2 spectra has been Doppler shifted to how it would appear in the S1 windows.  The vertical dashed lines delineate the strongest (middle) \ion{Ca}{ii} line seen in both the S1 and S2 spectra. }
\label{fig:transit_spectrum}
\end{figure*}

Doppler shifting the S2 transit spectra by the aforementioned radial velocity offset ($-621.6$ \kms) moves the S2 transit spectra from their own window-specific wavelength scale to the wavelength scale in S1's windows.  This aligns the spectra approximately in pixel space i.e. how S2's flux looks in S1's window.  Figure \ref{fig:transit_spectrum} shows the strongest (middle) \ion{Ca}{ii} lines in the S1 and S2 spectra are aligned with each other in both transits.  B19 have already confirmed that S1's DR2 radial velocity is spurious, which means the absorption lines used to measure its DR2 radial velocity actually came from S2.  This plot was used to confirm B19's findings and was announced on the DR2 Known Issues webpage\footnote{\url{https://www.cosmos.esa.int/web/gaia/dr2-known-issues#RadialVelocitiesCrowdedRegions}}, which was timed to appear when B19 appeared on astro-ph: 28 January 2019.

In the first transit (top plot of \figref[fig:transit_spectrum]), S1's spectrum has more flux than S2's spectrum at the blue end.  This suggests that more of S2's spectrum is in S1's window than is in S2's window.  In the second transit (bottom plot of \figref[fig:transit_spectrum]), S1's spectrum has a similar amount of flux as S2's spectrum at the blue end.  Nevertheless, S2's strongest \ion{Ca}{ii} line is still seen in S1's spectrum.  The flux levels are different between the two transits. This is most likely because they may include straylight, which has not all been removed by the time-independent straylight map used to subtract the background in the CU6-DR2 pipeline \citep{sartoretti2018}.

\citet{katz2019} visually examined the combined spectrum of each DR2 source with an absolute radial velocity larger than 500~\kms, one by one, to check the location of their \ion{Ca}{ii} lines was consistent with their measured radial velocity.   S1 passed this test because the RVS spectra of  S1 Doppler shifted to rest and combined shows \ion{Ca}{ii} lines at rest (\figref[fig:combined_spectrum]).  These \ion{Ca}{ii} lines actually belong to  S2 but are consistent with  S1's contaminated DR2 radial velocity. 

\begin{figure*}
\centering
\includegraphics[width=0.9\columnwidth]{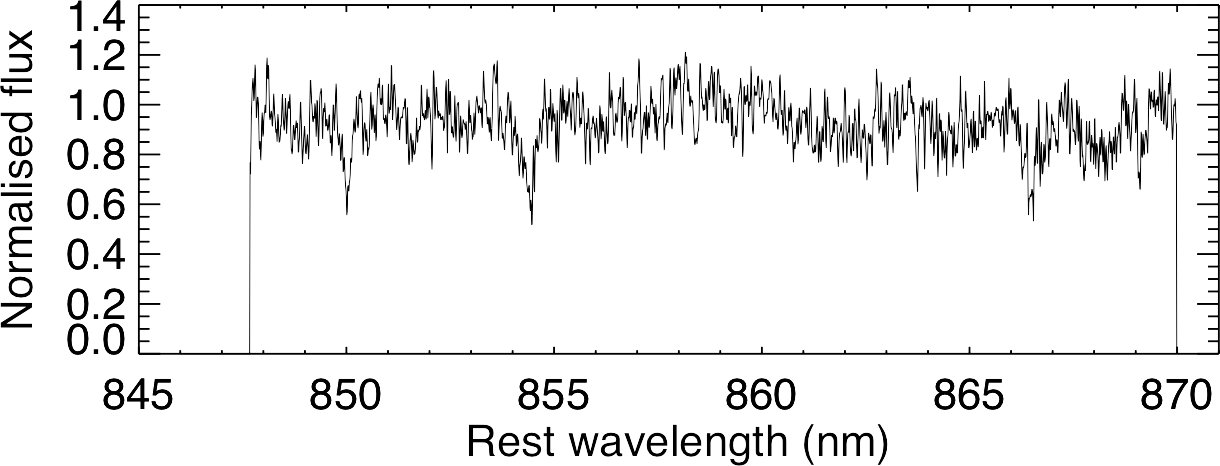}
\hspace{0.5cm}
\includegraphics[width=0.9\columnwidth]{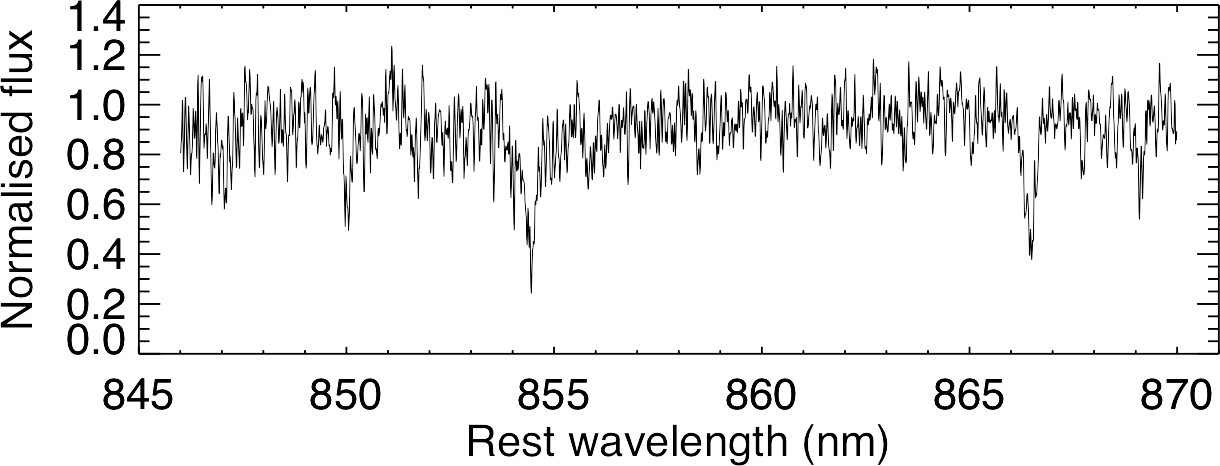}
\includegraphics[width=0.9\columnwidth]{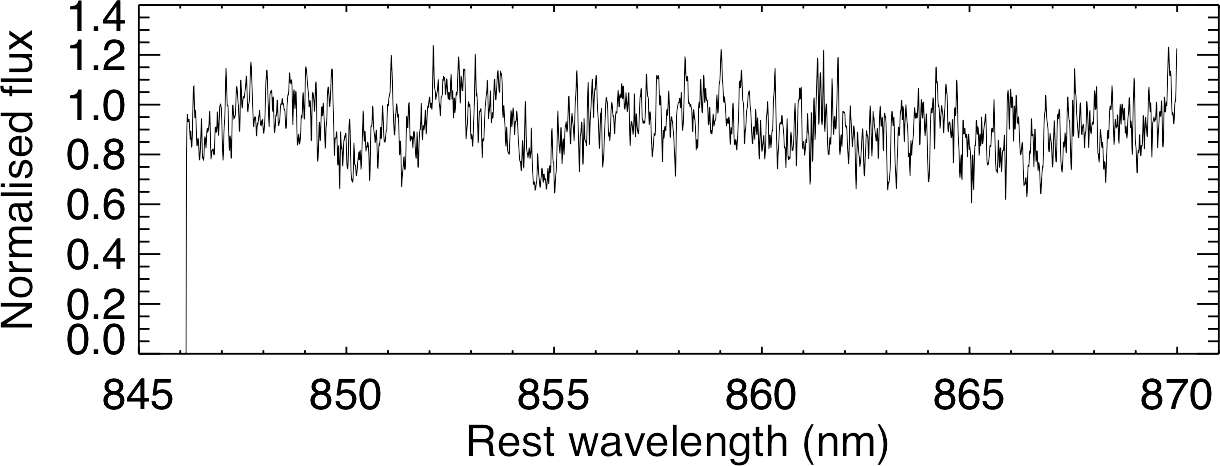}
\hspace{0.5cm}
\includegraphics[width=0.9\columnwidth]{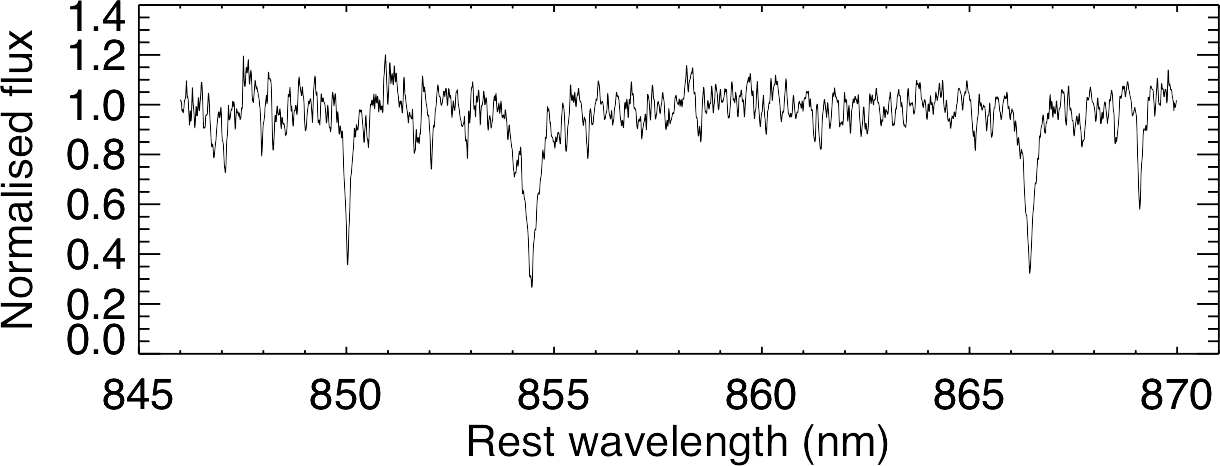}
\caption[]{Normalised RVS spectra, Doppler shifted to rest, resampled and combined into a single spectrum:  S1 (left panels),  S2 (right panels) from the CU6 pipelines: DR2 (top panels) and DR3 practice run (bottom panels) to show how deblending reveals S1's uncontaminated spectrum in the bottom left panel.  The number of combined CCD spectra are 21 (top left and top right), 26 (bottom left) and 43 (bottom right).}
\label{fig:combined_spectrum}
\end{figure*}

\subsection{\citet{boubert2019} list}
\label{sec:b19list}

B19 used GOST to assess the potential contamination of  S1 but doing this for each  DR2 source with a radial velocity is not feasible.  Therefore to identify cases like  S1, B19 searched for DR2 sources with a radial velocity and have a companion in the full  DR2 catalogue within 6.4 arcsec that either itself has a radial velocity or that is brighter in $\grp$ or $G$ magnitudes.  The resulting list of 70,365  DR2 sources with potentially contaminated radial velocities was made publicly available with B19.  The majority of DR2 hypervelocity candidates are in the B19 list, including  S1 (\figref[fig:b19hist]).  The fraction of stars in DR2 with radial velocities that are also in the B19 list is greater at the faint end than the bright end (\figref[fig:b19hist_g]). The B19 list has a similar sky distribution to the stellar density of DR2 sources with radial velocities (cf. \citealt{katz2019} fig. 4) because chance alignment of windows is most likely to occur in crowded regions.

\begin{figure}
\centering
\includegraphics[width=\columnwidth]{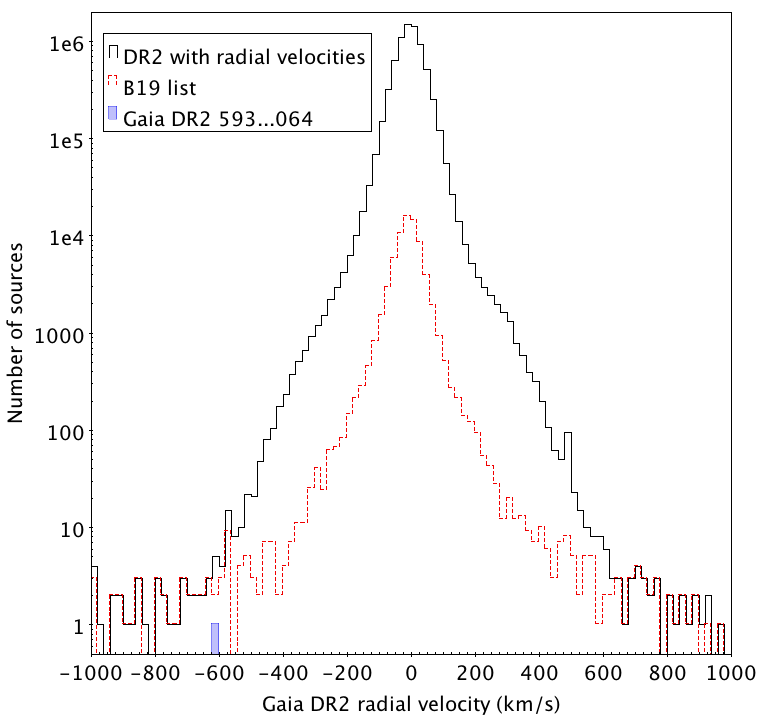}
\caption[]{Histogram of all the DR2 radial velocities per 25 \kms interval with the B19 list and  S1 overlaid.}
\label{fig:b19hist}
\end{figure}

\begin{figure}
\centering
\includegraphics[width=\columnwidth]{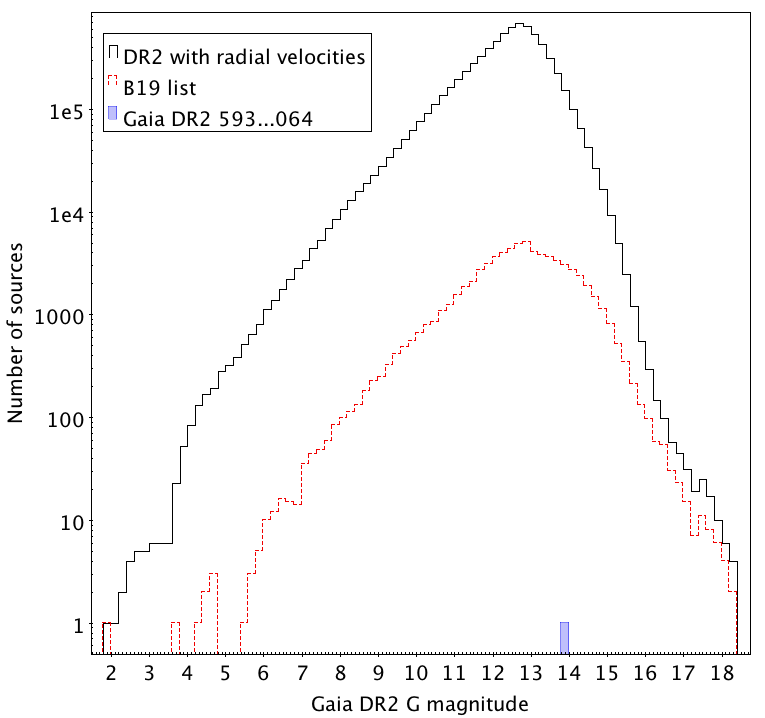}
\caption[]{Histogram of DR2 magnitudes of  DR2 sources with radial velocities per 0.2 $G$ mag intervals with the B19 list and  S1 overlaid.}
\label{fig:b19hist_g}
\end{figure}

\subsection{CU6-DR3 pipeline}
\label{sec:dr3pipe}

Unlike the DR2 version, the CU6-DR3 pipeline \emph{does} include the functionality to `deblend' overlapping windows.  The fluxes in all the overlapping windows are simultaneously `deblended' by modelling the contribution of each source to each window using the AC line spread function (LSF) and AC peak position as a function of AL:  

\begin{equation}
S_\mathrm{W} = I_\mathrm{N}M_\mathrm{WN} 
\label{equ:deblend1}
\end{equation}

\noindent where $S_\mathrm{W}$ is the total signal in one AL sample in window W, $I_\mathrm{N}$ is the integrated signal under the AC profile of source N and $M_\mathrm{WN}$ is the fraction of signal from source N contributing to window M.  Applying equation \ref{equ:deblend1} to \figref[fig:deblend]:

\begin{equation}
\begin{split}
S_\mathrm{1} &= I_\mathrm{1}M_\mathrm{11} + I_\mathrm{2}M_\mathrm{12}\\
S_\mathrm{2} &= I_\mathrm{1}M_\mathrm{21} + I_\mathrm{2}M_\mathrm{22} + I_\mathrm{3}M_\mathrm{23}\\
S_\mathrm{3} &= I_\mathrm{2}M_\mathrm{22} + I_\mathrm{3}M_\mathrm{32}
\end{split}
\label{equ:deblend2}
\end{equation}
 
\noindent where $S_\mathrm{1}$, $S_\mathrm{2}$ and $S_\mathrm{3}$ are the total signal in one AL sample of the left, middle and right windows, respectively, in \figref[fig:deblend].

In matrix notation, equation \ref{equ:deblend1} becomes ${\bf S = I.M}$, where ${\bf S}$ are the measured, blended fluxes and ${\bf I}$ are the deblended fluxes, which are solved for by inverting the matrix.  Deblending will be fully described and validated in Seabroke et al. (2022, in prep. for DR3).

\begin{figure}
\centering
\includegraphics[width=\columnwidth]{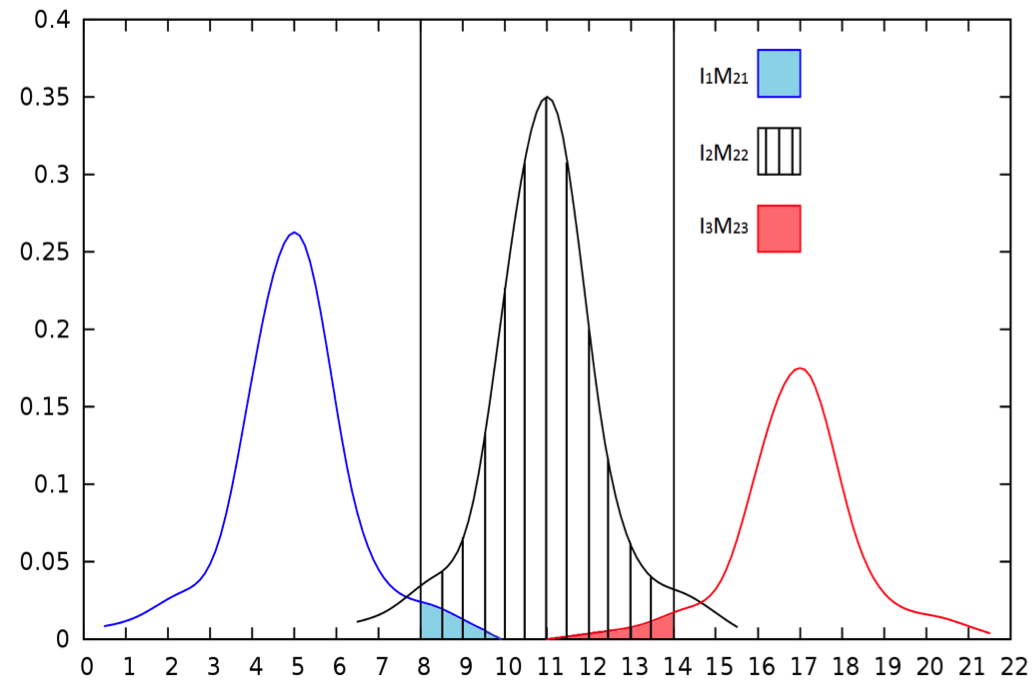}
\caption[]{Illustration of the flux profile (in arbitrary flux units) in one AL sample in three RVS windows centred on three sources, as a function of AC pixel.}
\label{fig:deblend}
\end{figure}

There were two runs of the CU6-DR3 pipeline.  A practice run and the operational run, from which CU6 data products will be published after validation.  The practice run $\grvs$ magnitudes of  S1 and  S2 were 12.7 and 12.0 mag, respectively, confirming that  S2 is brighter and that its spectra are capable of dominating  S1's spectra.  

 S1's practice run radial velocity was based on 11 transits, all of which were blended with other windows.  Six of them have spectra with more than half their length blended.  All these windows were successfully deblended by the CU6-DR3 pipeline.  \figref[fig:combined_spectrum] shows the contaminating flux from  S2, that was present in the DR2-combined spectrum (top left panel), has been removed from the DR3 practice run combined spectrum (bottom left panel).  The latter is noisier and without the sharp \ion{Ca}{ii} lines of the former.  This resulted in  S1's DR3 practice run radial velocity,  $-65.2 \pm 11.5$ \kms.  It is consistent with the B19 value of $-56.5 \pm 5.3$ \kms, validating deblending at least in this case.

There is no overlap in the transit IDs between the 11 DR3 practice run transits and the seven DR2 transits.  The former are all non-rectangular, truncated windows and the latter are all rectangular, truncated windows.  The latter were all excluded from the DR3 practice run radial velocity determination because they could not be deblended successfully.  This was most likely the result of the track of the AC peaks of the spectra (illustrated in \figref[fig:window]) of  S1 and  S2 being too close for the deblend algorithm to yield a numerically stable solution.

The CU6-DR3 pipeline operational run did not output a radial velocity for  S1.  This was because every  S1 window was found to have one or more point background sources within 3 magnitudes of  S1.  All these windows were excluded to prevent potential contamination.  This functionality was not fully switched on in the DR3 practice run.  

{More transits and more functionality preventing contaminating flux means that the DR3 radial velocities should be more reliable than the DR2 ones (Sect. \ref{sec:rv}), such that the complication found by B19 should not be repeated in DR3, and these can be used to check the DR2 ones.  Although the DR3 radial velocities have their own features that are being examined during validation, we are looking for large radial velocity differences between DR2 and DR3.  For example, although  S1 is not in DR3, comparing its DR3 practice run radial velocity ($-65.2$ \kms) with its DR2 value ($-614.3$ \kms) reveals that the DR2 value is incorrect.   

\section{Method}

\subsection{Overview}

\begin{figure*}
\centering
\includegraphics[width=\textwidth]{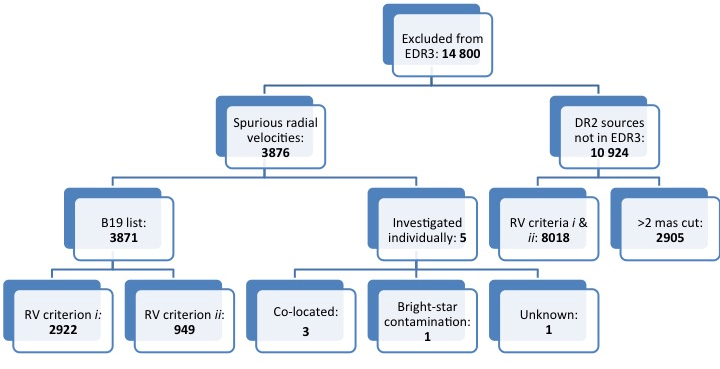}
\caption[]{Flow chart of why sources with a radial velocity (RV) in DR2 do not have radial velocities in EDR3.} 
\label{fig:summary}
\end{figure*}

\begin{table}
\caption{Summary of the radial velocity (RV) criteria used to compare DR2 and DR3.}             
\label{table:criteria}      
\centering          
\begin{tabular}{ll}     % 7 columns 
\hline\hline 
Criteriion & Description \\
\hline
$i$ & Reject if no DR3 RV\\
$ii_a$ & Keep if $|\textrm{RV}_{\textrm{DR2}} - \textrm{RV}_{\textrm{DR3}}| \leq 4$ \kms\\
$ii_b$ & Reject if $|\textrm{RV}_{\textrm{DR2}} - \textrm{RV}_{\textrm{DR3}}| > 25$ \kms\\
$ii_c$ & Reject if $ 4 < |\textrm{RV}_{\textrm{DR2}} - \textrm{RV}_{\textrm{DR3}}| \leq 25$\\
           & $\&$~$|\textrm{RV}_{\textrm{DR2}} - \textrm{RV}_{\textrm{DR3}}|/\sigma_{\textrm{RV,diff}} > 5$ \\
\hline             
\end{tabular}
\end{table}

Figure \ref{fig:summary} provides an overview of the methods used to remove DR2 radial velocities from EDR3, which is described in more detail in the following sections.  The primary method is comparing DR2 radial velocities to unpublished, preliminary DR3 radial velocities.  The radial velocity criteria are summarised in Table \ref{table:criteria} and described in detail in Sect. \ref{sec:rv}.  The criteria are developed to identify erroneous DR2 radial velocities, such as the one found by B19.  If there is not a DR3 radial velocity (2922 sources, Sect. \ref{sec:i}: radial velocity criterion $i$) or the radial velocity difference between DR2 and DR3 significantly differs (949 sources, Sect. \ref{sec:ii}: radial velocity criterion $ii$), it is assumed that the DR2 radial velocity is the one in error.  There are three advantages to using DR3 radial velocities.  Firstly and crucially, blended windows were not deblended in the DR2 pipeline but they are in the DR3 pipeline.  Secondly, the survey duration is increased from 1.8 years (DR2) to 2.8  years (DR3).  Additionally, the DR3 radial velocities are derived from an upgraded pipeline compared to DR2.   

EDR3 is based on a new astrometric solution and a new source list, which means sources in DR2 may not be in EDR3.  Proper motions and epoch position propagation are used to match sources with radial velocities in DR2 to EDR3 sources.  This is the other method used in this work, which is described in detail in Sect. \ref{Sect:Xmatch}.  It results in 10\,924 DR2 sources with radial velocities not being matched to an EDR3 source and so these radial velocities are not able to be included in EDR3 (summarised in \figref[fig:summary]).   This is not because of the quality of the radial velocities.  Their DR2 astrometry (and thus radial velocity) is based on one component of a close binary pair and their (E)DR3 astrometry (and thus radial velocity) is based on the other component of the close binary pair.   If the radial velocities significantly differ (radial velocity criteria $i$ and $ii$), it suggests that DR2 astrometry and EDR3 astrometry are measuring different components of the binary and so these sources are excluded from EDR3 (8018 in total).  If the radial velocities agree, either DR2 astrometry and EDR3 astrometry are measuring the same component of the binary or the different components have similar radial velocities.  The latter cases have their DR2 radial velocities excluded from EDR3 by identifying them with greater than 2 mas separation (see Sect. \ref{Sect:source} for more details).

\subsection{Cross-matching sources with radial velocities in DR2 and sources in EDR3}
\label{Sect:Xmatch}

As explained in~\cite{torra2021}, the identifier of a specific object can change between subsequent data releases. As a consequence, the assignment of DR2 radial velocities to  EDR3 objects cannot be based exclusively on their  DR2 source identifier, as this could lead to erroneous associations between the radial velocity and the source. The general recommendation from DPAC in that respect is to use a positional cross-match between the two catalogues and a global  DR2 to EDR3 cone search neighbourhood is provided together with  EDR3 \citep{brown2021}. This global cross-match uses a 2 arcsec search radius, which provides several possible matches to a given  DR2 source within this radius. For the specific purpose of the  DR2 radial velocity propagation to  EDR3, a similar cross-match (limited to the sources holding radial velocities) was performed in order to select one unique and best possible source match to each relevant source.

For this purpose, a search radius of 85 mas was used to ensure one unique match to each  DR2 source. Prior to cross-matching, epoch propagation was applied from the  DR2 epoch (2015.5) to the  EDR3 epoch (2016.0) where  DR2 proper motions were available. Once the matches were identified, the source separation at the epoch of  EDR3 was computed. Following on this, the epoch propagation was then performed on the  EDR3 matches from 2016.0 to 2015.5 in order also to obtain a measure of the source separations at the epoch of  DR2. Based on this cross-match, there was no match for a little fewer than 2000 sources and an additional cross-match with a larger radius of 2000 mas was performed. When several matches are possible, the one with the smallest separation is kept.

Overall, matches for 7\,224\,630 sources having a radial velocity in  DR2 are assigned through this cross-match. This is one source less than the corresponding number in  DR2.  The missing source is Polaris Ab (DR2 576402619921505664), the fainter component of Polaris' close binary pair, because Polaris Ab is not in EDR3 at all, which is explained below.
Polaris Ab is not in the B19 list so the entire B19 list and high-velocity stars not in the B19 list have (E)DR3 source IDs, which are used in Sect. \ref{sec:rv} and \ref{sec:hvs_dr3}.

\begin{figure}
\centering
\includegraphics[width=\columnwidth]{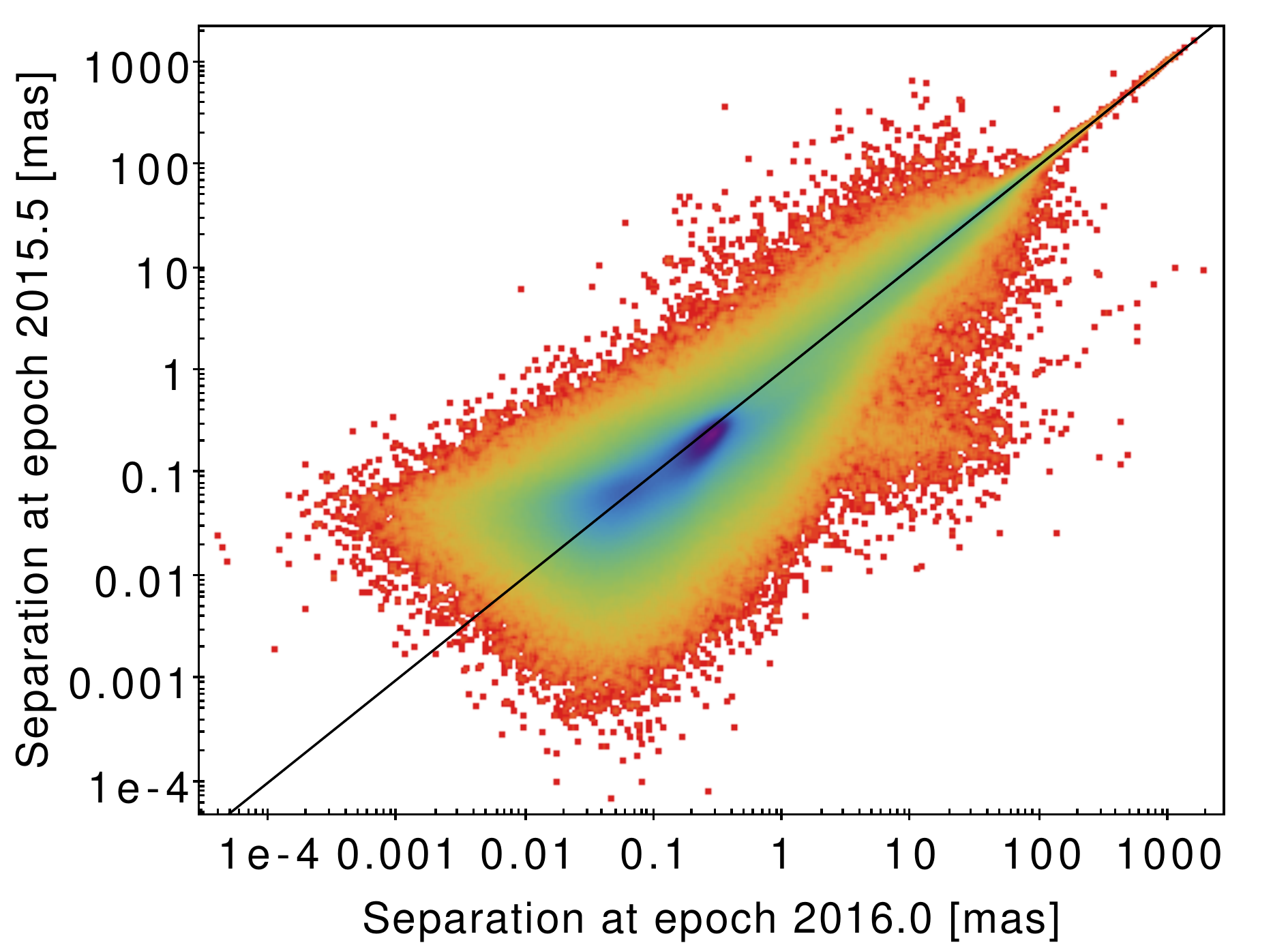}
\caption[]{Match distances for 7\,224\,630 sources calculated
using either EDR3 proper motions (epoch 2015.5 on the ordinate) or DR2 proper motions
(epoch 2016.0 on the abscissa).  Blue colours denote the densest regions.  Red squares denote individual sources.}
\label{fig:sep_sep}
\end{figure}

The match distances at the two epochs are illustrated in Fig.~\ref{fig:sep_sep}.
The vast majority of sources have matches at distances well below 1~mas, as
could be expected from bright sources where more than 99\% have proper motions
in both catalogues. The sharp diagonal at large separations corresponds to
cases without proper motion in either catalogue, and the small group with
small distances at the DR2 epoch but larger at the EDR3 epoch corresponds to sources
that had no proper motions in DR2 but do have them now.

In order to understand why some sources have match distances of
10--100~mas, Fig.~\ref{fig:ipd_multi_sep} shows the percentage of transits for each
source having more than one image in the astrometric field. For the vast
majority this fraction is 0--1\%, but it rises sharply as soon as match
distances are larger than a few tenths of mas, and for distances above
some 10~mas almost all sources have clear signs of duplicity (close binary pairs).
In EDR3, secondary images are better suppressed \citep{lindegren2021} than in DR2 \citep{lindegren2018}
resulting in a slightly different astrometric solution.  

Polaris Ab is a secondary component and its absence from EDR3 is likely as a result of its suppression in Polaris' EDR3 astrometric solution.  Polaris Aa is missing from both DR2 and EDR3 because its brightness varies between 1.86--2.13 in $V$ magnitude \citep{samus2009} and {\it Gaia}'s detection efficiency starts to drop at $V \approx G \approx 3$ mag because of saturation that is too strong \citep{gaia2016}.

\begin{figure}
\centering
\includegraphics[width=\columnwidth]{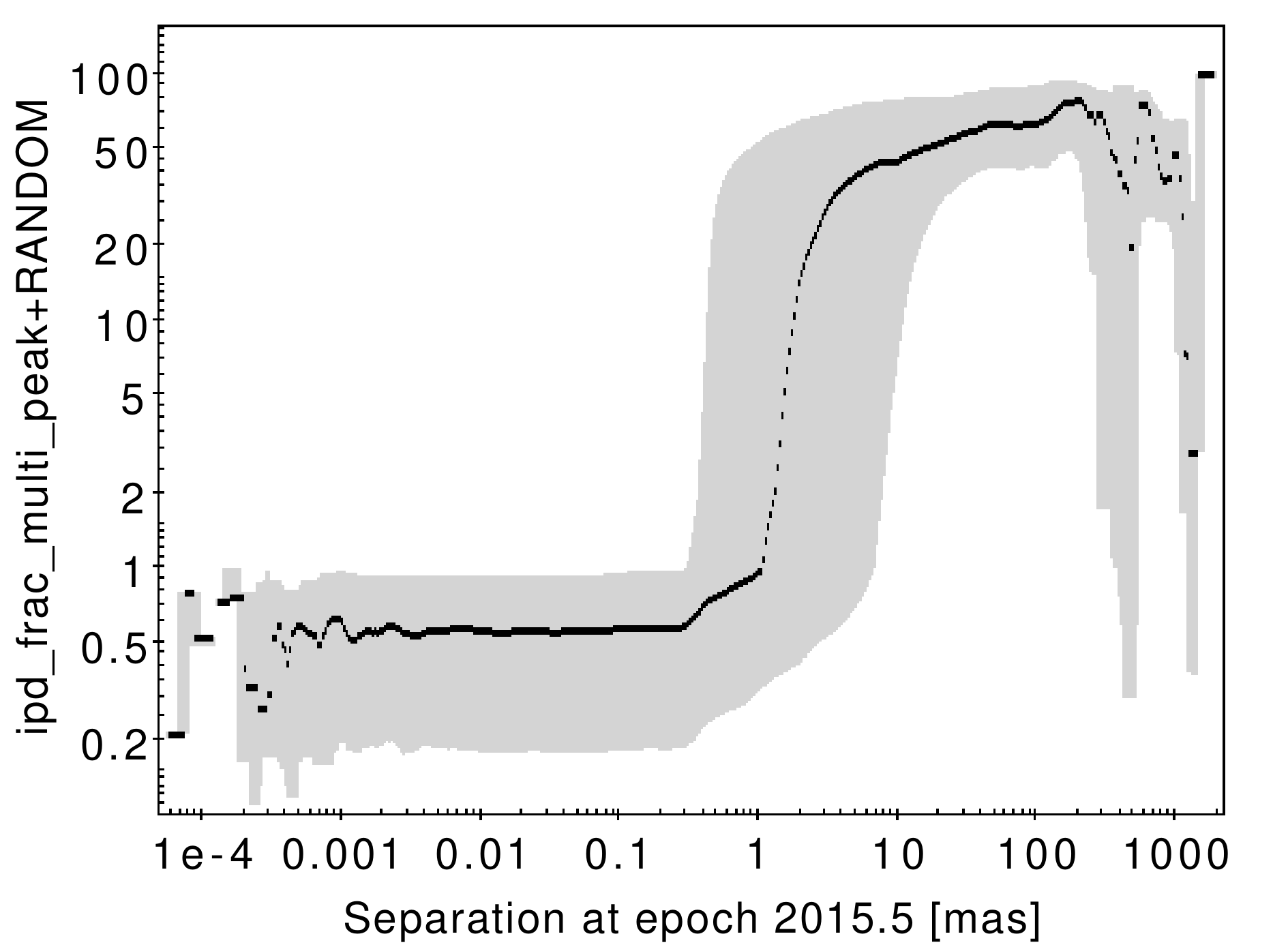}
\caption[]{Percentage of transits for each source having more than one image in
the astrometric field as a function of the match distance, calculated
with proper motions from EDR3. A random number in the range 0--1 was added
because \url{ipd_frac_multi_peak} is stored as an integer.  The black line is a running median and the shaded region delineates the 16th-84th percentiles.} 
\label{fig:ipd_multi_sep}
\end{figure}
}

\subsection{Comparing DR2 and DR3 radial velocities: B19 list}
\label{sec:rv}

\begin{figure}
\centering
\includegraphics[width=\columnwidth]{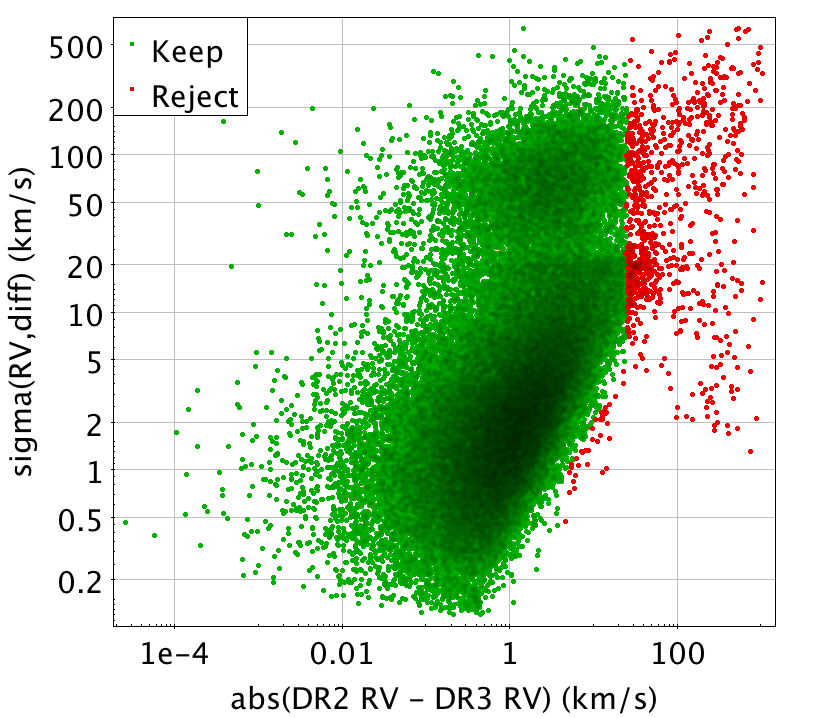}
\caption[]{Comparison of DR2 and DR3 radial velocity and uncertainties: $\sigma_{\textrm{RV,diff}}$ is defined in Equation \ref{equ:rv_diff}. The red dots with $ 4 < |\textrm{RV}_{\textrm{DR2}} - \textrm{RV}_{\textrm{DR3}}| \leq 25$ \kms are selected by Equation \ref{equ:rv_diff2}.} 
\label{fig:rvError_vs_absDiff}
\end{figure}

\subsubsection{Radial velocity criterion $i$: missing DR3 radial velocities}
\label{sec:i}

 DR2 radial velocities are the median of the individual transit radial velocities.  The  CU6-DR3 pipeline uses two methods: the aforementioned DR2 one and it also combines the transit cross-correlation functions from which it measures the radial velocity.  Requiring both of these methods to produce valid measurements gives 37\,499\,608 DR3 sources with radial velocities.  Matching the  DR2 source IDs with the (E)DR3 source IDs in the B19 list (Sect. \ref{Sect:Xmatch})  finds 2922 (out of 70\,365, 4\%) that are missing a  DR3 radial velocity, including  S1 (as explained in Sect. \ref{sec:dr3pipe}).   It is suspicious that a source is in the B19 list and it is doubly suspicious that the source does not have a  DR3 radial velocity.  Therefore, it was decided to exclude the radial velocity of these sources from  EDR3 (hereafter referred to as radial velocity criterion $i$).  It could be that the  DR2 radial velocity is correct and the complication is with the  DR3 CU6 pipeline, but, given the small number of these sources, this decision was considered acceptable.

\subsubsection{Radial velocity criterion $ii$: comparing DR2 and DR3 radial velocities}
\label{sec:ii}

67\,433 (out of 70\,365, 96\% of the B19 sources) have radial velocities in both  DR2 and  DR3.  A radial velocity comparison was made on the median of the transit radial velocities from both pipelines.  All the sources with an absolute radial velocity difference of less than 4 \kms~had their radial velocities kept in EDR3 (hereafter referred to as radial velocity criterion $ii_a$).  This is because there may be systematic differences in the radial velocities in  DR2 and  DR3 that are not taken into account by their random uncertainties.  In  DR2, the typical cool star radial velocity precision is 0.5 \kms (\citealt{katz2019} fig. 19).  Assuming a similar precision in  DR3 (which is conservative), this multiplies 0.5 \kms~by\,$\sqrt{2}$ to give the 1-sigma combined random uncertainty as 0.7 \kms.  The 5-sigma combined random uncertainty is about 3.5 \kms, rounded up to 4 \kms to take potential systematics (e.g. wavelength, template mismatch, etc.) into account.

As an additional guiding parameter, we used the uncertainty on the radial velocity difference between DR2 and DR3 ($\sigma_{\textrm{RV,diff}}$) given by:

\begin{equation}
\sigma_{\textrm{RV,diff}} = \sqrt{\sigma_{\textrm{RV,DR2}}^2  + \sigma_{\textrm{RV,DR3}}^2},
\label{equ:rv_diff}
\end{equation}

\noindent where $\sigma_{\textrm{RV,DR2}}$ is the radial velocity uncertainty from DR2, calculated using \citet{sartoretti2018} equation 19, and $\sigma_{\textrm{RV,DR3}}$ is the radial velocity uncertainty from DR3, calculated using \citet{sartoretti2018} equation 18, i.e. before the DR3 calibration floor had been calculated. $\sigma_{\textrm{RV,DR2}}$ was filtered to be $<$20 \kms~in DR2 but an equivalent filter has not yet been applied to DR3.  Consequently, \figref[fig:rvError_vs_absDiff] shows that $\sigma_{\textrm{RV,diff}}$ extends to large values.  It was decided to keep the radial velocities in EDR3 of all the sources with an absolute radial velocity difference of less than 4 \kms, regardless of $\sigma_{\textrm{RV,diff}}$ because outlying radial velocity values (caused by uncorrected instrumental effects such as cosmic rays) can inflate $\sigma_{\textrm{RV,DR3}}$ without invalidating the DR3 radial velocity itself, especially if it is in good agreement with the DR2 radial velocity.

It was decided to remove from  EDR3 the radial velocities  of all the sources with an absolute radial velocity difference of greater than 25 \kms (hereafter referred to as radial velocity criterion $ii_b$).  This prevents large radial velocity uncertainties, caused by outliers (owing to uncorrected instrumental effects), allowing an agreement within the uncertainties between DR2 and DR3.  The worst  DR2 radial velocity precision is obtained for the template $T_{\textrm{eff}} = 6500$~K stars and it is 3.5~\kms (\citealt{katz2019} fig. 19). Assuming a similar precision in  DR3 (again conservative), this multiplies 3.5~\kms~by $\sqrt{2}$ to give the 1-sigma combined random uncertainty as 4.9~\kms.  The 5-sigma combined random uncertainty is about 24.7~\kms, rounded up to 25~\kms, again to take potential systematics into account.

For absolute radial velocity differences between 4 and 25~\kms, it was decided to remove  the radial velocities from  EDR3 of all the sources with

\begin{equation}
|\textrm{RV}_{\textrm{DR2}} - \textrm{RV}_{\textrm{DR3}}|/\sigma_{\textrm{RV,diff}} > 5
\label{equ:rv_diff2}
\end{equation}

\noindent (hereafter referred to as radial velocity criterion $ii_c$).    Applying radial velocity criteria $ii_a$, $ii_b$ and $ii_c$ (collectively referred to as radial velocity criterion $ii$), removes the radial velocities of 949 sources out of 67\,433 (1\%) that have radial velocities in both  DR2 and  DR3 (\figref[fig:rvError_vs_absDiff]).

\subsubsection{Combining radial velocity criteria $i$ and $ii$}

\begin{figure}
\centering
\includegraphics[width=\columnwidth]{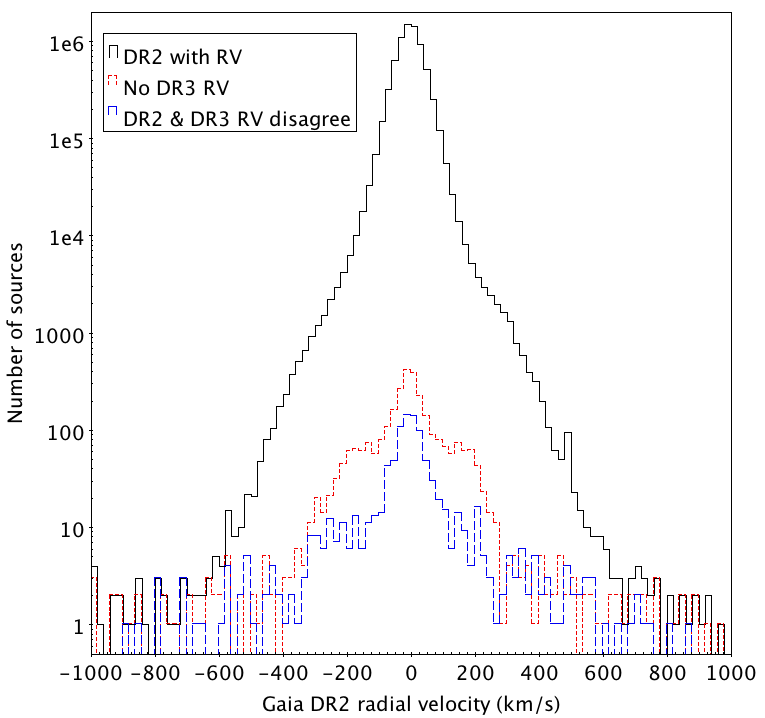}
\caption[]{Histogram of all the  DR2 radial velocities.   The distribution of sources with radial velocities missing in DR3 (radial velocity criterion $i$) and the distribution of sources that failed radial velocity criterion $ii$ are overlaid.}
\label{fig:hist_reject}
\end{figure}

\begin{figure}
\centering
\includegraphics[width=\columnwidth]{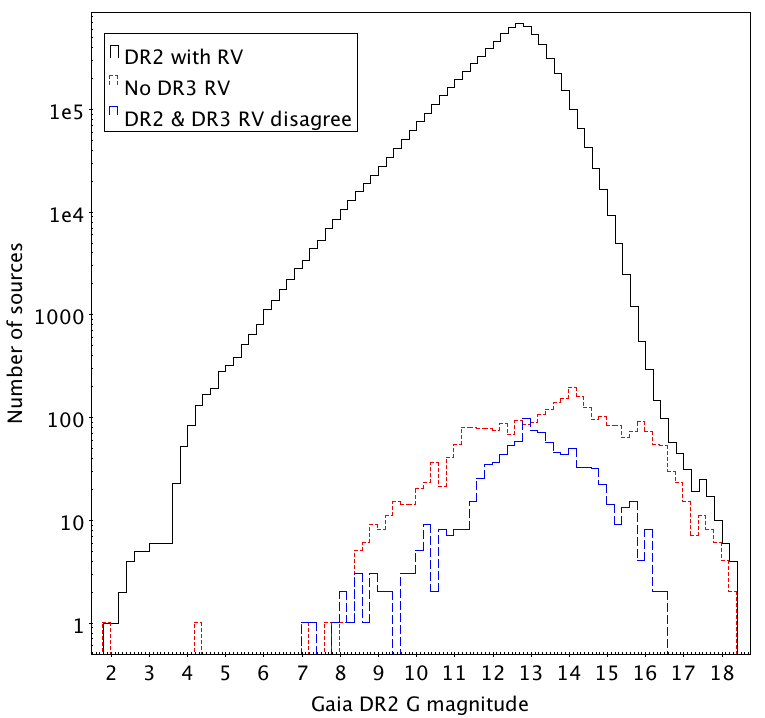}
\caption[]{Histogram of  DR2 magnitudes of  DR2 sources with radial velocities.  The distribution of sources with radial velocities missing in DR3 (radial velocity criterion $i$) and the distribution of sources that failed radial velocity criteria $ii$ are overlaid.}
\label{fig:b19hist_g_reject}
\end{figure}

Combining radial velocity criteria $i$ and $ii$ removes 3871 DR2 radial velocities  from EDR3. Figure \ref{fig:hist_reject} shows that both of these exclusion criteria remove the majority of the hypervelocity star candidates. Figure \ref{fig:b19hist_g_reject} reveals that the distribution of sources that failed criteria $i$ ({\it No DR3 RV}) reaches fainter than the distribution of sources that failed criteria $ii$ ({\it DR2 \& DR3 RV disagree}).  Even though the CU6-DR2 pipeline selection was based on $G_{\mathrm{RVS}} < 12$ mag \citep{sartoretti2018}, there are sources with a DR2 radial velocity that were considered too faint to be selected by the CU6-DR3 pipeline ($G_{\mathrm{RVS}} < 14$ mag, Sartoretti in prep.), hence they fail criteria $i$ ({\it No DR3 RV}).  This is because of the difference between the photometry used to select sources to process in the CU6 DR2 and DR3 pipelines. 

The CU6-DR2 pipeline used external $G_{\mathrm{RVS}}$ ($G_{\mathrm{RVS}}^{\mathrm{ext,1}}$) from the Initial \textit{Gaia} Source List (IGSL, \citealt{smart2014}) to select sources with $G_{\mathrm{RVS}}^{\mathrm{ext,1}} < 12$ mag.    The CU6-DR3 pipeline uses external $G_{\mathrm{RVS}}$ ($G_{\mathrm{RVS}}^{\mathrm{ext,2}}$) derived from preliminary EDR3 photometry, $G$ and $G_{\mathrm{RP}}$, using \citet{gaiabrown2018} equations 2 and 3, to select sources with $G_{\mathrm{RVS}}^{\mathrm{ext,2}} < 14$ mag.  $G_{\mathrm{RVS}}^{\mathrm{ext,2}}$ was not available at the start of the CU6-DR2 pipeline processing, even using DR2 photometry, hence the need for $G_{\mathrm{RVS}}^{\mathrm{ext,1}}$.  The reasons for $G_{\mathrm{RVS}}^{\mathrm{ext,1}}$ being much brighter than $G_{\mathrm{RVS}}^{\mathrm{ext,2}}$ are explored in Sect. \ref{sec:hvs}.
 
Because B19 searched around 6.4 arcsec around each DR2 source with a radial velocity, the sky distribution of their list of potentially contaminated DR2 radial velocities traces stellar density.  In particular, the Ophiuchus cloud complex is seen as an under-density of stars, starting in the Galactic plane at $l\approx30^{\circ}$ and extending above the Galactic centre.  The sky distribution of sources with radial velocities missing in DR3 (\figref[fig:rejected_aitoff]) delineates fewer stellar density features.  This is presumably where the window density is too high for blended windows to be successfully deblended.

\begin{figure}
\centering
\includegraphics[width=\columnwidth]{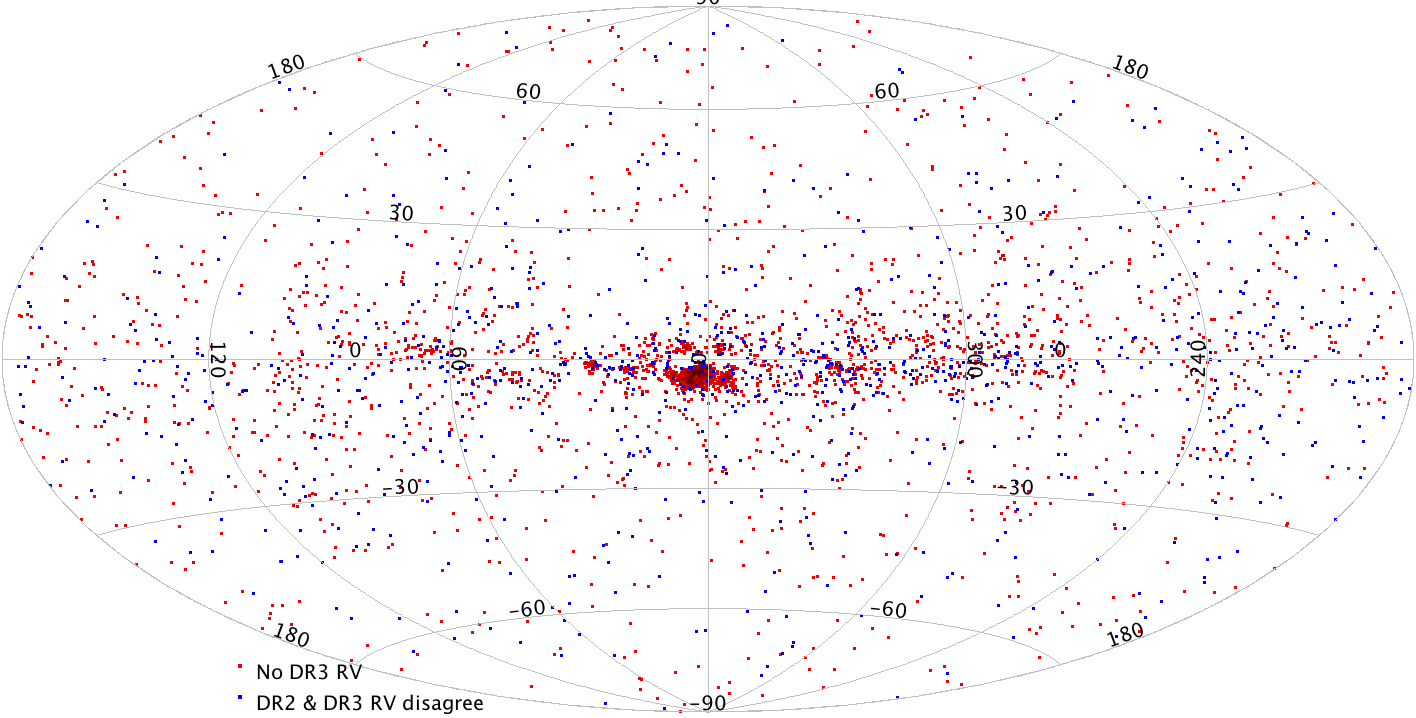}
\caption[]{Galactic Aitoff projection of the sources with radial velocities missing in  DR3 (radial velocity criterion $i$) and the distribution of sources that failed radial velocity criteria $ii$.  The large red over-density south of the Galactic centre is coincident with the Large Sagittarius Star Cloud, which contains Baade's Window.  The small red over-density at $(l,b) = (27,-3)^{\circ}$ is coincident with the Scutum Star Cloud.  Two red over-densities are also visible north of the Galactic centre in the Ophiuchus region.}
\label{fig:rejected_aitoff}
\end{figure}

\subsection{Comparing DR2 and DR3 radial velocities: high-velocity stars not in B19 list}
\label{sec:hvs_dr3}

It was decided to limit the comparison between DR2 and DR3 radial velocities to the B19 list, where the reliability is already in question (Sect. \ref{sec:rv}), and to high-velocity stars (this section), where the B19 complication was originally found.  We use the same definition of high velocity as \citet{katz2019}: absolute DR2 radial velocity greater than 500 \kms.

\begin{figure}
\centering
\includegraphics[width=\columnwidth]{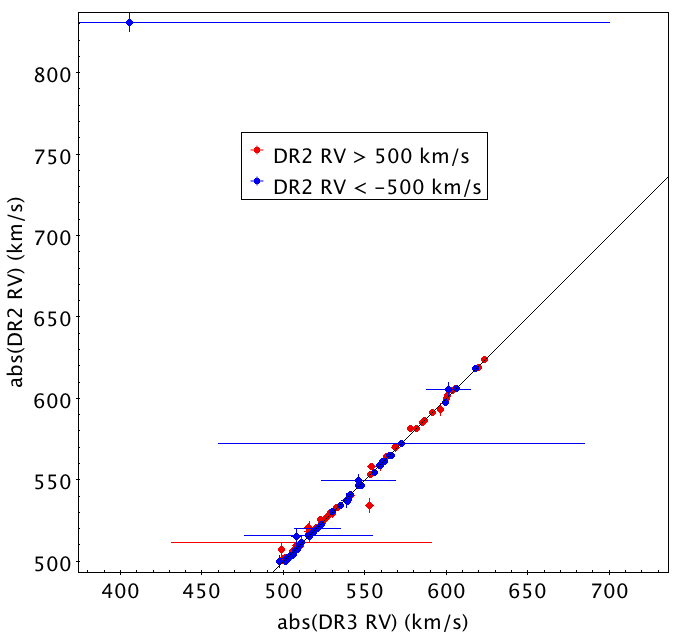}
\includegraphics[width=\columnwidth]{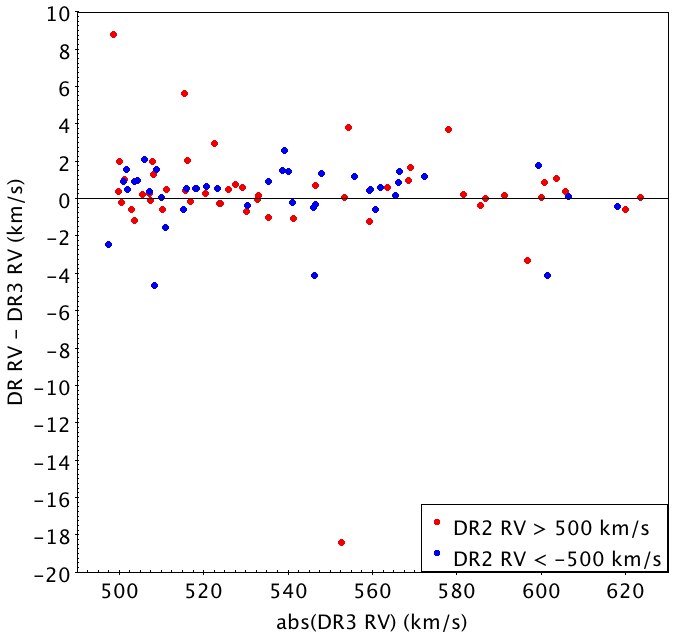}
\caption[]{Comparing the DR2 and DR3 radial velocities of the high-velocity stars not in the B19 list.  The outlier in the top left corner of the upper plot is not plotted in the lower plot.}
\label{fig:dr2_v_dr3}
\end{figure}

 \figref[fig:dr2_v_dr3] shows there is good agreement between the DR2 and DR3 radial velocities of the high-velocity stars not in the B19 list.  Not all sources agree.  This is likely to be because of multiplicity, which we are not investigating in this paper.  We are looking for larger differences indicative of contamination and there is one such star, which is not shown in the top panel of  \figref[fig:dr2_v_dr3].  It is DR2 5305975869928712320: its radial velocity in DR2 is $-830.6 \pm 5.6$ \kms~(\figref[fig:hist_rv_remaining]) but its CU6-DR3 pipeline value is $-405.4 \pm 294.6$ \kms.  
  
 There are five sources with absolute DR2 radial velocities greater than 500 \kms~and not in the B19 list, that do not have a DR3 radial velocity.  Their DR2 radial velocities are also displayed in \figref[fig:hist_rv_remaining].  These five sources and the aforementioned DR2 5305975869928712320 are investigated collectively in Sect. \ref{sec:hvs} and individually in Appendix \ref{sec:appendix_hvs}.  
 
\begin{figure}
\centering
\includegraphics[width=\columnwidth]{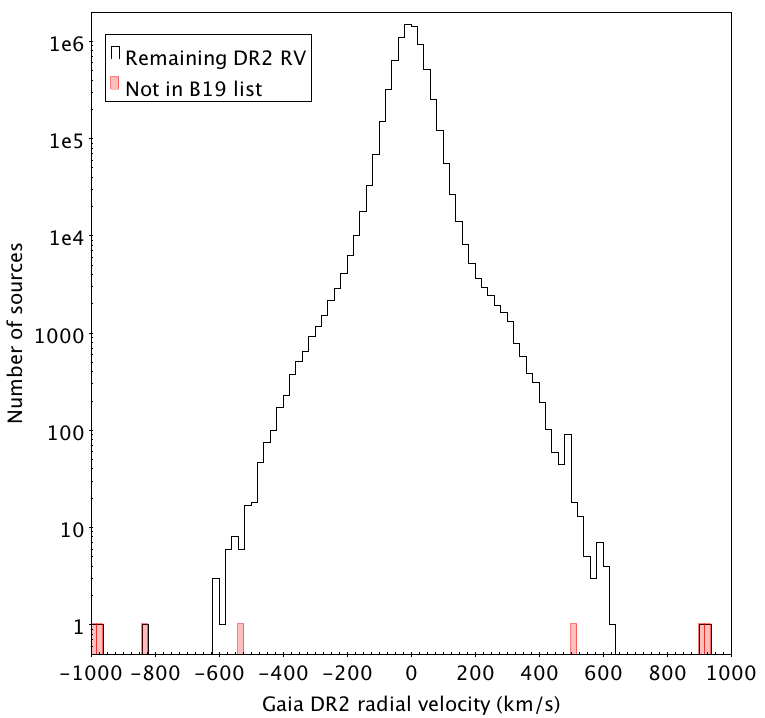}
\caption[]{Histogram of remaining  DR2 radial velocities with the distribution of the sources not in B19 and with absolute radial velocities greater than 500 \kms overlaid.}
\label{fig:hist_rv_remaining}
\end{figure}

\subsection{Investigating individual high-velocity stars}
\label{sec:hvs}

Table \ref{table:notB19} summaries details of the six sources, identified in the previous section, publicly available from IGSL3 \citep{smart2013}.  It gives their DR2 source ID but only 4092328917916154368's source ID has not changed between IGSL3 and DR2.  All the others have different IGSL3 source IDs because of cross-match improvements leading up to DR2.

All six sources have $G_{\mathrm{RVS}}^{\mathrm{ext,1}} < 12$ mag, which is why they were processed by the CU6-DR2 pipeline.
If $G_{\mathrm{RVS}}^{\mathrm{ext,2}}$ had been available at the start of the CU6-DR2 pipeline processing, five of the six sources would not have been processed by the CU6-DR2 pipeline because they have $G_{\mathrm{RVS}}^{\mathrm{ext,2}} > 12$ mag.  Two have $G_{\mathrm{RVS}}^{\mathrm{ext,2}} < 14$ mag and so are bright enough to be processed by the CU6-DR3 pipeline (but radial velocities were not output for these two sources).  The other four have $G_{\mathrm{RVS}}^{\mathrm{ext,2}} > 14$ mag and so will not be processed until the CU6-DR4 pipeline.

\begin{table*}
\caption{Details of the sources with absolute DR2 radial velocities (RV in \kms) greater than 500 \kms, not in the B19 list and with (one source) and without (five sources) a DR3 radial velocity, where: \# is the number of DR2 transits; $A$ and $B$ give the origin of $B_J$ and $R_F$, respectively, using the \citet{smart2014} table A.3 photometric transformation numbers\tablefootmark{a}; and $\alpha$ and  $\beta$ give the fraction of RP transits that are contaminated and deblended, respectively (see Section \ref{sec:hvs} for more details).}         
\label{table:notB19}      
\centering          
\begin{tabular}{lrcllrrcc|lr}  \hline\hline 
DR2 source ID & RV & \# & $A$ & $B$ & $G_{\mathrm{RVS}}^{\mathrm{ext,1}}$ & $G_{\mathrm{RVS}}^{\mathrm{ext,2}}$ & $\alpha$ & $\beta$  & Contaminated by & $G_{\mathrm{RVS}}^{\mathrm{ext,2}}$\\
\hline
4658865791827681536 & $-$987.5 & 4 & 2MASS\tablefootmark{b} & PPMXL & 10.7 & 16.3  & 0.00 & 0.00 & 4658865791827521408 & 11.8 \\
5966712023814100736 & $-$967.7 & 2 & PPXML & GSC23 & 11.1 & 14.9  & 0.00 & 0.00 & ? & 11.9 \\
5305975869928712320 & $-$830.6 & 2 & 2MASS\tablefootmark{c} & 2MASS\tablefootmark{c} & 10.0 & 12.8 & 0.05 & 0.66 & 5305975869928710912 & 5.8  \\
5413575658354375040 & 500.1 & 2 & 2MASS\tablefootmark{d}  & 2MASS\tablefootmark{d}  & 12.0 & 12.0  & 0.00 & 0.98 & Glob. clust. NGC 3210 & N/A \\
5827538590793373696 & 902.6 & 2 & PPXML & PPXML & 11.1 & 15.9  & 0.00 & 0.00 & 5827538590793371776 & 10.9 \\
4092328917916154368 & 937.5 & 2 & 2MASS\tablefootmark{e} & 2MASS\tablefootmark{e} & 11.4 & 14.4  & 0.70 & 0.00 & 4092328917911305984 & 8.6\\
\hline
\end{tabular}
\tablefoot{
\tablefoottext{a}{$A = 41$ and $B= 42$ denotes that 2MASS $J$ and $K$ was used to derive $B_J$ and $R_F$, respectively, although \citet{smart2013} mistakenly switch the definition to $A = 42$ and $B= 41$.}
  \tablefoottext{b}{2MASS $J$ and $K$ photometry is biased by a nearby star that has contaminated the background estimation: Cfg = c (see \citealt{cutri2003} for details of 2MASS flags).}
   \tablefoottext{c}{Source is not detected in 2MASS $J$ magnitude band so this magnitude is an upper limit: Qflg = U, Rflg = 0.  Source is detected in 2MASS $K$ magnitude band but it is biased by a nearby star that has contaminated the background estimation: Cfg = c.}
    \tablefoottext{d}{2MASS $K$ magnitude may be contaminated by a diffraction spike from a nearby star: Cfg = d.} 
     \tablefoottext{e}{Source is not detected in 2MASS $J$ and $K$ magnitude bands so these magnitudes are upper limits: Qflg = U, Rflg = 0. } 
  }
\end{table*}

Table \ref{table:notB19} reveals that the origin of $G_{\mathrm{RVS}}^{\mathrm{ext,1}}$ for all six sources is either the Second Guide Star Catalog version 2.3 (GSC23, \citealt{lasker2008}), the Positions and Proper Motions ``Extra Large'' Catalog (PPMXL, \citealt{roeser2010}) or the Two Micron All-Sky Survey Point Source Catalog (2MASS, \citealt{skrutskie2006}).  The order of these three catalogues is the priority order of the assignment of IGSL $G_{\mathrm{RVS}}^{\mathrm{ext,1}}$, with 2MASS the lowest priority  (\citealt{smart2014} table 1).  

Photographic blue ($B_J$) and red ($R_F$) magnitudes from these catalogues were used to derive $G_{\mathrm{RVS}}^{\mathrm{ext,1}}$ using \citet{smart2014} table A.3 photometric transformation number 51.  GSC23 measures $B_J$ and $R_F$ directly from photographic plates and PPMXL includes $B_J$ and $R_F$ measured directly by the U. S. Naval Observatory B Catalogue (USNO-B1.0, \citealt{monet2003}).  2MASS $J$ and $K$ are transformed to $B_J$ and $R_F$ first using \citet{smart2014} table A.3 photometric transformation numbers 41 and 42.  The uncertainty on $G_{\mathrm{RVS}}^{\mathrm{ext,1}}$ for all six sources is given by \citet{smart2013} as $\pm$0.50 mag.  The precision of $G_{\mathrm{RVS}}^{\mathrm{ext,2}}$ is 0.1 mag or better \citep{gaiabrown2018}.

\begin{figure*}
\centering
\includegraphics[width=0.8\columnwidth]{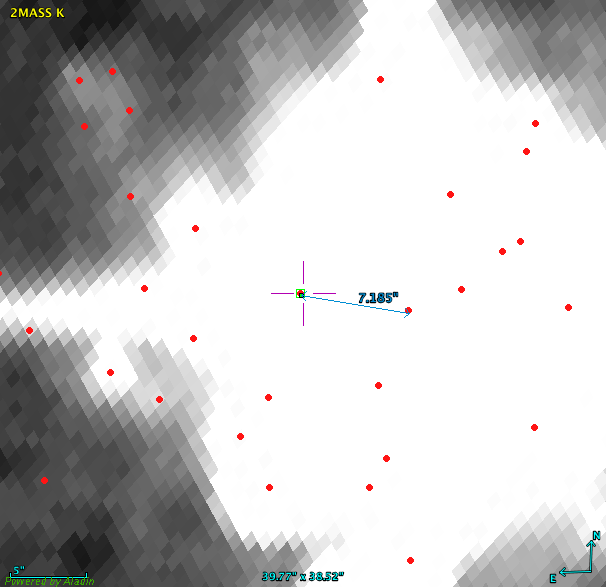}
\includegraphics[width=0.8\columnwidth]{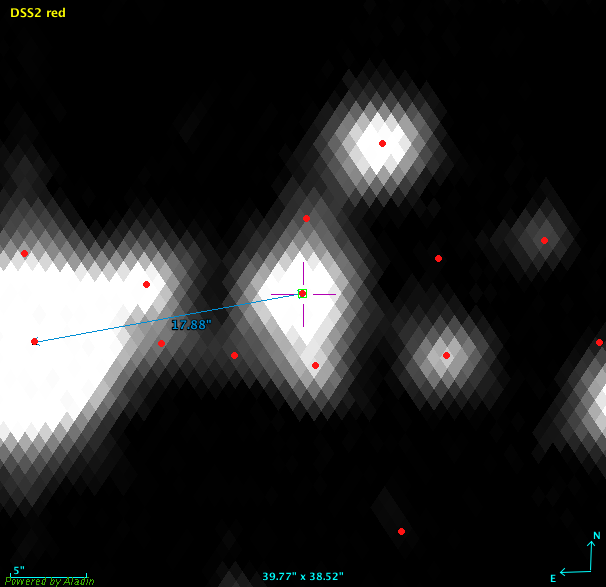}
\includegraphics[width=0.8\columnwidth]{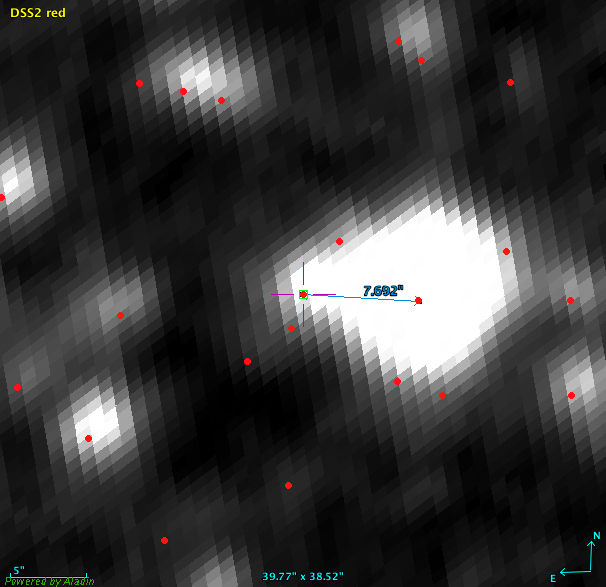}
\includegraphics[width=0.8\columnwidth]{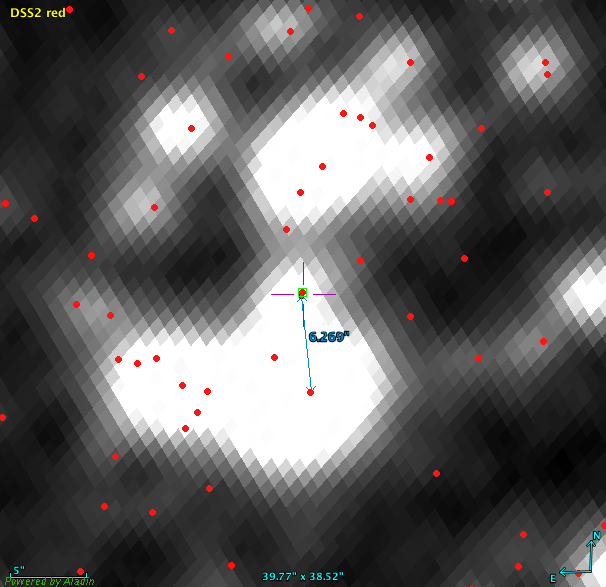}
\includegraphics[width=0.8\columnwidth]{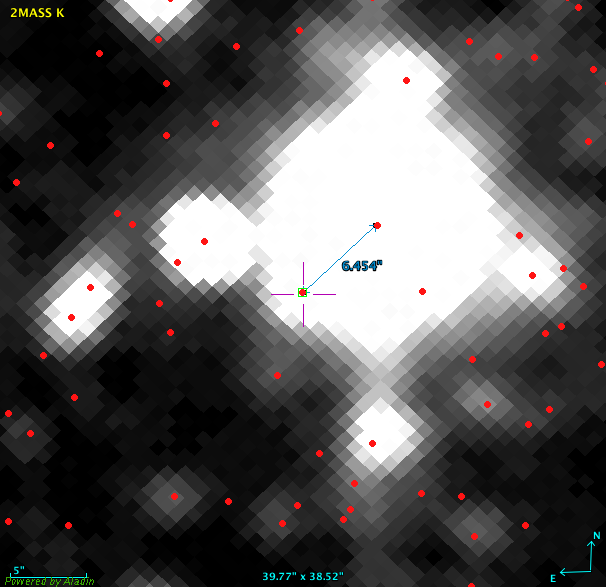}
\includegraphics[width=0.8\columnwidth]{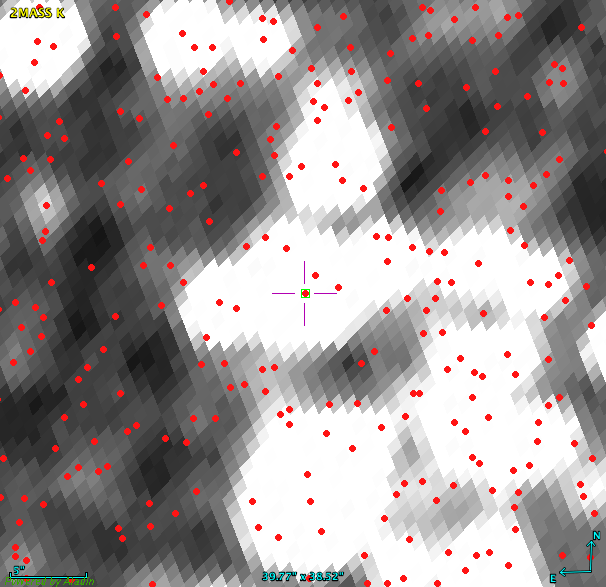}
\caption[]{Aladin sky atlas \citep{bonnarel2000} views of 2MASS $K$ (if both $B_J$ and $R_F$ were derived using 2MASS $J$ and $K$, \citealt{skrutskie2006}) or DSS2 red (Digital Sky Survey second generation, \citealt{lasker1996}).  All the six sources are in the southern hemisphere so {\it DSS2 red} refers to red photographic plates from the Anglo-Australian Observatory $F$-band Second Epoch Survey (AAO-SES, \citealt{morgan1992}), observed by the UK Schmidt Telescope, digitised by the Space Telescope Science Institute.  The same plates were scanned by USNO-B1.0 or GSC23 to measure $R_F$.  The six sources in Table \ref{table:notB19} are indicated with a purple cross-hair (and green square).  The blue arrow joins the target to a nearby brighter source, if suspected of contaminating its window.  The red dots denote EDR3 sources.  {\it Top left}: DR2 5305975869928712320; {\it top right}: DR2 5966712023814100736; {\it middle left}: DR2 4658865791827681536; {\it middle right}: DR2 5827538590793373696; {\it bottom left}: DR2 4092328917916154368; {\it bottom right}: DR2 5413575658354375040.}
\label{fig:sky}
\end{figure*}

Figure \ref{fig:sky} shows that all six sources have point spread functions (PSFs) that are overlapping neighbouring source's PSFs.  Faint objects in the halos of bright ones were often missed by GSC23, which explains why only one source gets $R_F$ from GSC23 in Table \ref{table:notB19}.  PSF reconstruction with multi-PSF image deconvolution (photometric deblending) was attempted in general in GSC23 but not in USNO-B1.0.  The photometric uncertainties in both $B_J$ and $R_F$ are 0.3 and 0.4 mag for GSC23 and PPMXL, respectively (\citealt{smart2014} table A.3).  These two considerations are presumably why GSC23 has a higher priority than PPMXL (USNO-B1.0).

Where 2MASS is the origin of $G_{\mathrm{RVS}}^{\mathrm{ext,1}}$ in Table \ref{table:notB19}, the blend flag in the 2MASS catalogue of these sources reveals that none of the 2MASS magnitudes used were deblended.  The footnotes in Table \ref{table:notB19} show the contamination or upper limit flags are often set.  The photometric uncertainties in both $B_J$ and $R_F$ are 0.5 mag for 2MASS (\citealt{smart2014} table A.3), which is larger than PPMXL (USNO-B1.0) and presumably why 2MASS has the lowest priority. 

Table \ref{table:notB19} displays the neighbours' $G_{\mathrm{RVS}}^{\mathrm{ext,2}}$ and confirms they are all brighter with the exception of DR2 5966712023814100736 and DR2 5413575658354375040.  Footnote (d) of \citet{smart2014} table A.3 states that photometric transformation number 51 should be used for objects with declination $>= 0$.  All the sources in Table \ref{table:notB19} have 51 as the origin of their $G_{\mathrm{RVS}}^{\mathrm{ext,1}}$ but they all have declination $<0$.  Footnote (e) indicates that photometric transformation number 59 should be used for objects with declination $< 0$.  Applying this gives brighter magnitudes than the $G_{\mathrm{RVS}}^{\mathrm{ext,1}}$ in Table \ref{table:notB19}, exacerbating the discrepancy between $G_{\mathrm{RVS}}^{\mathrm{ext,1}}$ and $G_{\mathrm{RVS}}^{\mathrm{ext,2}}$.

Table \ref{table:notB19} discloses that at least one and normally both of $B_J$ and $R_F$ were not deblended in the derivation of  $G_{\mathrm{RVS}}^{\mathrm{ext,1}}$ for all six sources.  {\it Gaia} EDR3 photometry is not deblended either.  $G_{\mathrm{RVS}}^{\mathrm{ext,2}}$ was derived from EDR3 $G$ and $G_{\mathrm{RP}}$, both of which can also suffer with contamination from other sources not in the same window and blending from sources that are in the same window.  EDR3 does not remove these sources from the mean photometry but it does flag when it occurs in $G_{\mathrm{RP}}$, such that the fraction of RP transits that are contaminated ($\alpha$) and blended ($\beta$) can be calculated using the following equations based on EDR3 fields:

\begin{equation}
\begin{split}
\alpha &= \frac{\textrm{phot\_rp\_n\_contaminated\_transits}\times 1.0}{\textrm{phot\_rp\_n\_obs}}\\
\beta  &= \frac{\textrm{phot\_rp\_n\_blended\_transits}\times 1.0}{\textrm{phot\_rp\_n\_obs}}
\end{split}
\label{equ:cu5}
\end{equation}

\noindent where \url{phot_rp_n_contaminated_transits} is the number of RP transits that contributed to the mean photometry and were considered to be contaminated by one or more nearby sources, \url{phot_rp_n_blended_transits} is the number of RP transits that contributed to the mean photometry and were flagged to be blends of more than one source (i.e. more than one source is present in the observing window), \url{phot_rp_n_obs} is the number of observations (CCD transits) that contributed to the integrated RP mean flux  and the term 1.0 is required to ensure the floating point division. The contaminating and/or blending sources may come from the other field-of-view (FoV).

Table \ref{table:notB19} gives $\alpha$ and $\beta$.   \citet{riello2021} point out that such ratios do not take into account the flux ratio between the target source and the contaminating or blending source(s) but that users can, in principle, assess this effect, at least in the case where the contaminating and/or blending is from a source that is close to the target source on the sky, although not when it comes from the other FoV.

\citet{riello2021} suggest that sources greater than 1.75 arcsec apart will not have blended RP windows, regardless of {\it Gaia}'s scan angle.  DR2 5413575658354375040's $\beta$ = 0.98 means nearly every RP window is blended.  \figref[fig:sky] confirms it has neighbouring sources less than 1.75 arcsec away.   DR2 5305975869928712320's $\beta$ = 0.66 but \figref[fig:sky] does not show any sources within 1.75 arcsec, suggesting the blending sources come from the other FoV.  DR2 4092328917916154368's $\alpha$ = 0.70 is most likely as a result of contamination from its bright neighbour.  It may not be chance that the three sources with $\alpha > 0$ and/or $\beta > 0$ are brighter than the three with $\alpha = \beta = 0$.  It may suggest that the former three's {\it Gaia} magnitudes have been inflated by contamination and/or blending.  \citet{riello2021} warn that even $\alpha = \beta = 0$ does not necessarily mean the source is not affected by crowding.  $\alpha$ and $\beta$ are limited to sources in the {\it Gaia} catalogue so close pairs, with one source never resolved by {\it Gaia}, will not be identified by $\alpha$ or $\beta$.  Nevertheless, $G_{\mathrm{RVS}}^{\mathrm{ext,2}}$ is more likely to reflect reality than $G_{\mathrm{RVS}}^{\mathrm{ext,1}}$ because of {\it Gaia}'s superior space-based angular resolution, compared to ground-based, seeing-limited angular resolution.  EDR3 includes sources with angular separations of 0.18 arcsec, but relatively few sources are found with separations less than about 0.6 arcsec \citep{lindegren2021}.  {\it Gaia}'s narrower PSF limits the impact of blending and contamination compared to the broader on-ground PSFs.  

None of these six sources are in the B19 list because all of the brighter neighbours (linked by the blue arrow in \figref[fig:sky]) are further away than the B19 6.4 arcsec search radius and the sources within 6.4 arcsec either do not have a radial velocity or are fainter in DR2 $\grp$ or $G$ magnitudes.  \citet{marchetti2019} searched for high-velocity stars using \url{rv_nb_transits} greater than 5 and this cut has been used widely in the literature.  Table \ref{table:notB19} reports that all six sources have \url{rv_nb_transits} less than 5, implying they were not used in high-velocity star studies.  Nevertheless, Appendix \ref{sec:appendix_hvs} investigates the provenance of each radial velocity in Table \ref{table:notB19} to assess whether they should be included in EDR3.  Five of the six are found to have contaminated DR2 radial velocities and are thus excluded from EDR3. 

\subsection{Source identifier ambiguity between DR2 and EDR3 and radial velocity assignment}
\label{Sect:source}

The cross-match from Sect. \ref{Sect:Xmatch} has the following characteristics:
\begin{itemize}
\item For 99.5\% of the  DR2 sources, the best match was in agreement with the source mapping expected from the source identifier evolution between  DR2 and  EDR3 \citep{torra2021}. Of those, the vast majority corresponds to sources that maintained their source identifier between the two releases. 
\item The other 0.5\% of the  DR2 sources were matched to a different source.  Particular attention was paid to this sub-sample as explained in the following.
\end{itemize}
 
The $\sim$36\,000 sources from the second point above are further analysed in order to decide whether the  EDR3 source could safely hold the radial velocity of the  DR2 sources it was matched to. After removing sources from the B19 list (Sect. \ref{sec:rv}) and high-velocity stars not in the B19 list (Sect. \ref{sec:hvs_dr3} and \ref{sec:hvs}), the remaining sources are split in two categories:
\begin{itemize}
\item Those that did not pass the radial velocity criteria $i$ (7807) and $ii$ (211) used in Sect. \ref{sec:rv} are also discarded, leading to the removal of another 8018 radial velocities (\figref[fig:claus_dpace_talk_8018]).
\item Of the other 26\,208 passing this quality criterion (\figref[fig:claus_dpace_talk_2905]), a further cut is applied based on the positional separation between the  DR2 sources and their  EDR3 match. Sources separated by more than 2 mas (\figref[fig:ipd_multi_sep]) at both of the considered epochs are ignored, discarding another 2905 radial velocities.
\end{itemize}

\begin{figure}
\centering
\includegraphics[width=\columnwidth]{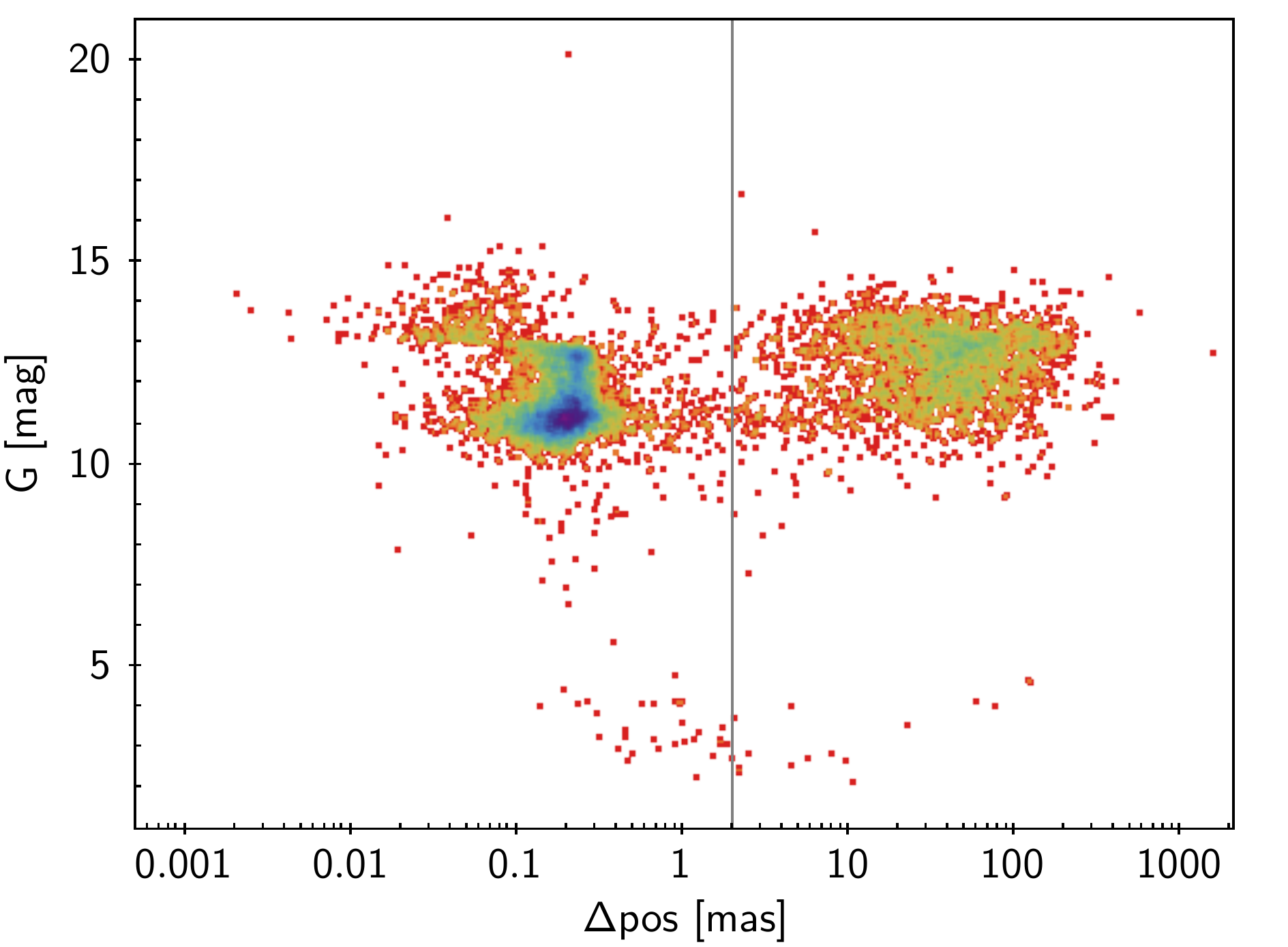}
\caption[]{EDR3 $G$ magnitude versus positional separation between EDR3 and DR2 of the 8018 sources that did not pass the radial velocity criteria $i$ and $ii$ from Sect. \ref{sec:rv}.}
\label{fig:claus_dpace_talk_8018}
\end{figure}

\begin{figure}
\centering
\includegraphics[width=\columnwidth]{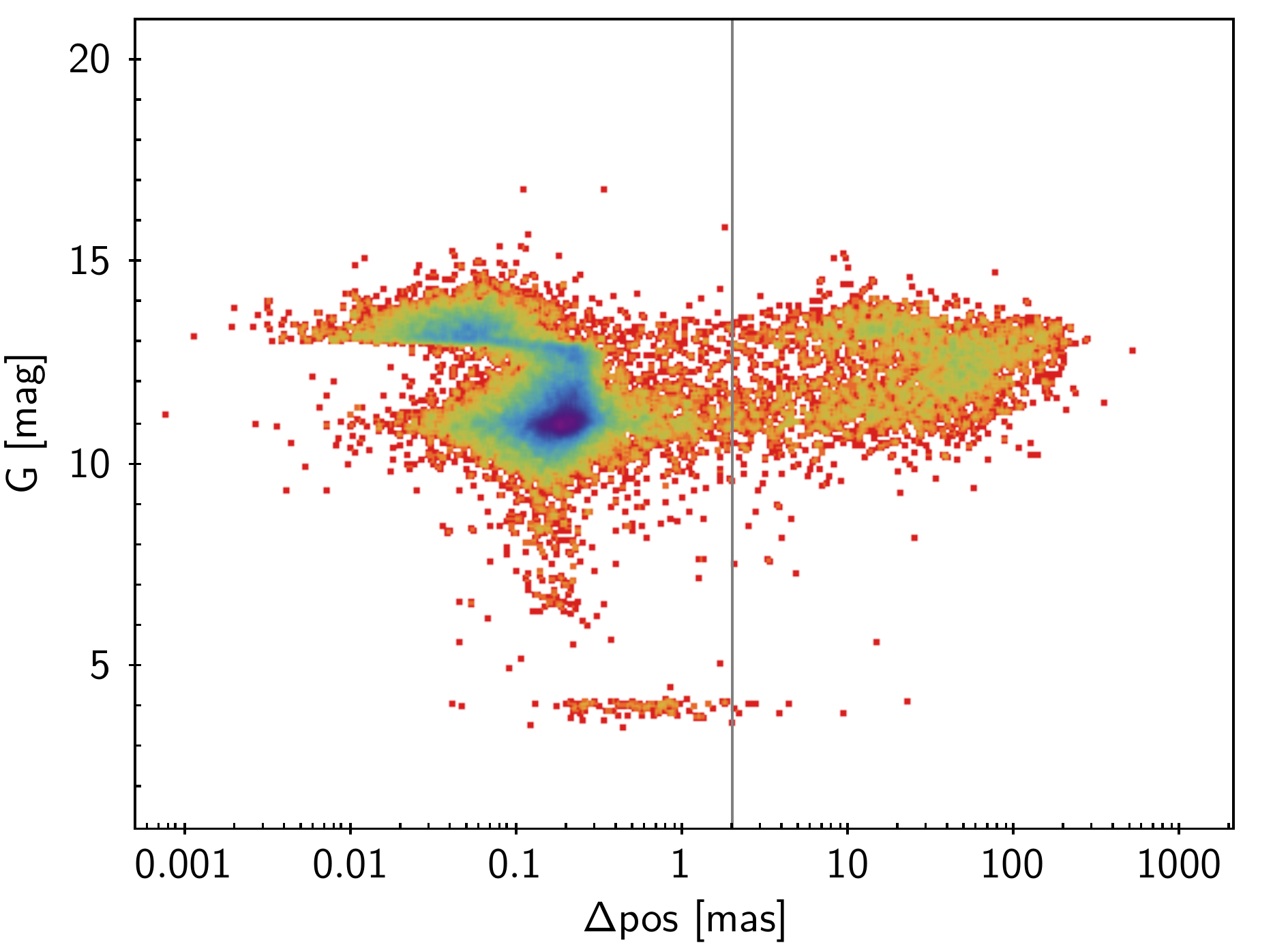}
\caption[]{EDR3 $G$ magnitude versus positional separation between EDR3 and DR2 of the 26\,208 sources passing the radial velocity criteria $i$ and $ii$ from Sect. \ref{sec:rv}.}
\label{fig:claus_dpace_talk_2905}
\end{figure}

Removal of sources from EDR3 owing to source ambiguity does not remove any high-velocity sources (\figref[fig:hist_rv_pos_sep]) but it does remove most of the brightest sources from DR2 (\figref[fig:hist_g_pos_sep]).  It is mainly the radial velocity criteria that is responsible for this.  Most of the sources with EDR3 $G<6$ mag were excluded from the CU6-DR3 pipeline owing to saturated samples, while they yielded radial velocities in DR2.  Sect. \ref{Sect:Xmatch} explained that these sources are close binary pairs and that the EDR3 astrometric processing suppressed the secondary images better than in DR2.  This suggests that the CU6-DR2 pipeline processed the secondary images of these sources, which did not have saturated samples and so were not excluded, and the CU6-DR3 pipeline processed the primary images of these sources, which did have saturated samples and so were excluded.  As expected for the brightest stars, these sources are distributed uniformly over the sky (black points in \figref[fig:sky_match]). 

The CU6-DR3 pipeline did not yield radial velocities for 7807 of the 8018 sources (radial velocity criterion $i$).  The 7807 radial velocities were excluded mainly as a result of point background contamination being flagged.  This suggests that both sources in the close pair were observed in the same RVS window, where the secondary image has been modelled as a point background source in the CU6-DR3 pipeline, found to be contaminating because it was within three magnitudes of the primary and hence the window was excluded.

The CU6-DR3 pipeline produced radial velocities for 211 of the 8018 sources but these disagreed with their DR2 radial velocities (radial velocity criterion $ii$).  The CU6-DR3 pipeline did not record point background magnitudes but given these sources are close binary pairs, it is likely the other source in the pair contributes to the point background contamination.  The windows were not excluded suggesting that the magnitude difference was larger than three magnitudes.

The 211 sources are included in the top plot in \figref[fig:sky_match] but they do not coincide with any of the overdensity of {\it Gaia} scans visible.  Instead, it is the 7807 sources that trace these out.  The North and South Ecliptic Poles are at (96,$+$30)$^{\circ}$ and (276,$-$30)$^{\circ}$, respectively.  The overdensity of scans connecting these points are the Ecliptic Pole Scanning Law (EPSL), which was observed in the first month of the nominal mission before changing to the Nominal Scanning Law (NSL).

As shown in \figref[fig:sky_match], EPSL is characterised by repeated scans of the same source with slowly changing scan directions. In this scenario, it is less plausible that the components of the close binary pair will be switched when assigning transits to sources (cross-match) during the EPSL.  It is the primary component that was most likely assigned a RVS window but the CU6-DR3 pipeline excludes these windows because of the point background contamination caused by the secondary.

During NSL scans, when the scan direction changes more rapidly, it is more plausible that the cross-match may switch between the primary and secondary.  When the single close binary pair source ID is assigned to the secondary and not the primary, the CU6-DR3 pipeline considers the primary as a point background source.   Point background contamination is flagged only if it is fainter than the target.  The primary is generally brighter than the secondary, meaning windows in this scenario are not excluded and so may produce a radial velocity that agrees with its DR2 value.  

Comparing the top plot of \figref[fig:sky_match] with \citet{boubert2020a} fig. 4 confirms the EPSL scans and suggests that the more diagonal and vertical overdensities of scans are related to the two decontamination campaigns, conducted to sublimate contaminating water ice from the
optics \citep{gaia2016}.  Both the DR2 and DR3 CU6 pipelines exclude the decontamination periods but data leading up to these periods will be increasingly contaminated.  The exclusion of these windows from the CU6-DR3 pipeline, suggests the contamination has the same effect as EPSL i.e. making it less likely cross-match switches occur even though it is NSL, meaning the secondary is consistently identified as point background contamination.  This consistently removes these windows from the CU6-DR3 pipeline, preventing a radial velocity measurement for these close binary pairs so these sources fail radial velocity criterion $i$.

\citet{boubert2020a} figs. 4 and 7 suggest the more horizontal overdensity of scans in the top plot of \figref[fig:sky_match] is related to the $G$ epoch photometry gaps.  Some $G$ epoch photometry gaps remove outliers originating from cross-match problems \citep{riello2018}.  Neither the DR2 nor DR3 CU6 pipelines exclude $G$ epoch photometry gaps but if the horizontal one in the top plot of \figref[fig:sky_match] is due to cross-match issues, this could also cause this period to behave like the EPSL period as well.

26\,208 sources passed the radial velocity criteria $i$ and $ii$ from Sect. \ref{sec:rv}, either because the radial velocities in DR2 and DR3 are measuring the same binary component or because they are measuring different components but their radial velocities are sufficiently close (less than 25 \kms).  Therefore, the astrometry is required to identify when the same binary component is observed in DR2 and EDR3 (less than 2 mas separation) and when different binary components are observed in DR2 and EDR3 (greater than 2 mas separation).  The resulting 2905 sources have distributions similar to that of the 8018 sources in radial velocities (\figref[fig:hist_rv_pos_sep]) and magnitudes (\figref[fig:hist_g_pos_sep]).  The sky distribution of the 2905 sources (bottom plot in \figref[fig:sky_match]) is different from the 8018 (top plot in \figref[fig:sky_match]) in that there is no hint of {\it Gaia}'s scans but there is a preference for the 2905 sources to be found in the Galactic plane.

\begin{figure}
\centering
\includegraphics[width=\columnwidth]{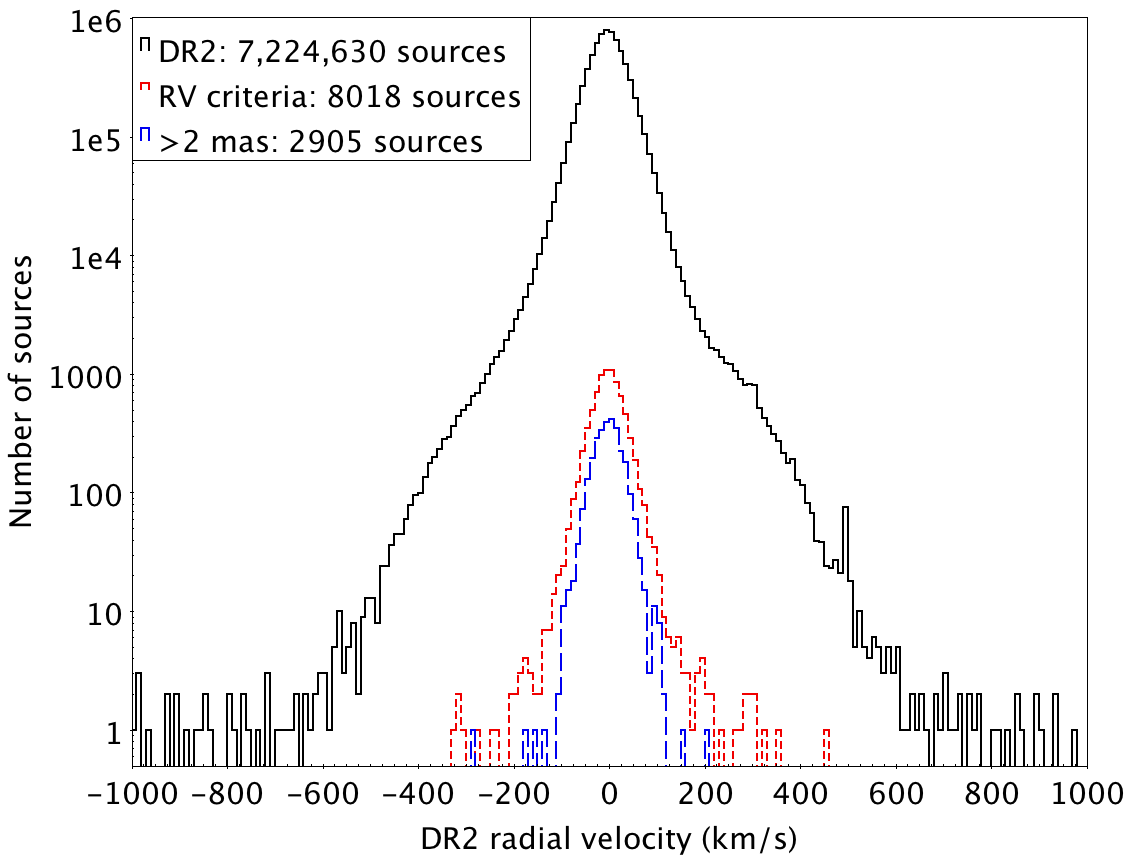}
\caption[]{Histogram of DR2 radial velocities of the two different groups of sources with DR2 radial velocities removed from EDR3 because of match difficulties.}
\label{fig:hist_rv_pos_sep}
\end{figure}

\begin{figure}
\centering
\includegraphics[width=\columnwidth]{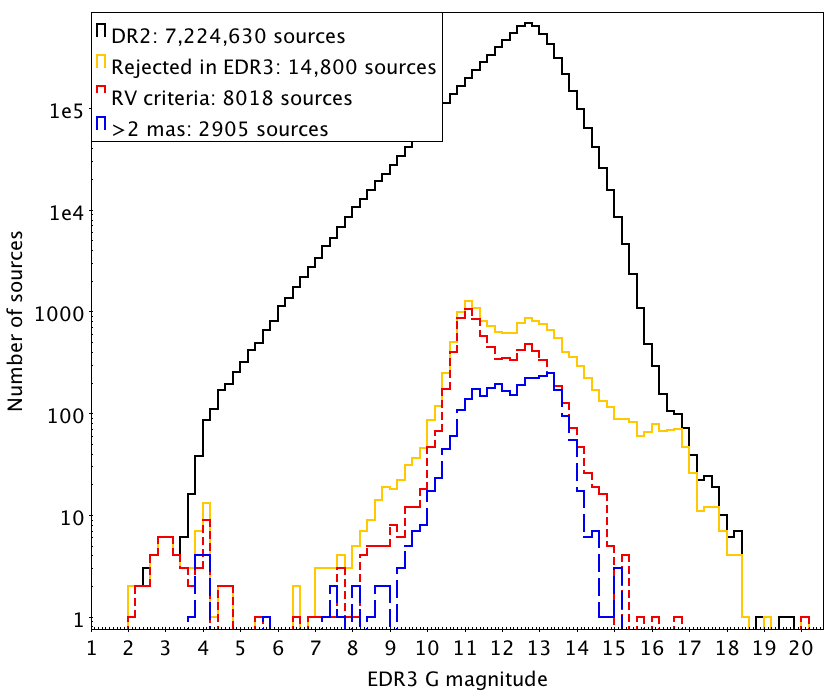}
\caption[]{Histogram of EDR3 $G$ magnitudes of the two different groups of sources with DR2 radial velocities removed from EDR3 because of match difficulties, as well as all the sources with DR2 radial velocities removed from EDR3.}
\label{fig:hist_g_pos_sep}
\end{figure}

\begin{figure*}
\centering
\includegraphics[width=\textwidth]{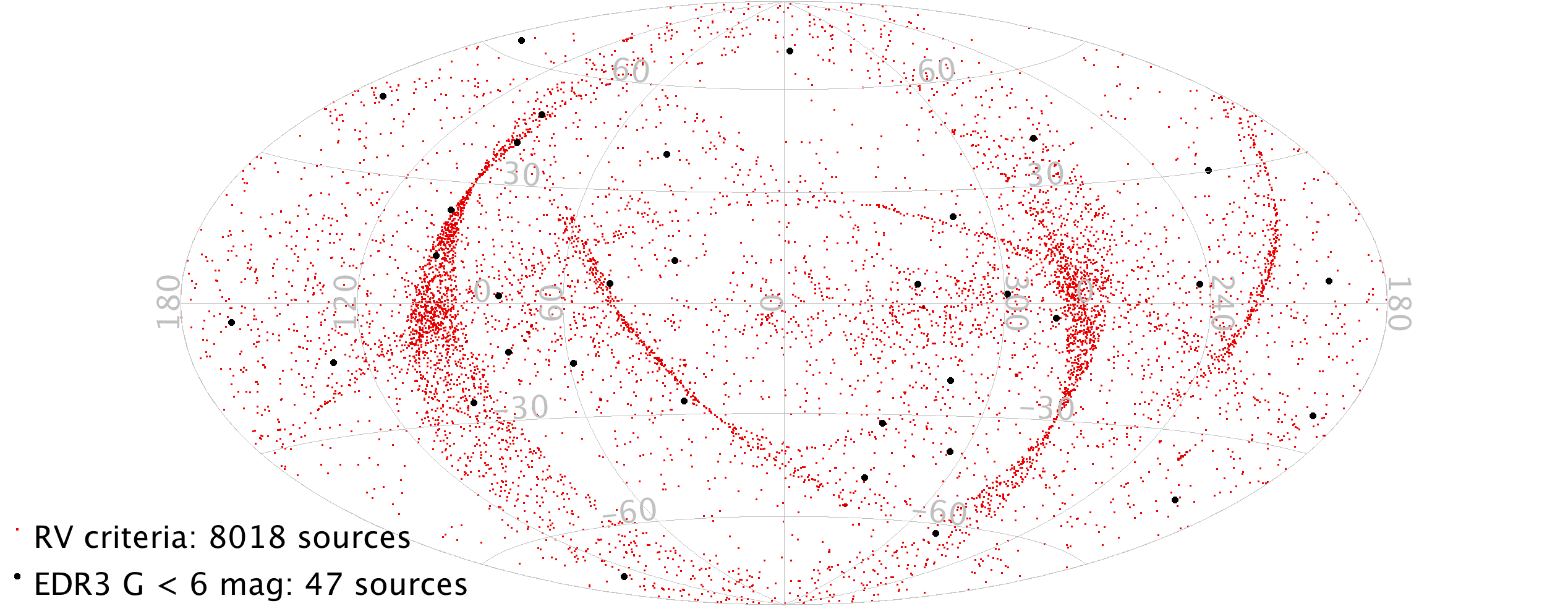}\\
\vspace{0.5cm}
\includegraphics[width=\textwidth]{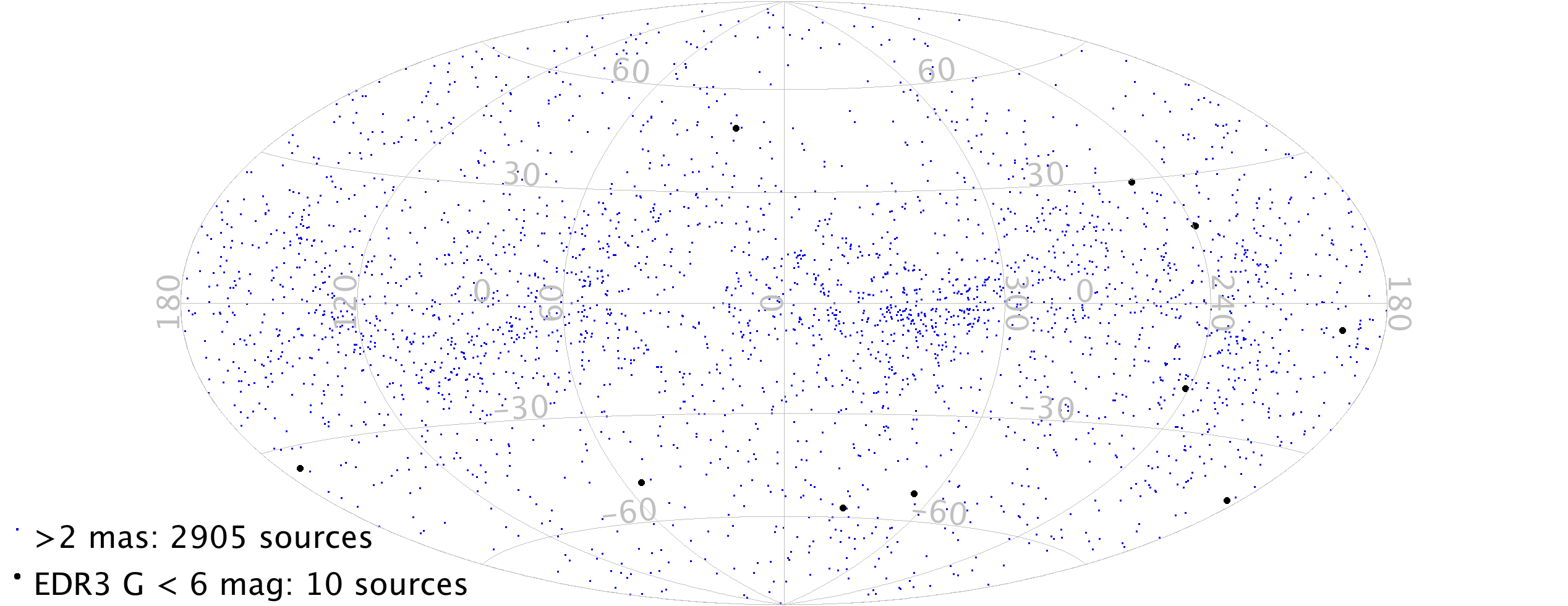}	
\caption[]{Galactic Aitoff projection of the two different groups of sources with DR2 radial velocities removed from EDR3 because of match difficulties.}
\label{fig:sky_match}
\end{figure*}

\section{Results}
\label{Sect:results}

\subsection{Overview}

The filters described in the previous sections were applied to the DR2 radial velocities to improve them before they were deposited into the EDR3 archive.  Overall, 14\,800 radial velocities from  DR2 have not been propagated to any  EDR3 sources. This leads to 7\,209\,831 sources with a radial velocity in  EDR3, which is 99.8\% of sources with a radial velocity in DR2 (7\,224\,631).  97\% of EDR3 sources with a DR2 radial velocity have exactly the same source identifier as they had in DR2 i.e. 3\% of DR2 radial velocities appear in EDR3 with a different source identifier.

\begin{figure}
\centering
\includegraphics[width=\columnwidth]{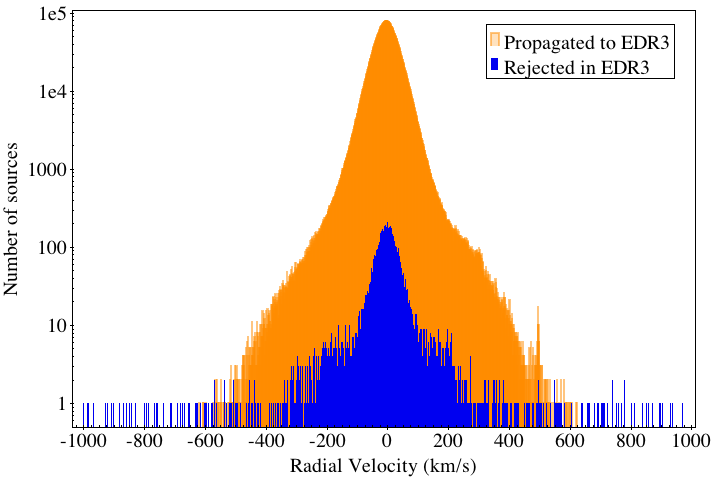}
\caption[EDR3 RV distribution]{Distribution of the  DR2 radial velocities propagated to  EDR3 (orange) and those rejected (blue).}
\label{fig:cu6spe_edr3rv_distribution}
\end{figure}

Figure \ref{fig:cu6spe_edr3rv_distribution} compares the radial velocity distribution of sources propagated to EDR3 and those rejected.  The DR2 radial velocity range is [$-999.3$,$+970.6$] \kms, which is a consequence of the DR2 radial velocity grid ($-1000$ to $+1000$ \kms) and filtering of results \citep{sartoretti2018,arenou2018}.  The EDR3 radial velocity range is [$-618.5$,$+623.6$] \kms, which is based on the same DR2 grid and filtering but with the extra filters described in this paper applied.  Appendix \ref{sect:appendix2} gives the EDR3 status of high-velocity stars in the extreme negative and positive tails of DR2's radial velocity distribution.  

Figure \ref{fig:hist_g_pos_sep} shows that the $G$ magnitude distribution of sources with DR2 radial velocities that remain in EDR3 has a long tail towards the faint end.  These sources have survived by not being in the B19 list or having absolute DR2 radial velocities less than 500 \kms.  They were not removed from DR2 because their radial velocity uncertainties were less than 20 \kms, the DR2 filter on this uncertainty \citep{sartoretti2018}.

\citet{schonrich2019} find a strong distance underestimate at the faint end of their {\it DR2 RV sample}.  Rather than a failure of {\it Gaia} parallaxes, they argue the most likely explanation for this is that the DR2 radial velocity uncertainties are underestimated at the faint end.  This is probable, given that the number of DR2 transits with a radial velocity tends to become fewer towards the faint end and the DR2 radial velocity uncertainty is a function of the number of transits and the standard deviation of the individual transit radial velocities \citep{sartoretti2018}.  However, there is no evidence that these radial velocities are contaminated so they remain in EDR3, albeit with the caveat identified by \citet{schonrich2019}.

\begin{figure}
\centering
\includegraphics[width=\columnwidth]{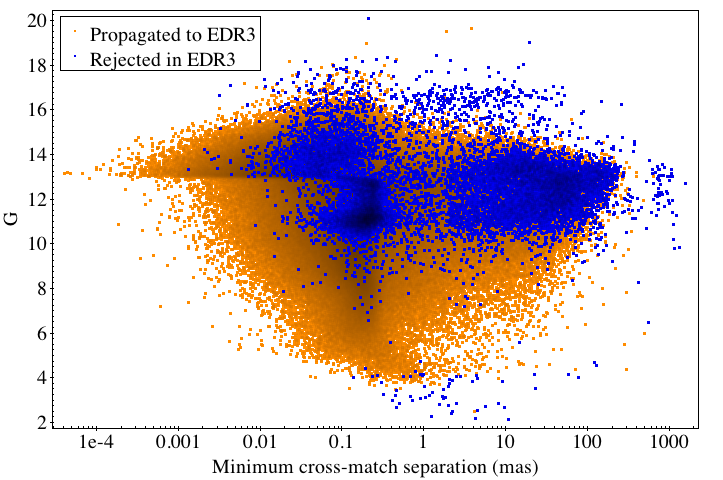}
\caption[EDR3 RV source match separation]{Angular separation between the original DR2 sources and their best match in  EDR3 for sources having a radial velocity in  DR2. The orange dots represent sources for which a radial velocity is propagated into  EDR3, while blue dots correspond to those sources where the  DR2 radial velocity is rejected.}
\label{fig:cu6spe_edr3rv_separation}
\end{figure}

Figures \ref{fig:hist_g_pos_sep} and \ref{fig:cu6spe_edr3rv_separation} show that the radial velocities of most sources brighter than $G$ = 4 mag have been removed from EDR3.  This is specific to EDR3 only because of the source ambiguity complication and consequently does not affect whether bright sources in DR3 have radial velocities.

\subsection{Comparison with the literature}
\label{sec:lit}

This section explores whether radial velocities from the literature validate the removal of radial velocities from EDR3.  The comparison is limited to the 3871 removed from EDR3 as a result of failing radial velocity criteria $i$ and $ii$ in Sect. \ref{sec:rv} and the five sources removed in Sect. \ref{sec:hvs} and Appendix \ref{sec:appendix_hvs}, which are assumed to have contaminated radial velocities, and not the 10\,923 sources removed from EDR3 as a result of source ambiguity in Sect. \ref{Sect:source}, which are assumed to not have contaminated radial velocities. 

\subsubsection{\citet{katz2019}}
\label{sec:katz}

\begin{figure}
\centering
\includegraphics[width=\columnwidth]{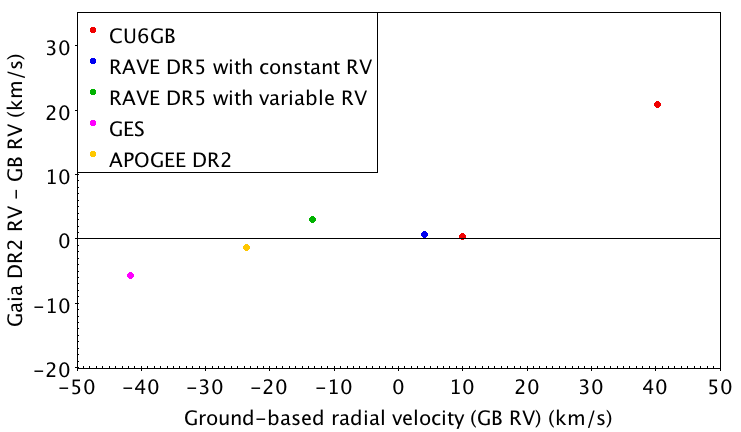}\\
\vspace{0.5cm}
\includegraphics[width=\columnwidth]{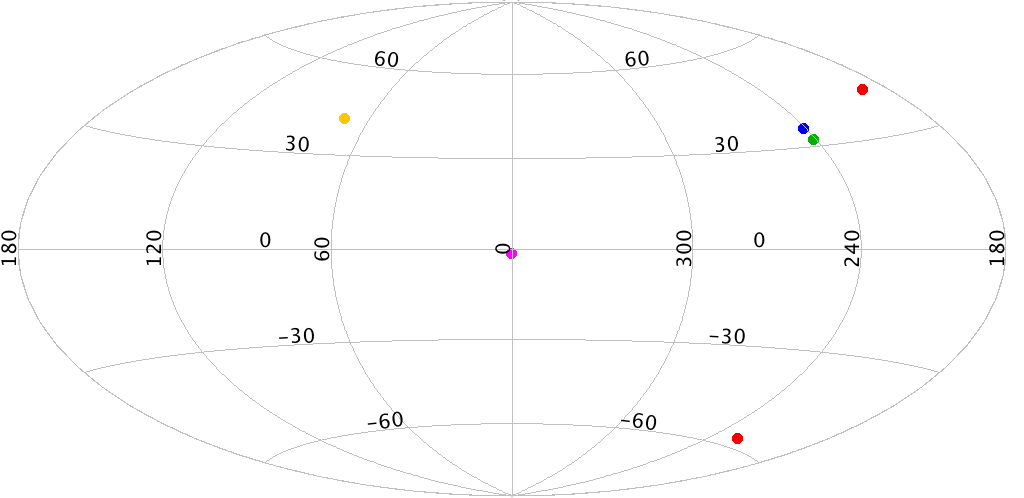}\\
\caption[]{{\it Top}: Residuals of the DR2 radial velocities excluded from EDR3 as a function of ground-based radial velocities. {\it Bottom}: Galactic Aitoff projection of DR2 sources with ground-based radial velocities that were excluded from EDR3, colour-coded the same as the top plot.}
\label{fig:gbs}
\end{figure}

\citet{katz2019} validated DR2 radial velocities against the following ground-based catalogues: CU6 ground-based standards (CU6GB, \citealt{soubiran2018}),
%radial velocities for the cancelled Space Interferometry Mission (SIM, \citealt{makarov2015}), 
Radial Velocity Experiment (RAVE DR5, \citealt{kunder2017}), Apache Point Observatory Galactic Evolution Experiment (APOGEE DR2, \citealt{abolfathi2018}) and the Gaia-ESO-Survey (GES, \citealt{gilmore2012}) and found good agreement.  This compilation was cross-matched against the 3876 sources and six were found in common.  The top of Figure \ref{fig:gbs} compares their radial velocities and shows that this good agreement also applies to most of these sources.  The radial velocity constancy has not been classified for the two most outlying sources.  Therefore, the reason for the difference could be binarity or variability or it could be the result of an erroneous DR2 radial velocity caused by a lack of deblending.

\subsubsection{RAVE DR6}
\label{sec:rave}

Previous studies have found excellent agreement in general between RAVE and {\it Gaia} DR2 \citep{deepak2018,steinmetz2018}.
RAVE DR6 \citep{steinmetz2020} provides a cross-match with {\it Gaia} DR2: 59 sources were found out of the 3876.  \citet{steinmetz2020} use the $R$ correlation coefficient \citep{tonry1979} as a quality indicator and define their core sample as $R>10$.  

Figure \ref{fig:rave_lamost} (top left) is colour-coded by $R$ and shows that sources with $R$ close to 10 are in good agreement.  The most radial velocity discrepant sources have much higher $R$ values, suggesting the RAVE value is correct and the {\it Gaia} one is spurious.  Out of the 59 RAVE sources in the 3876, there are 42 (71\%) sources with no DR3 radial velocity (triangles) and 17 (29\%) sources where the DR2 and DR3 radial velocities disagree according to the radial velocity criteria $ii$ from Sect. \ref{sec:rv}.

The source with the largest radial velocity difference between RAVE DR6 and {\it Gaia} DR2 is source ID 3754815441303370496 (hereafter S14, Table \ref{table:symbol}).  It is investigated individually in Appendix \ref{sect:3754815441303370496}.

\subsubsection{LAMOST DR6}
\label{sec:lamost}

The Large Sky Area Multi-Object Fiber Spectroscopic Telescope (LAMOST) DR6\footnote{\url{http://dr6.lamost.org/v2/}} Low Resolution Spectral Survey A, F, G and K Star Catalog was cross-matched with the 3876 {\it Gaia} DR2 sources. {\it Gaia} DR2's epoch 2015.5 positions were propagated to LAMOST's input catalogue epoch 2000 using {\it Gaia} DR2's astrometry.  The closest sources from each catalogue within 1 arcsec were taken as the match: 79 sources are found out of the 3876.

In the absence of another quality measure, \figref[fig:rave_lamost] (top right) colour-codes sources by their separation at epoch 2000. The most radial velocity discrepant source has the smallest separation, suggesting the LAMOST value is correct and the {\it Gaia} one is spurious.  This is source ID 1803504050895768704 (hereafter S16, Table \ref{table:symbol}).  It is investigated individually in Appendix \ref{sect:1803504050895768704}.

\begin{figure*}
\centering
\includegraphics[width=0.9\columnwidth]{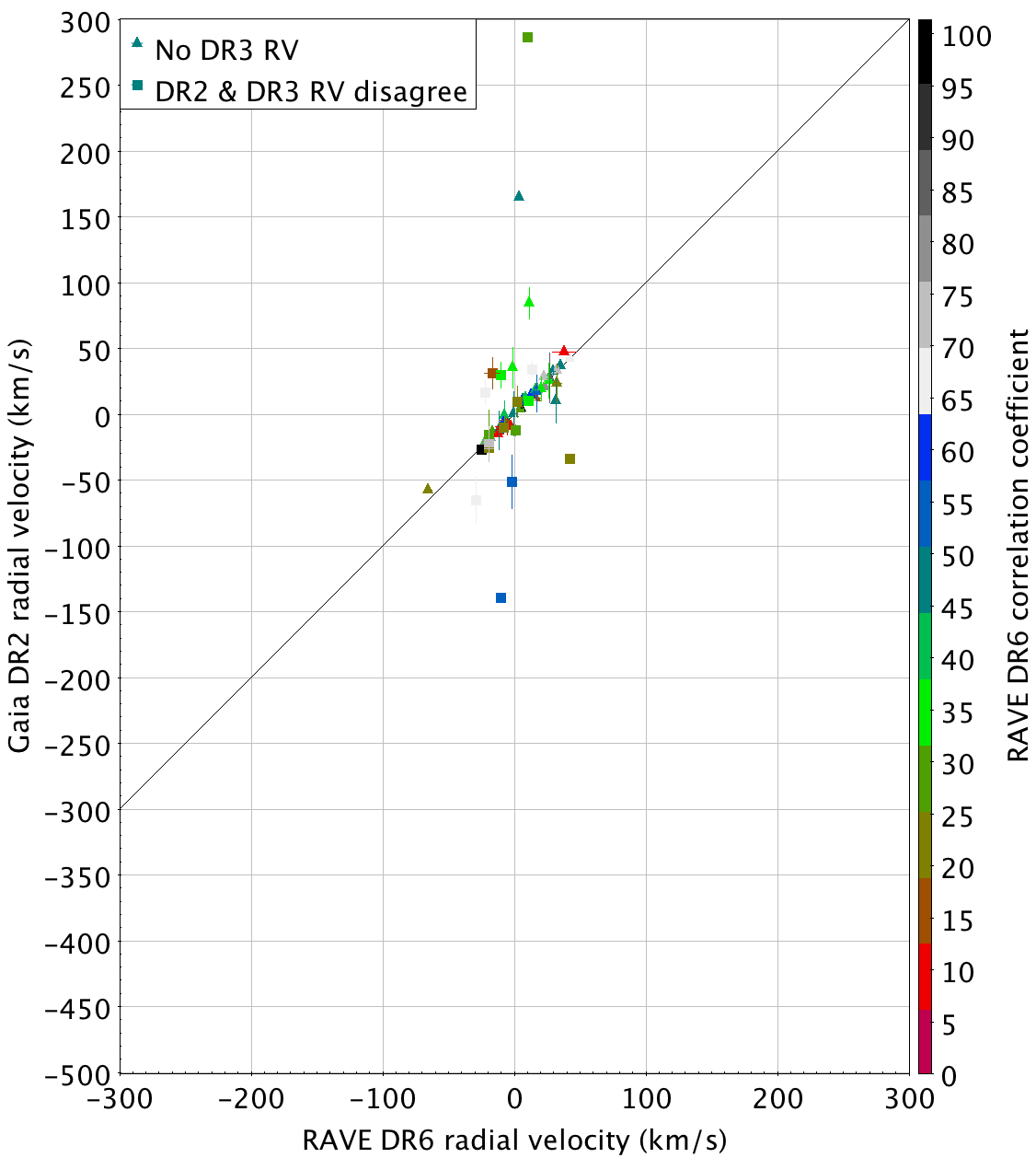}
\hspace{1cm}
\includegraphics[width=0.9\columnwidth]{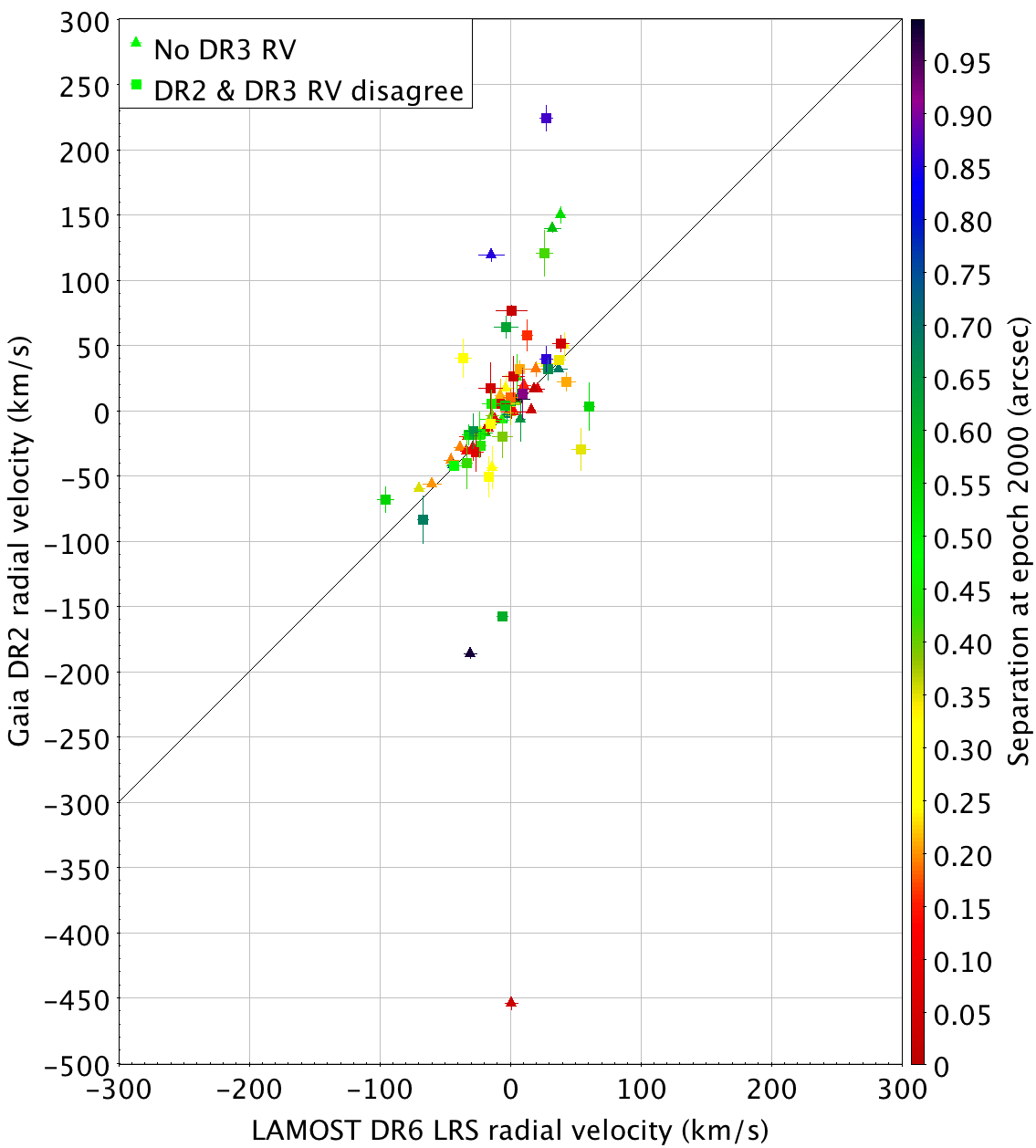}\\
\vspace{1cm}
\includegraphics[width=0.9\columnwidth]{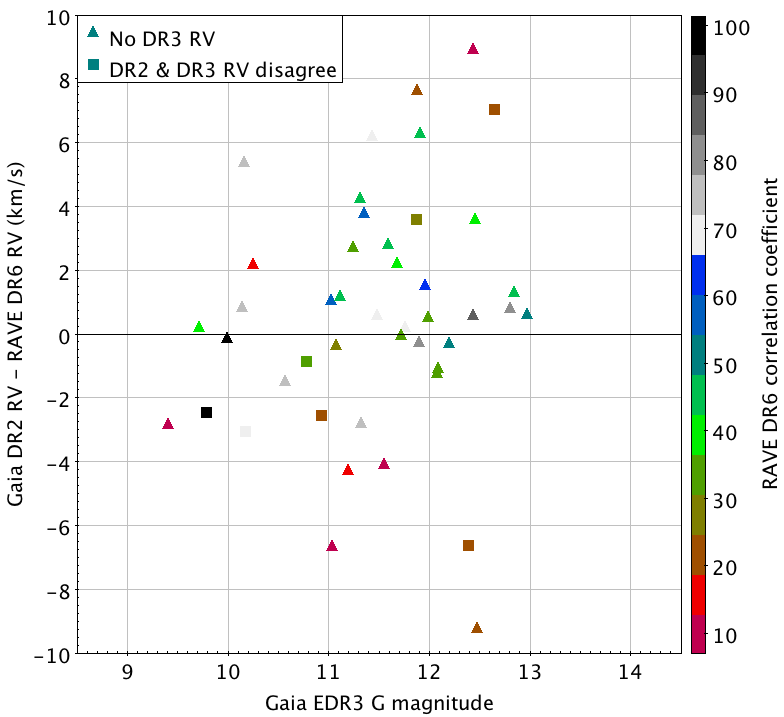}
\hspace{1cm}
\includegraphics[width=0.9\columnwidth]{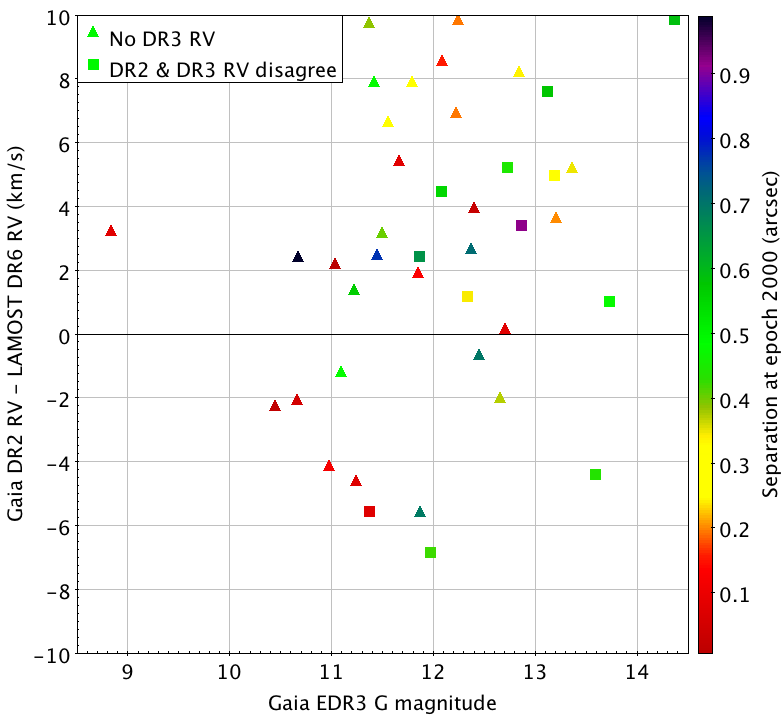}
\caption[]{{\it Top row}: {\it Gaia} DR2 radial velocities versus radial velocities from RAVE DR6 ({\it left}),  colour-coded according to the $R$ correlation coefficient from RAVE DR6, and LAMOST DR6 ({\it right}), colour-coded according to the separation at epoch 2000.   {\it Bottom row}: Radial velocity differences as a function of {\it Gaia} EDR3 $G$ magnitude with the same colour-coding as the top row.}
\label{fig:rave_lamost}
\end{figure*}

\begin{figure*}
\centering
\includegraphics[width=\textwidth]{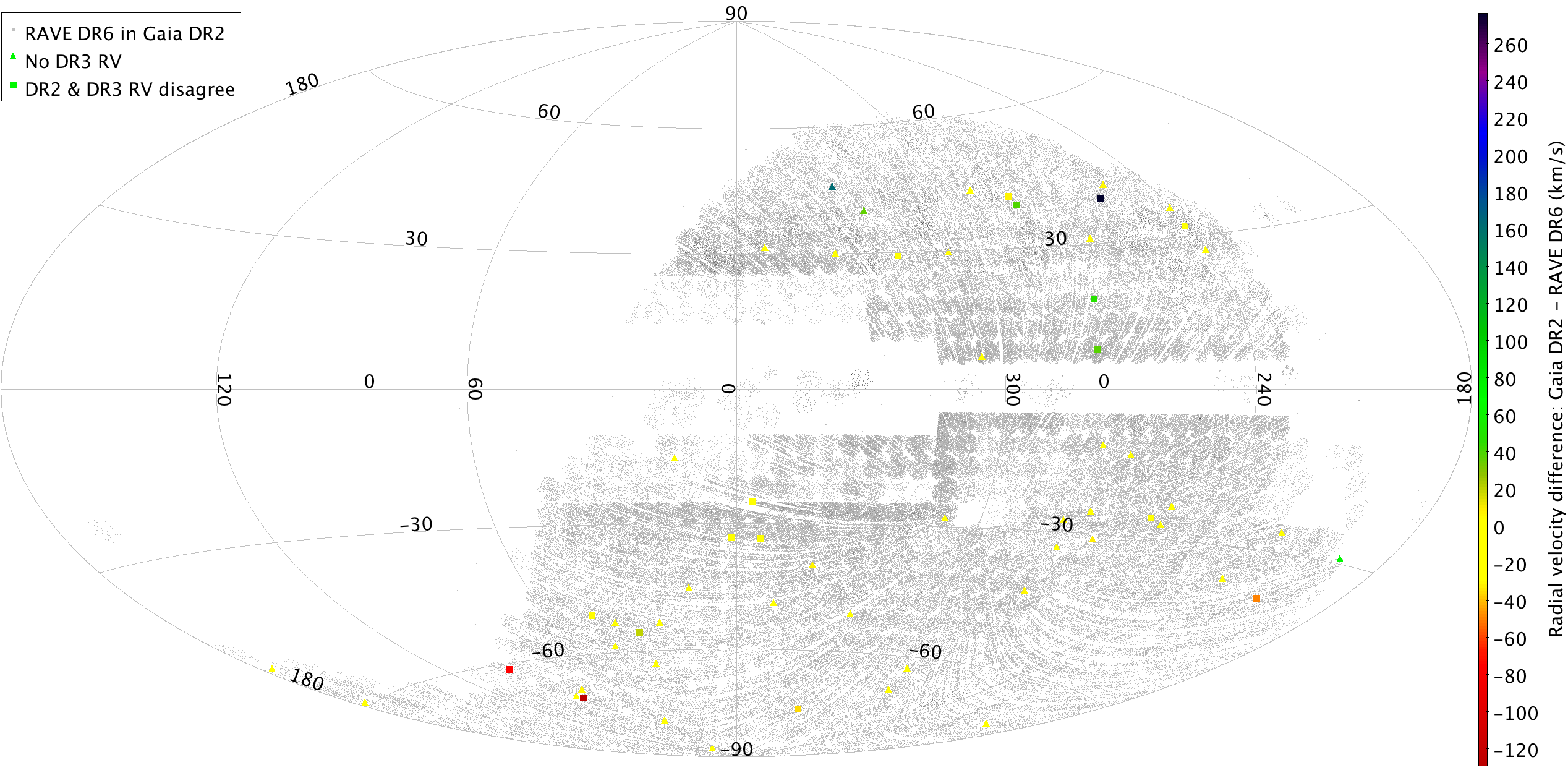}\\
\vspace{1cm}
\includegraphics[width=\textwidth]{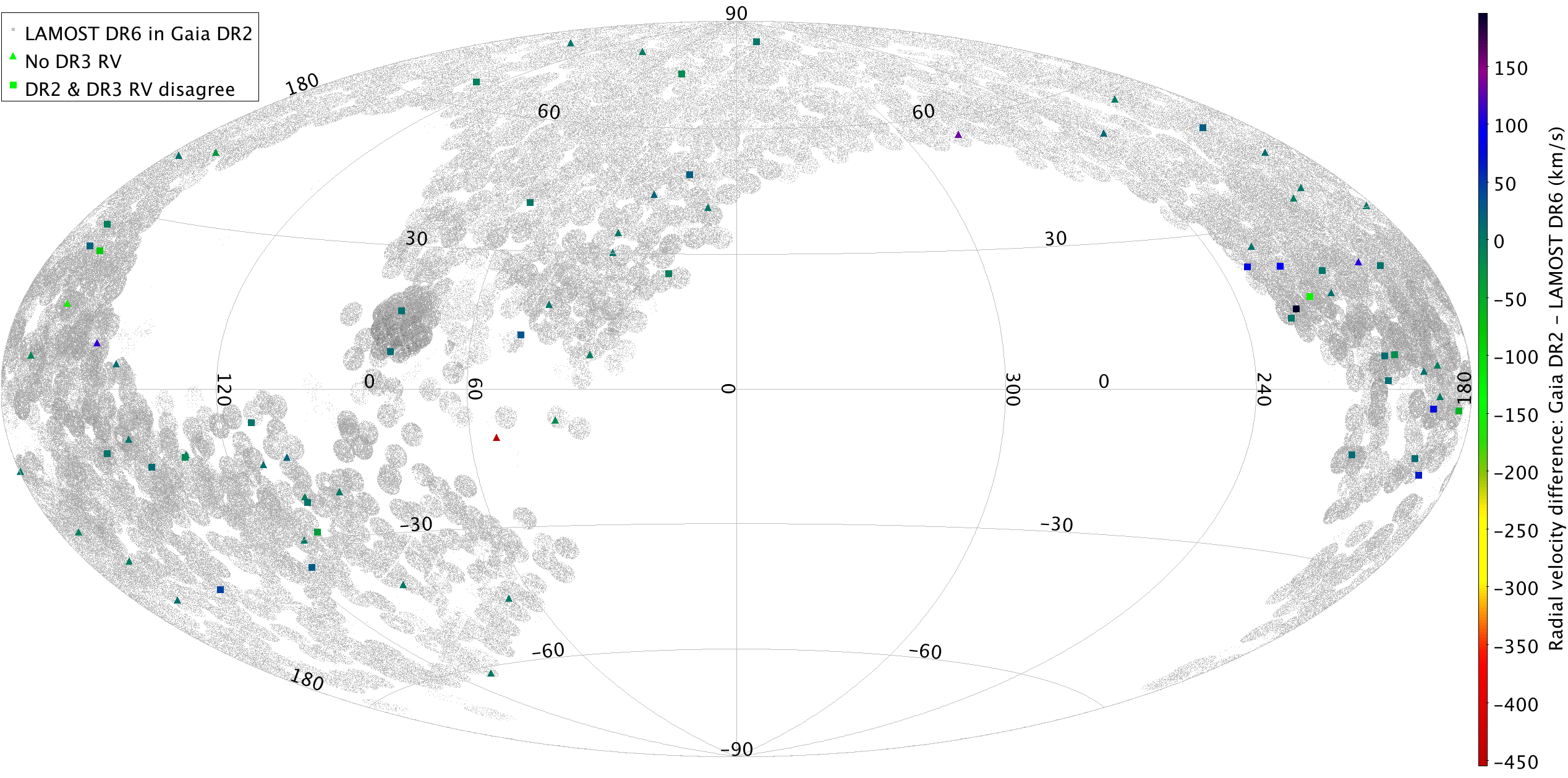}\\
\vspace{1cm}
\caption[]{Galactic Aitoff projection of RAVE DR6 ({\it top}) and LAMOST DR6 ({\it bottom}) stars in {\it Gaia} DR2, both colour-coded according to their respective radial velocity differences.}
\label{fig:rave_lamost2}
\end{figure*}

Figure \ref{fig:rave_lamost} (top right) shows approximate agreement between the LAMOST and {\it Gaia} radial velocities but with some large outliers.  Figure \ref{fig:rave_lamost} (bottom right) hints at the systematic radial velocity difference of 5.38 \kms, reported by \citet{tian2019}.  Out of the 79 LAMOST sources in the 3876, there are 43 (54\%) sources where there is no DR3 radial velocity (triangles) and 36 (46\%) sources where the DR2 and DR3 radial velocities disagree (squares).

\subsubsection{\citet{li2020}}
\label{sec:li2020}

The ``Three New Late-type Hypervelocity Star Candidates from Gaia DR2 by Refined Selection Criteria'' found by \citet{li2020} have the DR2 source IDs 5716044263405220096, 5850309098637075328 and 5966712023814100736.  The first two stars are in EDR3 i.e. their DR2 radial velocities, $-453.5 \pm 2.4$ and $-486.9 \pm 5.0$ \kms, are considered reliable.  However, the last star is not in EDR3 i.e. its DR2 radial velocity, $-967.7 \pm 5.8$ \kms, is no longer considered reliable (Table \ref{table:summary_neg}).

DR2 1995066395528322560 is in the B19 list and is excluded from EDR3 because its DR3 radial velocity not being consistent with the DR2 one:  $-799.1 \pm 1.1$ \kms.   \citet{li2020} obtained a radial velocity of $5 \pm 34$ \kms, which is consistent with the preliminary DR3 value.  This confirms the decision to exclude its DR2 radial velocity from EDR3.

\subsubsection{Comparison with the literature: summary}

Comparing {\it Gaia} DR2 radial velocities with the literature used by \citet{katz2019} (six sources in Sect. \ref{sec:katz}), RAVE (59 sources in Sect. \ref{sec:rave}) and LAMOST (79 sources in Sect. \ref{sec:lamost}) reveals mainly good agreement with a few large outliers.  The bottom of \figref[fig:gbs] and \figref[fig:rave_lamost2] reveal that there is a paucity of literature values towards the Galactic bulge.  However, this is where the majority of the 3876 sources (that had their DR2 radial velocities removed from EDR3) reside (\figref[fig:rejected_aitoff]).  The Large Sagittarius Star Cloud alone accounts for about 20\%.  This means the literature sources are not representative of the majority of the 3876 sources and hence it is not possible to fully verify the choice of filter for these sources.  Nevertheless, the literature sources do represent a minority of the 3876 sources, which generally agree with the {\it Gaia} DR2 radial velocities.  These are mainly false positives of the filter but not all are because some of the literature sources are large outliers.

There is sufficient metadata persisted in the CU6-DR3 pipeline to identify the brightness of contaminating sources in blends.  The radial velocities from the literature presented here could have been used to determine the difference in brightness at which radial velocity contamination starts.  However, this is complicated by the window geometry being different in some or all of the transits, potentially leading to different amounts of contamination in each transit, which would need to be propagated to the final DR2 radial velocity.  This is beyond the scope of this paper but it will occur naturally as a consequence of deblending in the CU6-DR3 pipeline and so the DR3 radial velocities will supersede the EDR3 ones.

There are 2922 sources without a preliminary DR3 radial velocity, which is 4\% of the B19 list.  This includes S16 (Appendix \ref{sect:1803504050895768704}), which has the largest radial velocity difference between DR2 and the literature in \figref[fig:rave_lamost] and does not have a preliminary DR3 radial velocity.  Although it may be an inefficient way of finding spurious DR2 radial velocities, excluding the 2922 DR2 radial velocities of sources without a preliminary DR3 radial velocity ensures removing sources like S16 from EDR3.

When a source has a radial velocity in both DR2 and preliminary DR3 and also in the literature, these values may disagree owing to intrinsic radial velocity variations.  The larger the difference the more likely it is that the difference is because of contaminated DR2 radial velocities.  Nevertheless, the CU6-DR3 pipeline may not have removed all contamination from the preliminary DR3 radial velocities and the literature values may also be affected by their own features (such as S14 in RAVE, Appendix \ref{sect:3754815441303370496}).

\section{Discussion}
\label{sec:discussion}

DR2 radial velocities of sources in the B19 list, and high-velocity stars not in the B19 list, are identified as contaminated and removed from EDR3 by comparing their DR2 radial velocities with their corresponding DR3 radial velocities (Sect. \ref{sec:rv}).  Because the DR3 radial velocities are preliminary and unpublished, they have their own limitations and so come with the following caveats.

The first caveat is that this approach assumes all the DR3 radial velocities are correct.  They remain preliminary until they are validated and published.  This means it is possible that some DR3 radial velocities that are used to validate DR2 radial velocities will be later found to be invalid and removed from DR3.  It may mean that the DR2 radial velocity is actually correct and is excluded from EDR3 erroneously.  If this happens, it is likely to be a small number of sources affected.

The second caveat is that the DR3 pipeline includes deblending of windows but this is limited to those that are overlapping each other.  We investigate a high-velocity star not in the B19 list (DR2 5305975869928712320, referred to in Appendix \ref{sec:appendix_hvs} as S3).  Its DR2 radial velocity of $-830.6 \pm 5.6$ \kms~is found to be caused by contamination from a very bright neighbour (DR2 5305975869928710912,  referred to in Appendix \ref{sec:appendix_hvs} as S4).  S4's \ion{Ca}{ii} absorption lines are just visible in S3's spectra at the wavelengths expected to yield S3's DR2 radial velocity, even though there is a seven AC pixel gap between the S3 and S4 windows (1.26 arcsec) i.e. S3's and S4's windows are non-overlapping.  Therefore, S3's DR2 radial velocity is excluded from EDR3.  

S3's DR3 radial velocity uncertainty is sufficiently large (294.6 \kms) that it will be excluded from DR3 as well.  It is so large because S3's four DR3 transits have different scan angles, causing different angular separations between S3 and S4.  The projection of this separation in the AL direction determines the radial velocity of the contaminating flux in the contaminated window.  However, it is possible that much fainter sources ($G_{\mathrm{RVS}} \approx 14$ mag), closer to even brighter sources but with windows not overlapping, could experience contamination that would cause their radial velocity uncertainty to vary but at a level that is smaller than the expected DR3 filter on radial velocity uncertainty (40 \kms, Sartoretti et al., in prep.).  Their DR2 radial velocities may be similar enough to their DR3 ones that they passed the radial velocity criteria and so they remain in EDR3.  

Validation of this complication prior to the publication of DR3 may find that this can occur and such radial velocities can be excluded from DR3.  If they also invalidate DR2 radial velocities in EDR3, the EDR3 Known Issues webpage\footnote{\url{https://www.cosmos.esa.int/web/gaia/edr3-known-issues}} will be updated.  This limitation of the DR3 pipeline will be upgraded in the DR4 pipeline by attempting to identify and flag for exclusion non-overlapping windows that are contaminated by bright, nearby sources.

The scientific potential of {\it Gaia} data close to bright stars is illustrated by the rediscovery of the \citet{auner1980} star cluster by \citet{koposov2017}, which is 10 arcminutes from the brightest star in the sky, Sirius.  Calibrating the RVS AC LSF wings outside RVS windows would allow non-overlapping windows close to bright sources to be deblended and their decontaminated radial velocities to be derived.  This will be challenging because it relies on the serendipitous acquisition of windows close to bright stars.  The most numerous are faint sources but these are read out as 1D windows.  There are also Virtual Objects (empty windows) but these are also 1D windows and less numerous.  Occasionally 1D windows are read out as 2D Calibration Faint Star windows \citep{cropper2018}.  It may be possible to calibrate the wings of the AC LSF from these observations in order to maximise the radial velocity output in future data releases. 

B19's search region of 6.4 arcsec assumed contamination from overlapping windows, which requires the scan angle to go through the positions of both sources.  This is not the case in S3 so the windows do not overlap.  While the angular distance is greater than 7 arcsec between S3 and S4 (which is why S3 is not in the B19 list), the scan angle not going through both sources means the angular distance between the windows can be foreshortened, in this case to 1.26 arcsec.  S4 is bright enough that its AC LSF can reach S3's window and contaminate it.  This is a limitation of the B19 list, and by limiting the radial velocity comparison of DR2 and DR3 to the B19 list and high-velocity stars, it is also a limitation of this paper.  S3 was excluded from EDR3 only because it was a high-velocity star not in B19 that was investigated individually (Appendix \ref{sec:5305975869928712320}).  It means that EDR3 sources greater than 6.4 arcsec away from very bright sources with less extreme DR2 radial velocities could still have contaminated radial velocities.  

Another limitation of both B19 and this paper is that contamination by sources from {\it Gaia}'s other field-of-view are not considered.  However, the other field-of-view is constantly changing, meaning that bright star contamination should affect only a small fraction of transits, which is further ameliorated by the higher number of transits in DR3.  

\section{Conclusions}
\label{sec:conc}

 {\textit{Gaia}'s Early Third Data Release (EDR3) does not contain new radial velocities because these will be published in \textit{Gaia}'s full third data release (DR3), expected in the first half of 2022.  EDR3 is based on a new astrometric solution \citep{lindegren2021} and a new source list \citep{torra2021}, which means sources in DR2 may not be present in EDR3 owing to the sources merging in with others or splitting into new ones.  To maximise the usefulness of EDR3, \textit{Gaia}'s second data release (DR2) radial velocities are propagated to EDR3 sources, which is the subject of this paper.  Two aspects of this propagation improve EDR3 over DR2.
 
The primary improvement of EDR3 over DR2 is that the contaminated DR2 radial velocities are removed from EDR3.  This improvement is limited to the 70\,365 sources with potentially contaminated DR2 radial velocities published by \citet{boubert2019}, hereafter the B19 list, and high-velocity stars not in the B19 list. The best radial velocity coverage of the B19 list is by the unpublished, preliminary DR3 radial velocities.  DR3 has the advantage of treating overlapping RVS windows by deblending their fluxes before deriving their radial velocities. DR2 did not include deblending but it did exclude the majority of overlapping windows from the radial velocity determination.  However, amongst the minority that were not excluded was DR2 5932173855446728064, the DR2 radial velocity of which was confirmed to be erroneous by B19 ground-based follow-up observations.  DR3 also has more transits to derive radial velocities from so it is assumed that DR3 radial velocities are uncontaminated and that if DR2 radial velocities differ significantly from the EDR3 values, then the DR2 radial velocities are contaminated and should be removed from EDR3.  

3871 DR2 radial velocities in the B19 list were removed from EDR3 either because they are significantly different from the DR3 radial velocities or because the CU6-DR3 pipeline did not produce a radial velocity.  All of these sources were cross-matched with Radial Velocity Experiment DR6 \citep{steinmetz2020} and Large Sky Area Multi-Object Fiber Spectroscopic Telescope DR6.  This found that most of the radial velocities excluded from EDR3 are in agreement with the literature values, which represent the lower limit of false positives of the filter.  However, it is not possible to quantify the true positives of the filter because the 3781 sources reside mainly towards the Galactic bulge, where there was a paucity of literature radial velocities.  Nevertheless, the filter has successfully removed some large discrepancies, which was the main aim of this work.  DR2 radial velocities have been successfully cleaned and improved for EDR3 at the expense of removing a small fraction of DR2 radial velocities.  The 3871 removed are 0.05\% of the 7\,224\,631 sources with a DR2 radial velocity, 5.5\% of the B19 list and 26\% of DR2 radial velocities excluded from EDR3.
 
The other improvement of EDR3 over DR2 is that the EDR3 astrometric processing suppresses the secondary images of close binary pairs better than in DR2.  Both DR2 and EDR3 fit single-star models in the astrometric processing.  The DR2 astrometry (and associated DR2 radial velocity) of close binary pairs may refer to either component.  The EDR3 astrometry (and associated DR2 radial velocity) of close binary pairs refers preferentially to the primary component.  Consequently, a match could not always be found between DR2 and EDR3 sources.  10\,924 DR2 sources could not be satisfactorily matched to any EDR3 sources so their DR2 radial velocities are also missing from EDR3.  This corresponds to 73\% of the DR2 radial velocities excluded from EDR3, which is not related to DR2 radial velocity quality.  Source ambiguity has arisen because of needing to match DR2 sources to (E)DR3 sources in order to transfer DR2 radial velocities to EDR3 sources.  This complication will not arise in DR3 because DR3 astrometry and source identification has been used consistently in the DR3 pipelines.

In total, 14\,800 radial velocities from DR2 are not propagated to any EDR3 sources, which is 0.2\% of the number of DR2 sources with radial velocities.  Such a small selection effect is unlikely to bias the comparison of Galaxy-level population models to the EDR3 sources with radial velocities.  However, about 20\% of the spurious DR2 radial velocities removed from EDR3 are towards the Large Sagittarius Star Cloud (\figref[fig:rejected_aitoff]), which coincides with the largest incompleteness of sources with DR2 radial velocities (\citealt{katz2019} fig. 7).  Here the selection effect may be more significant but it should be possible to model this with publicly available tools (e.g. \citealt{boubert2020a,boubert2020b,boubert2021,everall2021}).

{\it Gaia}'s radial velocities are expected to enable the discovery of hundreds of hypervelocity stars \citep{marchetti2018a}.  The aim of the processing improvements described in this paper is to clean EDR3 radial velocities to avoid false-positive hypervelocity candidates, as occurred with DR2 \citep{boubert2019}.  EDR3 has already been used to search for hypervelocity stars \citep{marchetti2021}.  As the number of sources with radial velocities increases from EDR3 ($\sim$7 million) to DR3 ($\sim$30 million) to DR4 ($\sim$150 million\footnote{\url{https://www.cosmos.esa.int/web/gaia/science-performance}}), the challenge of excluding spurious radial velocities to produce reliable hypervelocity candidates increases.  The lessons learnt from individually investigating high-velocity stars in DR2 are being applied to the validation of DR3 radial velocities, which will occur until they are published in 2022.  These lessons are also being applied to the design of the CU6-DR4 pipeline.  Continuing the symbiotic relationship between DPAC and the community, as exemplified by  \citet{boubert2019}, will continue to improve the reliability of {\it Gaia}'s radial velocities.

\begin{acknowledgements}
This work made use of data from the European Space Agency (ESA) mission {\it Gaia} (\url{https://www.cosmos.esa.int/gaia}), processed by the {\it Gaia} Data Processing and Analysis Consortium (DPAC, \url{https://www.cosmos.esa.int/web/gaia/dpac/consortium}). Funding for the DPAC has been provided by national institutions, in particular the institutions participating in the {\it Gaia} Multilateral Agreement. Most of the authors are current or past members of the ESA {\it Gaia}
mission team and of the {\it Gaia} DPAC and their work has been supported by 
the French Centre National de la Recherche Scientifique (CNRS), the Centre National
d' Etudes Spatiales (CNES), the L'Agence Nationale de la Recherche, the
R\'{e}gion Aquitaine, the Universit\'e de Bordeaux, the Utinam Institute of the Universit\'e
de Franche-Comt\'e, and the Institut des Sciences de l' Univers (INSU); the United Kingdom Particle Physics and Astronomy Research Council (PPARC), the United Kingdom Science and Technology Facilities Council (STFC), and the United Kingdom Space Agency (UKSA) (through the following grants to the University of Bristol, the University of Cambridge, the University of Edinburgh, the University of Leicester, the Mullard Space Sciences Laboratory of University College London, and the United Kingdom Rutherford Appleton Laboratory (RAL): PP/D006511/1, PP/D006546/1, PP/D006570/1, ST/I000852/1, ST/J005045/1, ST/K00056X/1, ST/K000209/1, ST/K000756/1, ST/L006561/1, ST/N000595/1, ST/N000641/1, ST/N000978/1, ST/N001117/1, ST/S000089/1, ST/S000976/1, ST/S001123/1, ST/S001948/1, ST/S002103/1, and ST/V000969/1) ; the Belgian Federal Science Policy Office (BELSPO) through various
Programme de D\'{e}veloppement d'Exp\'{e}riences Scientifiques (PRODEX) grants;
the German Aerospace Agency (Deutsches Zentrum fur Luft- und Raumfahrt
e.V., DLR); the Algerian Centre de Recherche 
en Astronomie, Astrophysique et G\'{e}ophysique of Bouzareah Observatory;
the Swiss State Secretariat for Education, Research,
and Innovation through the ESA PRODEX programme, the Mesures
d'Accompagnement, the Swiss Activit\'{e}s Nationales Compl\'{e}mentaires, and the
Swiss National Science Foundation; the Slovenian Research Agency (research
core funding No. P1-0188).\\
This work was supported by the Spanish Ministry of Science, Innovation and
University (MICIU/FEDER, UE) through grants RTI2018-095076-B-C21,
ESP2016-80079-C2-1-R, and the Institute of Cosmos Sciences University of
Barcelona (ICCUB, Unidad de Excelencia 'Mar\'{\i}a de Maeztu') through grants
MDM-2014-0369 and CEX2019-000918-M.\\
TA has received funding from the European Union's Horizon 2020 research and innovation programme under the Marie Sk\l{}odowska Curie grant agreement No. 745617 and from the Ramon y Cajal Fellowship RYC2018-025968-I.\\
The Digitized Sky Surveys were produced at the Space Telescope Science Institute under U.S. Government grant NAG W-2166. The images of these surveys are based on photographic data obtained using the Oschin Schmidt Telescope on Palomar Mountain and the UK Schmidt Telescope. The plates were processed into the present compressed digital form with the permission of these institutions. The National Geographic Society - Palomar Observatory Sky Atlas (POSS-I) was made by the California Institute of Technology with grants from the National Geographic Society. The Second Palomar Observatory Sky Survey (POSS-II) was made by the California Institute of Technology with funds from the National Science Foundation, the National Geographic Society, the Sloan Foundation, the Samuel Oschin Foundation, and the Eastman Kodak Corporation. The Oschin Schmidt Telescope is operated by the California Institute of Technology and Palomar Observatory. The UK Schmidt Telescope was operated by the Royal Observatory Edinburgh, with funding from the UK Science and Engineering Research Council (later the UK Particle Physics and Astronomy Research Council), until 1988 June, and thereafter by the Anglo-Australian Observatory. The blue plates of the southern Sky Atlas and its Equatorial Extension (together known as the SERC-J), as well as the Equatorial Red (ER), and the Second Epoch [red] Survey (SES) were all taken with the UK Schmidt. Supplemental funding for sky-survey work at the ST ScI is provided by the European Southern Observatory.\\
This publication makes use of data products from the Two Micron All Sky Survey, which is a joint project of the University of Massachusetts and the Infrared Processing and Analysis Center/California Institute of Technology, funded by the National Aeronautics and Space Administration and the National Science Foundation.\\
Funding for Rave has been provided by: the Leibniz Institute for Astrophysics Potsdam (AIP); the Australian Astronomical Observatory; the Australian National University; the Australian Research Council; the French National Research Agency; the German Research Foundation (SPP 1177 and SFB 881); the European Research Council (ERC-StG 240271 Galactica); the Istituto Nazionale di Astrofisica at Padova; The Johns Hopkins University; the National Science Foundation of the USA (AST-0908326); the W. M. Keck foundation; the Macquarie University; the Netherlands Research School for Astronomy; the Natural Sciences and Engineering Research Council of Canada; the Slovenian Research Agency; the Swiss National Science Foundation; the Science \& Technology Facilities Council of the UK; Opticon; Strasbourg Observatory; and the Universities of Basel, Groningen, Heidelberg and Sydney.\\
Guoshoujing Telescope (the Large Sky Area Multi-Object Fiber Spectroscopic Telescope LAMOST) is a National Major Scientific Project built by the Chinese Academy of Sciences. Funding for the project has been provided by the National Development and Reform Commission. LAMOST is operated and managed by the National Astronomical Observatories, Chinese Academy of Sciences.\\
Most of the analysis in this paper was done in the software TOPCAT \citep{taylor2005}.  This research has made use of ``Aladin sky atlas" developed at CDS, Strasbourg Observatory, France \citep{bonnarel2000} and the DPAC tool, FoVInspector, developed by Berry Holl (CU7, University of Geneva).\\
We are very grateful to Josh Cooke, who worked on this project in summer 2019 as a physics undergraduate student from the University of York.  His work was superseded by the DPAC-internal availability of the preliminary DR3 radial velocities in spring 2020.\\
We thank the anonymous referee for their constructive feedback that helped to clarify the paper.
\end{acknowledgements}

 \bibliographystyle{aa} % style aa.bst
 \bibliography{seabroke}

\begin{appendix}

\section{Investigating individual high-velocity stars}
\label{sec:appendix_hvs}

\subsection{Introduction}

Section \ref{sec:hvs_dr3} identifies six high-velocity stars not in the B19 list to investigate further.  One of the sources (DR2 5305975869928712320) has a radial velocity in both DR2 and (preliminary) DR3 but with a very large DR3 uncertainty (294.6~\kms).   The other five do not have a DR3 radial velocity.  Their photometry, leading to their selection for the CU6-DR2 pipeline, is collectively investigated in Sect. \ref{sec:hvs}.  The following sections investigate each source individually to assess the provenance of their DR2 radial velocities and whether they should be included in EDR3.  Table \ref{table:symbol} provides a reference for the shorthand used to refer to these six sources and their neighbours in the following sections.  Table \ref{table:transit} provides a reference for the transit IDs mentioned in the following sections and their decoding. 

\begin{table}
\caption{DR2 source IDs and how they are referred to in this work.  Their EDR3 source IDs are the same as their DR2 ones.  S1-2 are discussed in the main text, S3-13 are discussed in this appendix and S14-17 are discussed in Appendix \ref{sect:appendix3}.}             
\label{table:symbol}      
\centering          
\begin{tabular}{lr}     
\hline\hline 
DR2 source ID & This work \\
\hline
5932173855446728064 & S1\\
5932173855446724352 & S2\\
5305975869928712320 & S3\\
5305975869928710912 & S4\\
5966712023814100736 & S5\\
5966713496979650304 & S6\\
4658865791827681536 & S7\\
4658865791827521408 & S8\\
5827538590793373696 & S9\\
5827538590793371776 & S10\\
4092328917916154368 & S11\\
4092328917911305984 & S12\\
5413575658354375040 & S13\\
3754815441303370496 & S14\\
3754815445598430080 & S15\\
1803504050895768704 & S16\\
1803504050895769472 & S17\\
\hline             
\end{tabular}
\end{table}

\begin{table}
\caption{Transit IDs and their decoding of individual high-velocity stars.}             
\label{table:transit}      
\centering          
\begin{tabular}{cll}     
\hline\hline 
Source & transit ID & Decoded transit ID\tablefootmark{a} \\
\hline
S3 & 16072612362908400 & 1162-069613365-2-7-0752\\
S4 & 16072612298814210 & 1162-069612874-2-7-0770\\
\hline 
S5 & 29710227536893472 & 2149-018703742-1-5-1568\\
S6 & 29710227419321968  &  2149-018702845-1-5-1648\\
\hline 
S5 & 45502366085051575 & 3291-057846718-1-7-1207\\ 
S6 & 45502366221628670  & 3291-057847760-1-7-1278\\
\hline 
\end{tabular}
\tablefoot{
\tablefoottext{a}{Decoded transit ID: {\it Revolution-Part\_of\_revolution-FoV-Row-AF1Column}, where {\it Revolution} is defined in \citet{gaia2016}, {\it Part\_of\_revolution} is in units of 0.2048 ms ($\approx$0.21 TDI) so each revolution is subdivided into exactly 105468750 such steps, {\it FoV} is field of view (1 or 2), {\it Row} is CCD row 1-7 (4-7 for RVS) and  {\it AF1Column} gives the AC position of the window in the first astrometric CCD strip.}
  }
\end{table}

\subsection{DR2 5305975869928712320}
\label{sec:5305975869928712320}

\begin{figure*}
\centering
\includegraphics[width=0.75\textwidth]{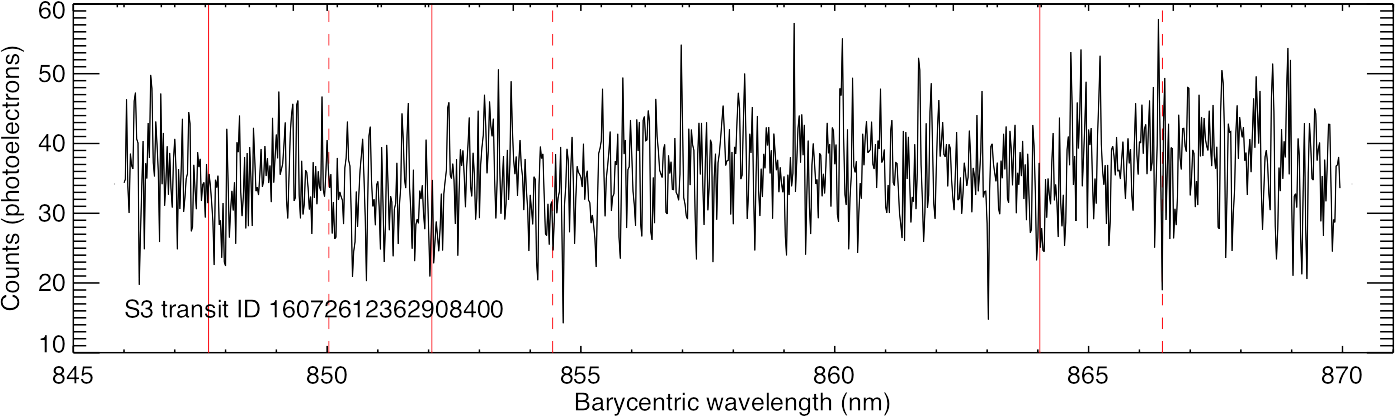}\\
\vspace{0.5cm}
\includegraphics[width=0.75\textwidth]{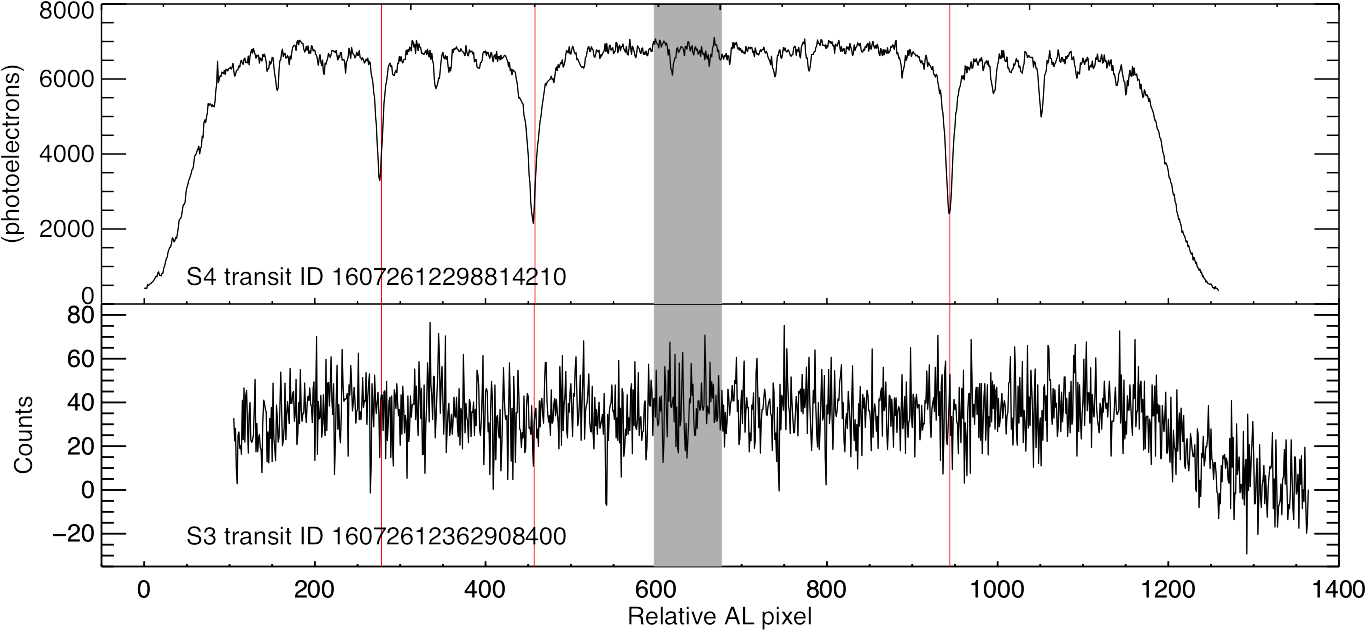}
\caption[]{The first DR2 transit (three-CCD averaged) spectrum of S3.  The red dashed lines delineate the \ion{Ca}{ii} absorption line rest wavelengths.  The red solid lines are the aforementioned rest wavelengths, Doppler shifted according to the transit radial velocity.  {\it Bottom plots}: Strip 16 (RVS strip 2) CCD spectra of the first DR2 transit of S3 and the corresponding transit of S4.  The red vertical lines are the same as for the top plot but plotted in relative AL pixel space.  The grey shaded area corresponds to the central waveband in the top sub-panel: 858-860 nm, which corresponds to waveband 856-858 nm in the bottom sub-panel.}
\label{fig:spectra5305975869928712320}
\end{figure*}

Figure \ref{fig:spectra5305975869928712320} plots the first DR2 transit (three-CCD averaged) spectrum of DR2 5305975869928712320 (hereafter S3, Table \ref{table:symbol}).\footnote{The CU6-DR2 pipeline did not persist CCD spectra but the CU6-DR3 pipeline did so \figref[fig:spectra5305975869928712320] is based on CU6-DR3 pipeline outputs.  These are expected to be the same as CU6-DR2 pipeline outputs, with the exception of more accurate straylight removal in the CU6-DR3 pipeline.  Even so, windows close to each other are expected to experience the same straylight pattern so the relative flux baseline is expected to be the same.}  It shows that the wavelengths of the middle and reddest \ion{Ca}{ii} absorption lines, Doppler shifted according to the transit's radial velocity, coincide with what appears to be absorption lines.  The question is whether these lines belong to S3 or to the close bright source in \figref[fig:sky] (top left panel), DR2 5305975869928710912 (HD 84676, hereafter S4, Table \ref{table:symbol}).  S4's DR2 $T_{\textrm{eff}} = 4661^{+15}_{-67}$ K \citep{rene2018} suggests that it is a K-type star with \ion{Ca}{ii}-dominated spectra. 

The bottom plots in \figref[fig:spectra5305975869928712320] are aligned in AL as the windows would have been seen simultaneously on the CCD i.e. S4 transit ID 16072612298814210 is observed one macrosample (105 AL pixels) earlier than S3 transit ID 16072612362908400 (TDI is clocking from right to left, Table \ref{table:transit}).  The solid red vertical lines from the top plot are overlaid in the bottom plots (unlike in the three-CCD averaged transit spectrum, the \ion{Ca}{ii} lines in the single-CCD spectrum of S3 are not very visible).  It shows they coincide with the relative AL positions of the strong \ion{Ca}{ii} lines in the S4 CCD spectrum, suggesting that S3's spectra have been contaminated with flux from S4.   

The position of the peak of the AC profile is not always at the AC centre of the window and so can be `decentred'.  AC decentring is measured by the CU6-DR3 pipeline at the AL centre of each window.  This is why the peak of each AC profile is not at the centre of either window in \figref[fig:ac_lsf5305975869928712320].  The AL centre is the only point where the AC decentring was calculated so we chose the AC LSF waveband that includes the AL centre of the brighter window (the wavebands are displayed in \figref[fig:spectra5305975869928712320]).  The AC LSFs were scaled to the mean flux of their corresponding spectra to reconstruct their AC profiles.  

The $G_{\mathrm{RVS}}$-predicted line is derived by using the trended $G_{\mathrm{RVS}}$ zero point to convert $G_{\mathrm{RVS}}^{\mathrm{ext,2}}$ to the average flux in a sample, and using this to scale the AC LSF.  Figure \ref{fig:ac_lsf5305975869928712320} shows the predicted and reconstructed lines are similar for S4.  This is not the case for S3.  Its reconstructed AC profile  is much larger than its predicted one (red dashed and solid lines in \figref[fig:ac_lsf5305975869928712320], respectively).  The predicted contribution of S3 and S4 to S3's window per AL sample is 7.6 and 22.6 electrons, respectively.  Thus the predicted total in S3's window is 30.2 electrons, which is close to the measured total of 32.5 electrons.   The reconstructed profile in S3's window does not assume any contamination.  Therefore, S3's actual AC profile is likely to look more like the S4's contribution to the window than S3's reconstructed one. 

AC LSFs are calibrated from the AC profiles of 2D windows that have all their CCD columns read out separately.  The AC decentring of profiles means that the AC LSF can be directly calibrated one or two pixels outside the nominal window limits ($\pm5$ pixels) when combining multiple observations.  However, outside of the directly calibrated region, the AC LSFs are extrapolated out to $\pm$20 AC pixels.  A further complication is that S3's $G_{\mathrm{RVS}}^{\mathrm{ext,2}}$ may be contaminated because more than half its RP transits are classified as blended (Table \ref{table:notB19}).  Therefore on their own, the AC profiles do not provide strong evidence that S4's AC profile can contaminate S3. 

The small uncertainty in S3's DR2 radial velocity (5.8 \kms) suggests that, if both transits are contaminated, it needs to happen in a very similar way in both transits.  Although not plotted here, the second transit is similar to the first in all respects.  The {\it Gaia} scan angle is very similar for these two transits and the direction is the same, from bottom right (south west) to top left (north east) of the top left panel in \figref[fig:sky], even though the transits are separated by 269 days.

There is enough evidence to suggest that S3's DR2 radial velocity is contaminated by S4 and so it has been excluded from EDR3.

\begin{figure}
\centering
\includegraphics[width=0.9\columnwidth]{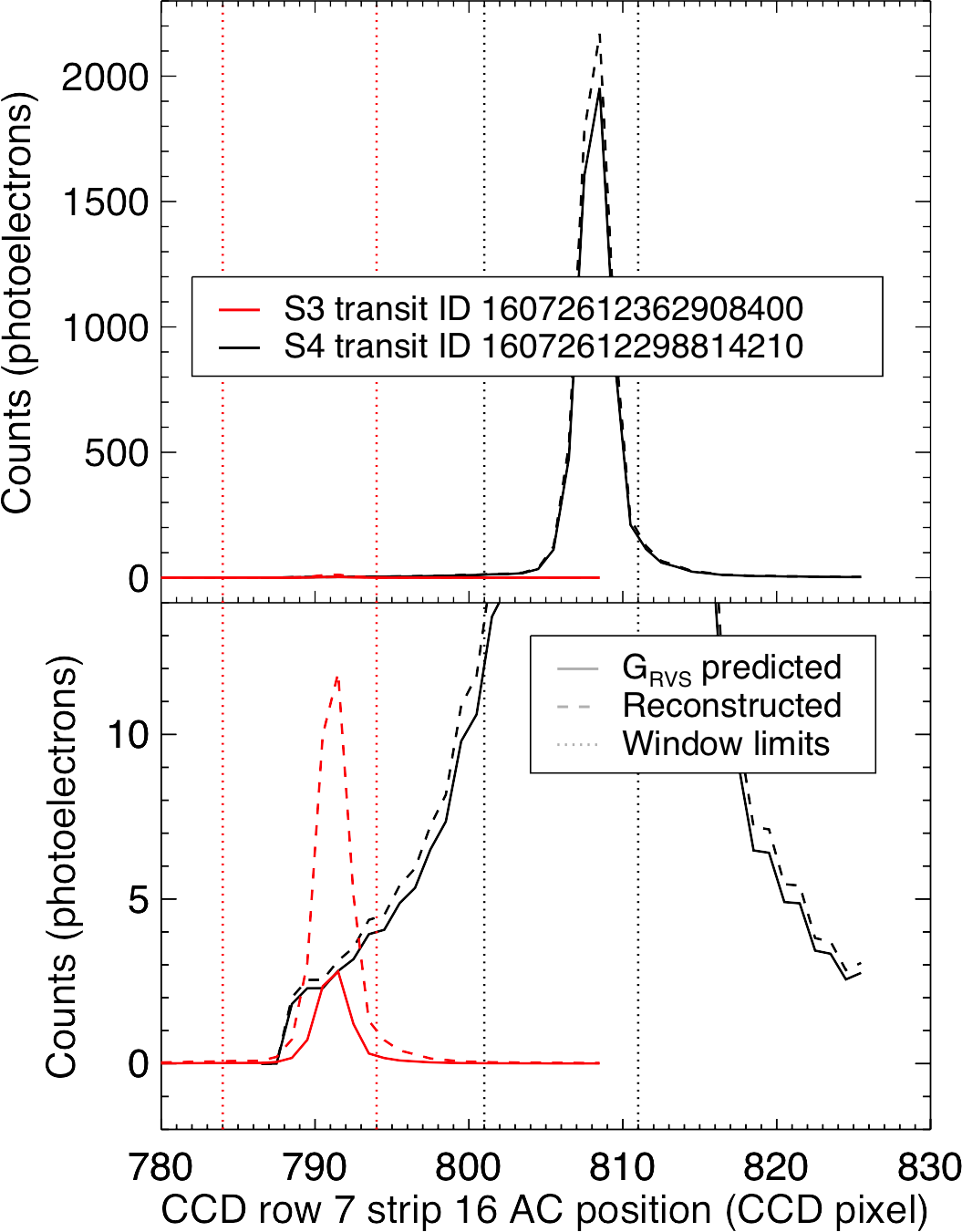}
\caption[]{$G_{\mathrm{RVS}}$-predicted and reconstructed AC profiles of the first DR2 transit of S3 and the corresponding transit of S4 and the AC extent of their windows.}
\label{fig:ac_lsf5305975869928712320}
\end{figure}

\subsection{DR2 5966712023814100736}
\label{Sect:5966712023814100736}

Figure \ref{fig:sky} (top right panel) shows that the PSF of DR2 5966712023814100736 (hereafter S5, Table \ref{table:symbol}) is not disturbed by the brighter DR2 5966713496979650304 (hereafter S6, Table \ref{table:symbol}).  This suggests that the S5's $R_F$ magnitude is not significantly contaminated.  The blue photographic plate from which S5's $B_J$ was measured is missing from the Aladin sky atlas and may be one of the original 11 missing plates.  If a default $B_J$ was used, it could explain why S5 has a much brighter $G_{\mathrm{RVS}}^{\mathrm{ext,1}}$ (11.1 mag) than $G_{\mathrm{RVS}}^{\mathrm{ext,2}}$ (14.9 mag) and why it was selected for the CU6-DR2 pipeline ($G_{\mathrm{RVS}}^{\mathrm{ext,1}} < 12$ mag) but not the CU6-DR3 pipeline ($G_{\mathrm{RVS}}^{\mathrm{ext,2}} < 14$ mag).   

\begin{figure*}
\centering
\includegraphics[width=0.75\textwidth]{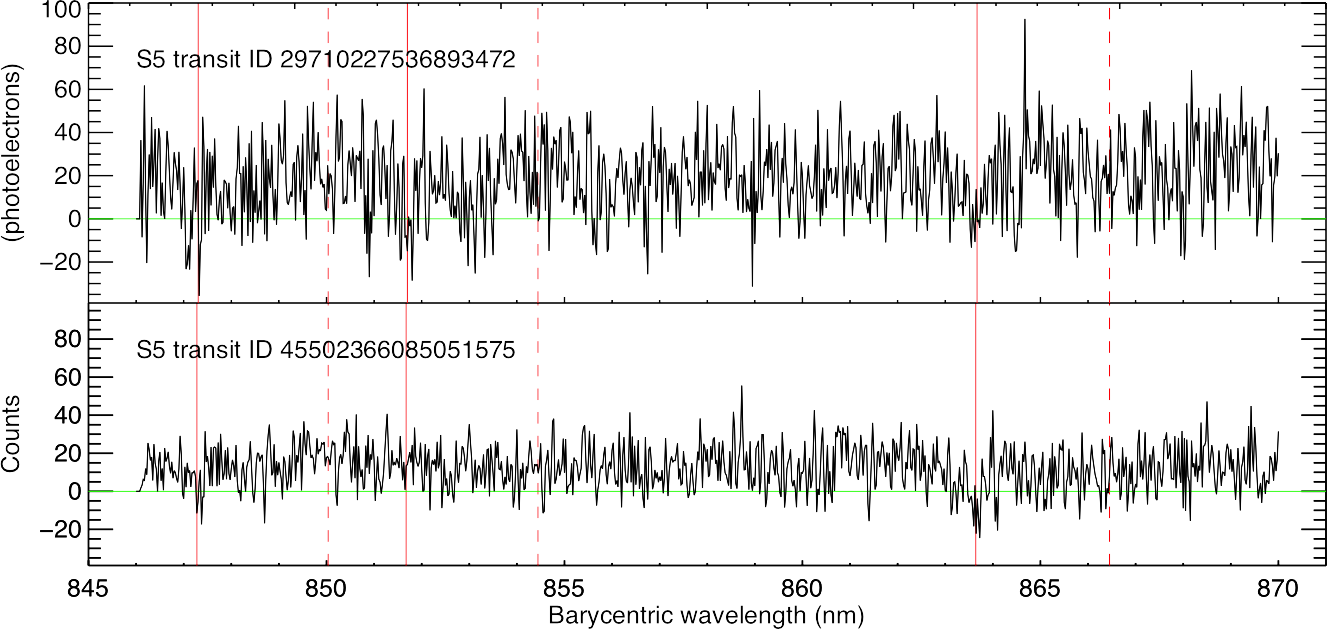}\\
\vspace{1cm}
\includegraphics[width=0.9\columnwidth]{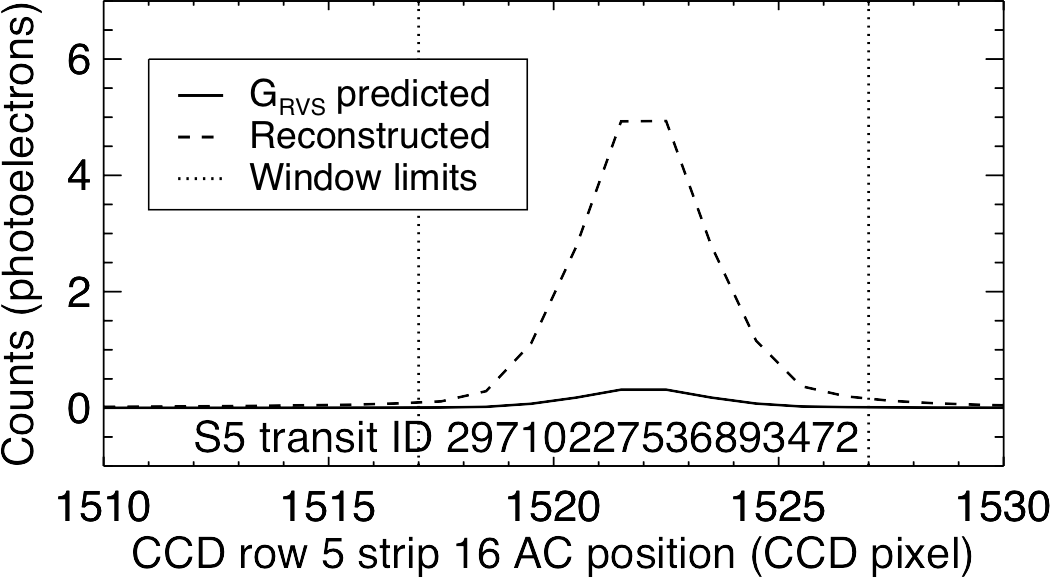}
\hspace{1cm}
\includegraphics[width=0.9\columnwidth]{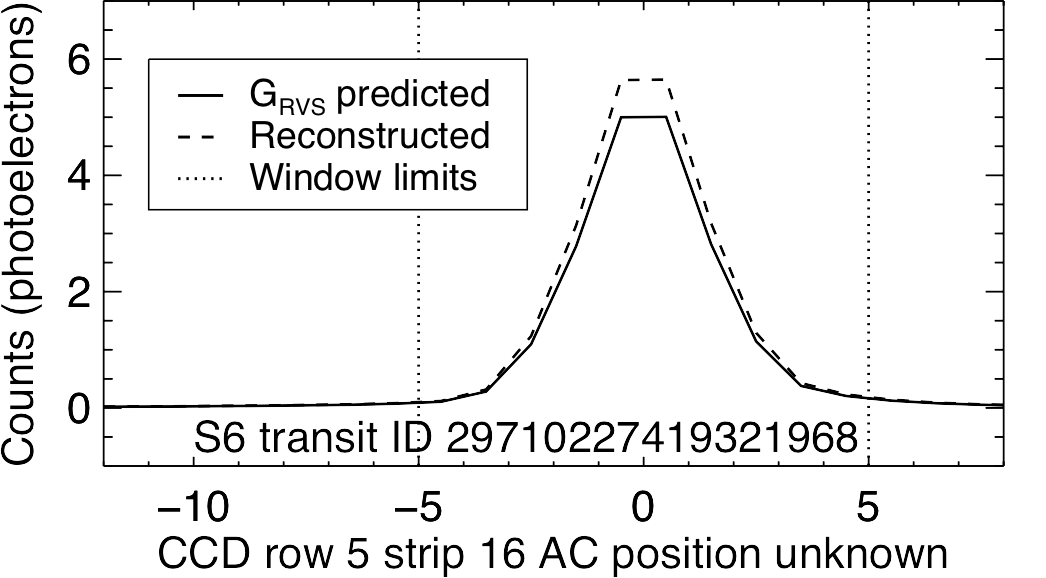}\\
\vspace{1cm}
\includegraphics[width=0.9\columnwidth]{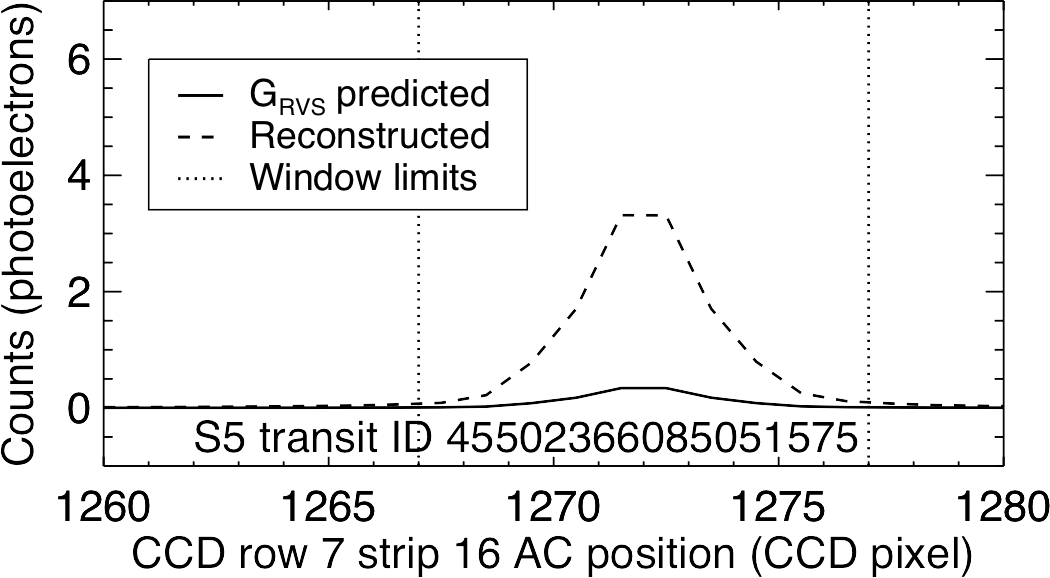}
\hspace{1cm}
\includegraphics[width=0.9\columnwidth]{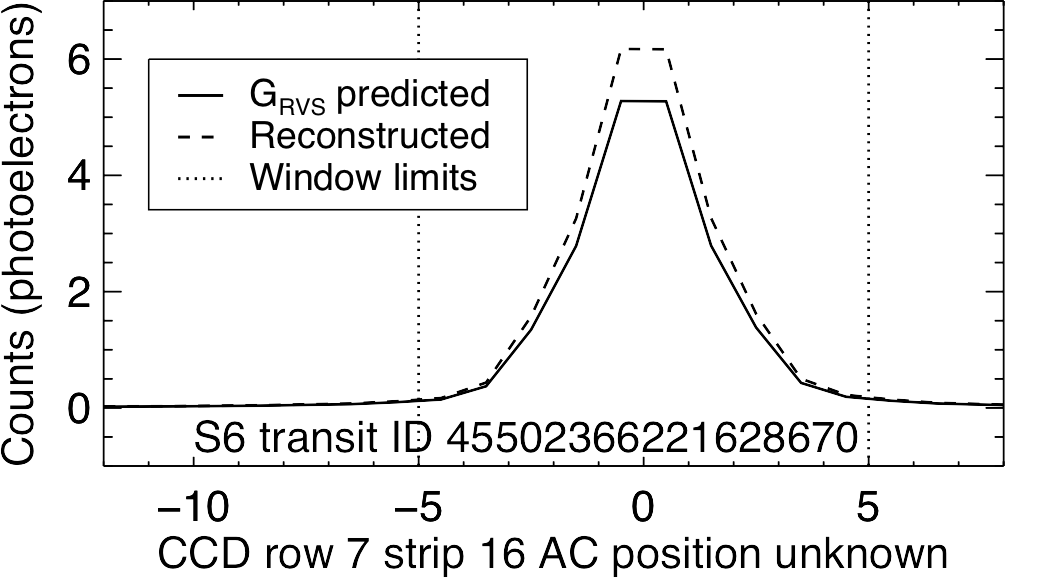}
\vspace{0.5cm}
\caption[]{{\it Top plot}: The two DR2 transit (three-CCD averaged) spectra of S5.  The red dashed lines delineate the \ion{Ca}{ii} absorption line rest wavelengths.  The red solid lines are the aforementioned rest wavelengths, Doppler shifted according to the transit radial velocity.  $G_{\mathrm{RVS}}$-predicted and reconstructed AC profiles of S5 and S6 in the first transit ({\it middle two plots}) and the second transit ({\it bottom two plots}). S6 was processed by the CU6-DR3 pipeline but the two transits closest in time to S5's two transits were excluded because it could not be successfully deblended.  Consequently, its window geometry information was not persisted, meaning its relative AL and AC positions and AC decentring are not known exactly.  S5's AC decentring is not known either because AC decentring is calculated in the CU6-DR3 pipeline and S5 was not processed by the CU6-DR3 pipeline.  Consequently, the AC profiles in the middle and bottom panels are centred in the centre of their windows but this may not be correct.}
\label{fig:ac_lsf5966712023814100736}
\end{figure*}

The CU6-DR2 pipeline CCD spectra of S5 were not persisted and were too faint to be processed by the CU6-DR3 pipeline.  Therefore \figref[fig:ac_lsf5966712023814100736] plots its DR2 transit (three-CCD averaged) spectra instead.  S5's DR2 $T_{\textrm{eff}} = 4166^{+351}_{-228}$ K \citep{rene2018} suggests that it is a K-type star with \ion{Ca}{ii}-dominated spectra. Figure \ref{fig:ac_lsf5966712023814100736} shows that the first transit (top panel) apparently has absorption lines where the middle and reddest \ion{Ca}{ii} lines are predicted by the measured radial velocity.  In the second transit only the reddest \ion{Ca}{ii} line is visible.  

  Neither S5 (Table \ref{table:notB19} in the main text) nor S6 have any contaminated or blended RP transits affecting their $G_{\mathrm{RVS}}^{\mathrm{ext,2}}$ magnitudes.  Nevertheless, the reconstructed AC profiles of S6 in \figref[fig:ac_lsf5966712023814100736] are larger than the $G_{\mathrm{RVS}}$-predicted ones.  The difference could be residual straylight not removed by the CU6-DR3 pipeline but it is not large enough to explain the larger difference between the reconstructed and $G_{\mathrm{RVS}}$-predicted AC profiles of S5.  This suggests {\it that} most of the S5 spectra is contaminating flux, not from S5 itself.  

\figref[fig:ac_lsf5966712023814100736] shows that in the first transit, S6's $G_{\mathrm{RVS}}$-predicted and reconstructed AC profiles are similar to S5's.  S6 can only contribute the required amount of contaminating flux if their windows are aligned in AC.  This could happen if the scan angles were along the arrow joining the two sources in \figref[fig:sky] (top right panel).  However, the scan angles are from top left (north east) to bottom right (south west) in the first transit and top right (north west) to bottom left (south east) in the second transit.  The spectra are dispersed along the scan direction (AL) and thus it is unlikely their windows overlap or are very close.  This is consistent with metadata from the CU6-DR2 pipeline, that the three DR2 windows of S5's first transit are not overlapping any other windows.   

The source south of S5 (DR2 5966712019514710272) is too faint to have a RVS window ($G_{\mathrm{RVS}}^{\mathrm{ext,2}} = 16.4$ mag) and too faint to be the source of contamination in both transits.  In S5's second transit, all three windows overlap other windows but are not truncated.  This means the windows were aligned in AC but not necessarily in AL.  Unlike the CU6-DR3 pipeline, the CU6-DR2 pipeline did not record which windows these were.  The scan angle of the second transit suggests that the windows belong to the source north-north-east of S5 (DR2 5966712775425205632).  Its $G_{\mathrm{RVS}}^{\mathrm{ext,2}} = 15.5$ mag, which is 0.6 mag fainter than S5.  While it is contributing some contaminating flux to the second transit, it does not explain both transits being contaminated.

The middle and reddest \ion{Ca}{ii} lines of K-type spectra normally have similar strength and are the two strongest lines.  This is not the case in S5's second transit spectrum.  Moreover, S5's first transit spectrum appears to have other features, e.g. between 864 and 865 nm, as strong as the supposed \ion{Ca}{ii} lines.  This argues that the features consistent with the \ion{Ca}{ii} lines are spurious.  The features could be because of random shot noise or an instrumental residual.  It may not be a coincidence that S5's first and second transits were observed on rows 5 and 7, respectively (Table \ref{table:transit}).  \citet{hambly2018} fig. 24 reveals that in the RVS CCDs, rows 5 and 7 have the worst and second worst performances in terms of bias non-uniformity.  This means the features could be uncorrected bias non-uniformity features.  

S5 is the source in Table \ref{table:notB19} closest to the Galactic centre: ($l,b$)=(345,0)$^{\circ}$.  The scan angles of S5's first and second transits are approximately parallel and perpendicular to the Galactic plane, respectively.   Therefore, the stellar density in the focal plane was higher for longer in the first transit compared to the second, causing the readout sequence to change more frequently, potentially accumulating more stray capacitance.  Stray capacitance, concluded to be the likely culprit for the bias non-uniformity \citep{hambly2018}, may be higher in S5's first transit than in its second one. Although the origin has not been definitely identified, S5's spectra appear to be contaminated so its DR2 radial velocity has been excluded from EDR3.

\subsection{DR2 4658865791827681536}
\label{sec:4658865791827681536}

Each of the four transits of DR2 4658865791827681536 (hereafter S7, Table \ref{table:symbol})  occurred within about two {\it Gaia} revolutions.  This means each scan angle and direction were very similar.  The scan always goes from right (west) to left (east) in \figref[fig:sky] (middle left plot) such that S7 was observed after its brighter neighbour, DR2 4658865791827521408 (hereafter S8, Table \ref{table:symbol}), each time.  In the first three transits, all the windows of S7 and S8 are offset in AL by one macrosample and aligned in AC (\figref[fig:window_geometry_sameAC_diffAL]).  This means that none of these windows are truncated.  

In the fourth transit, S8's window was excluded from both the CU6-DR2 and CU6-DR3 pipelines but the reason was not recorded.  There is only 0.3 revolutions between the third and fourth transit, suggesting that the fourth transit will have the same window geometry as the other three transits.  

\begin{figure}
\centering
\includegraphics[width=\columnwidth]{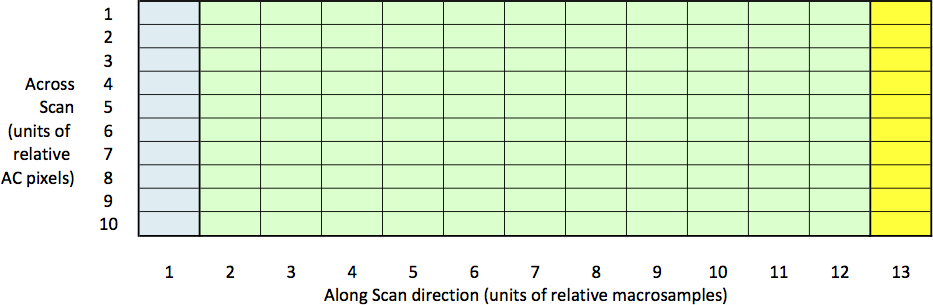}
\caption[]{Same as \figref[fig:window] but the two RVS windows are aligned in AC so do not truncate each other i.e. both windows are 10 AC pixels high.  The blue macrosample is the first leading macrosample, which is not overlapped.  The yellow macrosample is the other window's last trailing macrosample, which is not overlapped.  The green macrosamples belong to both windows and are overlapped but not truncated.}
\label{fig:window_geometry_sameAC_diffAL}
\end{figure}

The CU6-DR2 pipeline excluded windows that were not rectangular, which explains why all the windows of S7 and these four transits of S8 (out of the 16 in total) were not excluded and yielded radial velocities for both sources in DR2.  The difference in brightness between S7 and S8 (4.5 mag) means that S7's DR2 radial velocity is contaminated and so has been excluded from EDR3.  S8's DR2 radial velocity is not contaminated by S7 in these four transits and so remains in EDR3.

\subsection{DR2 5827538590793373696}
\label{sec:5827538590793373696}

Each of the two transits of DR2 5827538590793373696 (hereafter S9, Table \ref{table:symbol}) occurred within about 0.3 {\it Gaia} revolutions.  This means each scan angle and direction were nearly identical.  Both scans went from top (north) to bottom (south) in \figref[fig:sky] (middle right plot) such that S9 was observed before its brighter neighbour, DR2 5827538590793371776 (hereafter S10, Table \ref{table:symbol}), each time.  In both transits, all the windows of S9 and S10 are offset in AL by one macrosample and aligned in AC (\figref[fig:window_geometry_sameAC_diffAL]).

As in the previous section, this means that all of these windows are rectangular, which explains why all the windows of S9 and these two transits of S10 (out of nine) were not excluded and yielded radial velocities for both sources in DR2.  The difference in brightness between S9 and S10 (5.0 mag) means that S9's DR2 radial velocity is contaminated and so has been excluded from EDR3.  S10's DR2 radial velocity is not contaminated by S9 in these two transits and so remains in EDR3.	

\subsection{DR2 4092328917916154368}
\label{sec:4092328917916154368}

The two transits of DR2 4092328917916154368 (hereafter S11, Table \ref{table:symbol}) occurred about eight months apart.  By chance, the scan angles and directions were similar.  Both scans went from bottom left (south east) to top right (north west) in \figref[fig:sky] (bottom left panel) such that S11 was always observed before its brighter neighbour, DR2 4092328917911305984 (hereafter S12, Table \ref{table:symbol}), each time.  

In the first transit, each of the three S11 and S12 windows are offset in AL by one macrosample and aligned in AC (\figref[fig:window_geometry_sameAC_diffAL]).  In the second transit, each of the three S11 and S12 windows are aligned in AL and there are three AC pixels overlapping between each of the three S11 windows and the S12 windows, which leaves the S11 windows rectangularly truncated to an AC width of seven pixels, similar to the window geometry in \figref[fig:window] and S1 in B19.

As in the previous section, this means that all of these windows in both transits are rectangular, which explains why all the windows of S11 and S12 (also two transits only) were not excluded and yielded radial velocities for both sources in DR2.  The difference in brightness between S11 and S12 (5.8 mag) means that S11's DR2 radial velocity is contaminated and so has been excluded from EDR3.  S12's DR2 radial velocity is not contaminated by S11 and so remains in EDR3.	

\subsection{DR2 5413575658354375040}
\label{Sect:5413575658354375040}

Figure \ref{fig:sky} (bottom right panel) reveals that DR2 5413575658354375040 (hereafter S13, Table \ref{table:symbol}) is in a crowded field.  It is the outskirts of the globular cluster NGC 3201, which is responsible for the secondary peak in \figref[fig:hist_rv_remaining].  S13's $G_{\mathrm{RVS}}^{\mathrm{ext,1}} = 11.96$ mag is very close to $G_{\mathrm{RVS}}^{\mathrm{ext,2}} = 11.99$ mag, which is less than 12.0 mag, meaning it was correct for the CU6-DR2 pipeline to process it.  This suggests that S13's PSF is not disturbed. The majority of sources in the bottom right panel of \figref[fig:sky] are much fainter than S13, which is why it is not in the B19 list.

Even though S13 is bright enough, it does not have a DR3 radial velocity because each of its 32 transits either could not be deblended or are excluded owing to point background contamination.  Its two DR2 transits are excluded from the CU6-DR3 pipeline because of point background contamination.  Its DR2 radial velocity of $500.1 \pm 2.9$ \kms~is consistent with its literature value of $497.6 \pm 0.3$ \kms \citep{mucciarelli2015}.  The latter was measured with the UV-Visual Echelle Spectrograph (UVES) at the Very Large Telescope.  The UVES fibre diameter of 1.0 arcsecond \citep{pasquini2000} is sufficiently small to minimise contamination of the UVES spectrum and its radial velocity from other sources.  This suggests that point background contamination did not affect its DR2 radial velocity, either because the contaminating spectra  are too faint or, if they are bright enough to contaminate S13's spectra, they are sufficiently aligned in AL to not affect its radial velocity.  Therefore, S13's DR2 radial velocity remains in EDR3.

\subsection{Summary}
\label{Sect:summary}

S7, S9 and S11 are excluded from EDR3 because their windows are co-located with the windows of much brighter sources.  While these three sources have $G_{\mathrm{RVS}}^{\mathrm{ext,1}}$ bright enough to be processed by the CU6-DR2 pipeline ($G_{\mathrm{RVS}}^{\mathrm{ext,1}} < 12.0$ mag), their $G_{\mathrm{RVS}}^{\mathrm{ext,2}}$ are too faint for the CU6-DR3 pipeline to process them ($G_{\mathrm{RVS}}^{\mathrm{ext,2}} > 14.0$ mag).  The sources responsible for their contamination are all bright enough to have been processed by the CU6 DR2 and DR3 pipelines.   Their co-located windows were let through by design in the CU6-DR2 pipeline, as expected.  However, they are also let through by the CU6-DR3 pipeline.

The CU6-DR3 pipeline should treat co-located windows aligned in AL (assuming they are duplicate detections of the same source and keeping the brightest detection).  Co-located windows not aligned in AL are not correctly identified as requiring deblending and so are let through.  This is not a problem for these three sources because they are so much brighter than the sources with which they are co-located so the contaminating flux is negligible and their transit radial velocities are unaffected.  Neither is it a problem for the fainter sources because they are too faint to have been processed by the CU6-DR3 pipeline.  Nonetheless, it could be a problem for some faint sources that have been processed by the CU6-DR3 pipeline.  If a minority of their transits are affected, this should not affect their final DR3 radial velocity and combined spectrum.  If the majority of their transits are affected, it could affect their final DR3 radial velocity  and combined spectrum.  This complication is ameliorated for DR3 compared to DR2 because of DR3's higher number of transits.  Nevertheless, validation of DR3 may be able to find such, presumably rare, cases and exclude them from being published.  This complication will be identified at a transit level in the CU6-DR4 pipeline and either deblended or excluded from the final DR4 radial velocity and combined spectrum. 

Of the five sources with their extreme DR2 radial velocities excluded from EDR3 after individual investigation, one (S5) was found to be contaminated but the origin of its contamination could not be unambiguously identified.  Instrumental flux from bias non-uniformity was identified as a possibility.  A less extreme example is found comparing excluded DR2 radial velocities to LAMOST DR6 (Sect. \ref{sec:lamost}). 

The source with the largest radial velocity difference between LAMOST DR6 and {\it Gaia} DR2 is source ID 1803504050895768704 (hereafter S16, Table \ref{table:symbol}).  Sect. \ref{sect:1803504050895768704} finds there are fewer features in the RVS spectra of S16 than S5, raising the prospect that S16's DR2 radial velocity is spurious owing to noise only.
While S16's $G_{\mathrm{RVS}}^{\mathrm{ext,1}}$ is just bright enough to be processed by the CU6-DR2 pipeline ($G_{\mathrm{RVS}}^{\mathrm{ext,1}} = 11.98$ mag), its $G_{\mathrm{RVS}}^{\mathrm{ext,2}}$ is too faint for the CU6-DR3 pipeline to process it ($G_{\mathrm{RVS}}^{\mathrm{ext,2}} = 14.8$ mag).  The CU6-DR2 pipeline magnitude limit was chosen because at $G_{\mathrm{RVS}} < 12.0$ mag, RVS CCD spectra have sufficient signal-to-noise that the peaks in their transit (fixed exposure time) cross-correlation functions (CCFs) correspond to the source's radial velocity.  At $G_{\mathrm{RVS}} > 12.0$ mag, the spectra are more noisy such that the peaks in their transit CCFs could correspond to the source's radial velocity or be spurious.  This prevents an unambiguous determination of transit radial velocities at $G_{\mathrm{RVS}} > 12.0$ mag, which is why the CU6-DR3 pipeline combines all the transit CCFs to derive each source's radial velocity at $G_{\mathrm{RVS}} > 12.0$ mag.

All of the five sources with their extreme DR2 radial velocities excluded from EDR3 after individual investigation were selected to be processed by the CU6-DR2 pipeline based on their $G_{\mathrm{RVS}}^{\mathrm{ext,1}}$.  Their $G_{\mathrm{RVS}}^{\mathrm{ext,2}}$ was found to be much fainter ($G_{\mathrm{RVS}}^{\mathrm{ext,2}} - G_{\mathrm{RVS}}^{\mathrm{ext,1}} = [2.8,5.6]$ mag).  Although both $G_{\mathrm{RVS}}^{\mathrm{ext,2}}$ and $G_{\mathrm{RVS}}^{\mathrm{ext,1}}$ are based on blended and/or contaminated photometry, {\it Gaia}'s superior space-based angular resolution limits the effect of blending and contamination, suggesting that these five sources were indeed too faint to be processed by the CU6-DR2 pipeline.  Being fainter than intended left them more susceptible to noise and spectral contamination.  The same bright star that contaminated their $G_{\mathrm{RVS}}^{\mathrm{ext,1}}$ also contaminated their RVS spectra and thus DR2 radial velocity.

The derivation of $G_{\mathrm{RVS}}^{\mathrm{ext,1}}$ for these five sources was based on photographic magnitudes \citep{smart2014}, which are shown to be blended and/or contaminated.  According to the declination of these five sources, their magnitudes come from photographic plates observed between 1978 and 1998 \citep{monet2003}.  This photographic legacy has not propagated into the CU6-DR3 pipeline because it selects its stars based on $G_{\mathrm{RVS}}^{\mathrm{ext,2}}$.  Therefore, DR3 radial velocities should not be much fainter than intended ($G_{\mathrm{RVS}}^{\mathrm{ext,2}} < 14.0$ mag) and thus more robust to contamination from outside their windows (blending in their windows has been treated).  

\section{EDR3 status of high-velocity stars}
\label{sect:appendix2}

The EDR3 status of high-velocity stars in the negative and positive tails of DR2's radial velocity distribution are presented in Tables \ref{table:summary_neg} and \ref{table:summary_pos}, respectively.

\newpage 

\begin{table*}       
\centering   
\caption{EDR3 status of high-velocity stars in the negative tail of DR2's radial velocity (RV in \kms) distribution, given by the EDR3 column: Yes (DR2 RV {\it is} in EDR3) or No (DR2 RV {\it is not} in EDR3).}    
\label{table:summary_neg}            
\begin{tabular}{lcrrcccc}     
\hline\hline 
(E)DR3 source ID & This work & DR2 RV & DR2 $\sigma_{\textrm{RV}}$  & DR2 transits & EDR3 & B19 list & Section\\
\hline
5933266834310007808 &  & $-$999.3 & 2.6 & 2 & No & Yes &\\
4658865791827681536 & S7 & $-$987.5 & 2.8 & 4 & No & No & \ref{sec:4658865791827681536} \\
4103049637327213440 &  & $-$986.2 & 4.5 & 2 & No & Yes &\\
5951114420631264640 &  & $-$984.3 & 3.4 & 2 & No & Yes &\\
5966712023814100736 & S5 & $-$967.7 & 5.8 & 2 & No & No  & \ref{Sect:5966712023814100736}, \ref{sec:li2020}\\
2198292118993038464 &  & $-$928.0 & 5.2 & 3 & No & Yes &\\
4058210969029372928 &  & $-$923.2 & 0.1 & 2 & No & Yes &\\
4062883829092182144 &  & $-$909.3 & 0.9 & 2 & No & Yes &\\
5959019801816582272 &  & $-$902.3 & 2.8 & 2 & No & Yes &\\
4065791380145075072 &  & $-$882.4 & 3.4 & 2 & No & Yes &\\
5977687963063223552 &  & $-$869.6 & 6.4 & 2 & No & Yes &\\
4314772322247632512 &  & $-$858.3 & 3.9 & 3 & No & Yes &\\
5943772294301995264 &  & $-$849.4 & 14.0 & 2 & No & Yes &\\
4294462487060476800 &  & $-$842.3 & 8.5 & 2 & No & Yes &\\
5305975869928712320 & S3 & $-$830.6 & 5.6 & 2 & No & No & \ref{sec:5305975869928712320}\\
1995066395528322560 &  & $-$799.1 & 1.1 & 2 & No & Yes & \ref{sec:li2020}\\
5971934527953031296 &  & $-$792.6 & 11.7 & 4 & No & Yes &\\
5866215870791287936 &  & $-$789.0 & 8.9 & 2 & No & Yes &\\
2041630300642968320 &  & $-$766.1 & 18.3 & 3 & No & Yes &\\
5871770569191291904 &  & $-$762.3 & 4.1 & 2 & No & Yes &\\
4103096400926398592 &  & $-$757.0 & 0.7 & 2 & No & Yes &\\
4105689496051901440 &  & $-$737.3 & 10.4 & 2 & No & Yes &\\
5231593594752514304 &  & $-$715.8 & 1.0 & 2 & No & Yes &\\
5883971746674851840 &  & $-$712.7 & 0.8 & 2 & No & Yes &\\
5878409248569969792 &  & $-$711.9 & 3.7 & 2 & No & Yes &\\
2033855963202467456 &  & $-$692.6 & 16.8 & 4 & No & Yes &\\
4065480978657619968 &  & $-$680.7 & 1.9 & 2 & No & Yes &\\
3397895966721032960 &  & $-$674.7 & 0.8 & 2 & No & Yes &\\
5965303648201160448 &  & $-$668.0 & 3.6 & 2 & No & Yes &\\
4515902234779888512 &  & $-$654.0 & 0.1 & 2 & No & Yes &\\
1809832393157398016 &  & $-$653.1 & 4.3 & 2 & No & Yes &\\
5953456066818230528 &  & $-$633.7 & 4.6 & 2 & No & Yes &\\
4063258144073821184 &  & $-$631.0 & 0.9 & 2 & No & Yes &\\
4269007246718955008 &  & $-$627.3 & 2.7 & 4 & No & Yes &\\
1814359288672674560 &  & $-$618.5 & 1.7 & 5 & Yes & No &\\
5932173855446728064 & S1 & $-$614.3 & 2.5 & 7 & No & Yes &\ref{sec:b19}\\
4531308286776328832 &  & $-$606.2 & 1.1 & 6 & Yes & No &\\
4587905579084735616 &  & $-$605.7 & 4.2 & 8 & Yes & No &\\
4100838558128545664 &  & $-$601.1 & 5.9 & 2 & No & Yes &\\
5979630593954336768 &  & $-$598.4 & 1.2 & 2 & No & Yes &\\
1788324708749944448 &  & $-$597.5 & 1.6 & 9 & Yes & No &\\
4252356788030249728 &  & $-$597.4 & 9.2 & 2 & No & Yes &\\
4051880668297749760 &  & $-$588.7 & 11.7 & 3 & No & Yes &\\
5956359499060605824 &  & $-$579.3 & 1.1 & 2 & No & Yes &\\
5931224697615320064 &  & $-$577.7 & 3.7 & 2 & No & Yes &\\
1805949089882861568\tablefootmark{a}  &  & $-$576.9 & 1.4 & 5 & No & Yes &\\
2125678412572858240 &  & $-$572.1 & 1.4 & 7 & Yes & No &\\
5253575237405660160 &  & $-$570.3 & 7.6 & 2 & No & Yes &\\
5932591154473648128 &  & $-$568.8 & 0.8 & 4 & No & Yes &\\
2018250663356989440 &  & $-$568.8 & 3.1 & 2 & No & Yes &\\
1953616147184247808 &  & $-$565.2 & 0.7 & 8 & Yes & No &\\
1789332097623284736 &  & $-$565.2 & 0.6 & 8 & Yes & No &\\
1859067080734005504 &  & $-$564.7 & 1.0 & 15 & Yes & No &\\
5307479791350326400 &  & $-$564.3 & 2.8 & 2 & No & Yes &\\
6053634083049269888 &  & $-$563.3 & 3.6 & 2 & No & Yes &\\
5867537964809068672 &  & $-$563.0 & 6.5 & 3 & No & Yes &\\
4585522112754440576 &  & $-$561.3 & 0.4 & 13 & Yes & No &\\
1335426489060934272 &  & $-$561.2 & 0.8 & 6 & Yes & No &\\
1336408284224866432 &  & $-$559.1 & 1.1 &  10 & Yes& No &\\
\hline
\end{tabular}
\tablefoot{
\tablefoottext{a}{DR2 source ID 1805949089882861696.} 
  }
\end{table*}

\begin{table*}  
\centering   
\caption{EDR3 status of high-velocity stars in the positive tail of DR2's radial velocity (RV in \kms) distribution, given by the EDR3 column: Yes (DR2 RV {\it is} in EDR3) or No (DR2 RV {\it is not} in EDR3).}    
\label{table:summary_pos}            
\begin{tabular}{lcrrcccc}     
\hline\hline 
(E)DR3 source ID & This work & DR2 RV & DR2 $\sigma_{\textrm{RV}}$  & DR2 transits & EDR3 & B19 list & Section\\
\hline
4251939076692958336 & & 970.6 & 2.1 & 2  & No & Yes & \\
4092328917916154368 & S11 & 937.5 & 0.9 & 2  & No & No & \ref{sec:4092328917916154368}\\
4251811155377211776 & & 930.9 & 2.2 & 2  & No & Yes & \\
5827538590793373696 & S9 & 902.6 & 1.3 & 2  & No & No & \ref{sec:5827538590793373696}\\
4057347680607821696 & & 898.9 & 3.6 & 2  & No & Yes & \\
5959713078252724352 & & 895.7 & 1.1 & 2  & No & Yes & \\
6101369964484138752 & & 869.8 & 19.9 & 4  & No & Yes & \\
4291541561371020160 & & 857.8 & 1.0 & 2  & No & Yes & \\
4148502382868008448 & & 851.3 & 2.4 & 2  & No & Yes & \\
5939029580905945600 & & 837.9 & 3.2 & 2  & No & Yes & \\
6019186143235920256 & & 810.5 & 18.3 & 5 & No & Yes & \\
5999456197349672832 & & 802.6 & 1.9 & 2  & No & Yes & \\
4100844124407433984 & & 778.0 & 0.1 & 2  & No & Yes & \\
2028915440803837312 & & 777.3 & 2.1 & 2  & No & Yes & \\
5625102999536111872 & & 768.6 & 4.2  & 3 & No & Yes & \\
4296894160078561280 & & 760.0 & 1.9 & 2  & No & Yes & \\
4155983425621616896 & & 752.6 & 0.4 & 2  & No & Yes & \\
1827795080271815040 & & 738.6 & 1.1 & 2  & No & Yes & \\
2251311188142608000 & & 738.2 & 3.7 & 2  & No & Yes & \\
1732532430739244544 & & 724.4 & 7.2 & 3 & No & Yes & \\
4098291913093819776 & & 713.6 & 2.5 & 2  & No & Yes & \\
6101408687905214208 & & 708.2 & 2.9 & 2  & No & Yes & \\
5964046013059722624 & & 706.5 & 4.3 & 3  & No & Yes & \\
4154757199609510912 & & 703.7 & 2.2 & 2  & No & Yes & \\
4504337678883284096 & & 694.1 & 4.0 & 2  & No & Yes & \\
4051413483316567552 & & 689.0 & 10.9 & 2  & No & Yes & \\
4265444618530378240 & & 680.1 & 4.4 & 2  & No & Yes & \\
5954626393878439040 & & 662.2 & 1.3 & 2  & No & Yes & \\
5852515607983045248 & & 659.6 & 13.6 & 2  & No & Yes & \\
5892583362223572096 & & 656.9 & 4.8 & 2  & No & Yes & \\
1825842828672942208 & & 641.8 & 2.0 & 2  & No & Yes & \\
4094614738213555072 & & 638.3 & 0.1 & 2  & No & Yes & \\
5936466825463371904 & & 636.5 & 3.8 & 3  & No & Yes & \\
6077622510498751616 & & 623.6 & 0.8 & 22  & Yes & No & \\
5331557897713152640 & & 619.2 & 1.2 & 10 & Yes & No & \\
6080513603289501184 & & 606.0 & 0.8 & 13  & Yes & No & \\
5413949281841538688 & & 605.3 & 3.6 & 2  & No & Yes & \\
5544764712854179840 & & 604.6 & 1.1 & 8 & Yes & No & \\
5227401642246690432 & & 603.0 & 3.2 & 2  & No & Yes & \\
5325632011075073408 & & 601.7 & 0.8 & 9  & Yes & No & \\
6148919860947868160 & & 600.0 & 0.3 & 16  & Yes & No & \\
5843949763886962304 & & 593.3 & 3.6 & 9 & Yes & No & \\
2921543205513614208 & & 591.5 & 0.6 & 20 & Yes & No & \\
5382632652358260864 & & 586.7 & 0.3 & 13 & Yes & No & \\
5515555464906936192 & & 585.2 & 0.6 & 7  & Yes & No & \\
6019187556324501888 & & 581.9 & 8.2 & 2  & No & Yes & \\
5354656541072512000 & & 581.7 & 2.0 & 8 & Yes & No & \\
5654083549060480256 & & 581.7 & 0.5 & 9  & Yes & No & \\
4076739732812337536 & & 572.3 & 4.8 & 2  & No & Yes & \\
3752558526183725440 & & 570.6 & 1.1 & 6  & Yes & No & \\
6418433113222352000 & & 570.0 & 2.4 & 15 & Yes & No & \\
4118205030747323648 & & 569.1 & 3.9 & 2  & No & Yes & \\
939821616976287104 & & 568.2 & 0.8 & 2  & No & Yes & \\
4150939038071816320 & & 564.1 & 0.5 & 2  & No & Yes & \\
5430581735975161344 & & 564.0 & 1.9 & 9  & Yes & No & \\
4320112542823692288 & & 562.3 & 2.7 & 2  & No & Yes & \\
4267378285875942272 & & 559.1 & 0.9 & 2  & No & Yes & \\
5358307297639362304 & & 558.6 & 2.8 & 6  & Yes & No & \\
5352190302148489088 & & 558.1 & 1.5 & 5 & Yes & No & \\
4095117691720913152 & & 554.2 & 18.7 & 2  & No & Yes & \\
\hline
\end{tabular}
\end{table*}

\section{Comparison with the literature}
\label{sect:appendix3}

\subsection{Source ID 3754815441303370496}
\label{sect:3754815441303370496}

The source with the largest radial velocity difference between RAVE DR6 and {\it Gaia} DR2 is source ID 3754815441303370496 (hereafter S14, Table \ref{table:symbol}) with two radial velocity observations between 2014-07-25 and 2016-05-23, the median of which is 286.4 $\pm$ 0.5 \kms.  S14's $G_{\mathrm{RVS}}^{\mathrm{ext,2}} = 12.2$ mag so this source should not have been processed by the CU6-DR2 pipeline ($G_{\mathrm{RVS}}^{\mathrm{ext,1}} < 12.0$ mag) but was processed by both the CU6 DR2 and DR3 pipelines.  This source corresponds to RAVE ID J103500.4-113004 with a single radial velocity measurement of 9.8 $\pm$ 2.8 \kms~observed on 2005-02-21.  The preliminary DR3 radial velocity agrees with the RAVE value within their uncertainties but not the DR2 radial velocity, which is why it is excluded from EDR3.

The two DR2 transits are not included in the DR3 measurement for the following reasons.  S14's windows in the first transit are all untruncated but each are overlapped by another window aligned in AC (cf. \figref[fig:window_geometry_sameAC_diffAL]).  The CU6-DR3 pipeline can not deblend these window geometries so the first transit is excluded.  The overlapped source is DR2 3754815445598430080 (hereafter S15, Table \ref{table:symbol}), which does not have a DR2 radial velocity.  S15's windows truncate each of S14's CCD spectra in the second transit, reducing their width to AC = 4 pixels (cf. \figref[fig:window]).  The CU6-DR3 pipeline is able to deblend these window geometries but the second transit is excluded because it has systematically negative fluxes in the spectra, owing to oversubtraction of the straylight.

S15 is actually brighter ($G_{\mathrm{RVS}}^{\mathrm{ext,2}} = 11.9$ mag) than S14.  It is less than 2 arcseconds away from S14, which is why S14 is in the B19 list.  RAVE radial velocities were obtained with the UK Schmidt Telescope and 6dF spectrograph.  The 6dF fibres were 6.7 arcsec in diameter \citep{parker1998}.  This means both S14 and S15 were observed in the same fibre, making RAVE ID J103500.4-113004 a composite spectrum and meaning the radial velocity is in doubt.  Nevertheless, the preliminary DR3 radial velocities of S14 and S15 agree within their uncertainties.  Assuming these are correct, RAVE's radial velocity is some average from both stars but because the stars have similar radial velocities, this is sufficient to validate that S14's DR2 radial velocity is spurious and needed to be deblended and so was correctly excluded from EDR3.  

\subsection{Source ID 1803504050895768704}
\label{sect:1803504050895768704}

The source with the largest radial velocity difference between LAMOST DR6v2 and {\it Gaia} DR2 is source ID 1803504050895768704 (hereafter S16, Table \ref{table:symbol}) with two radial velocity observations between 2014-07-25 and 2016-05-23, the median of which is $-453.9 \pm 4.6$ \kms.  S16's $G_{\mathrm{RVS}}^{\mathrm{ext,2}} = 14.8$ mag means it should not have been processed by the CU6-DR2 pipeline ($G_{\mathrm{RVS}}^{\mathrm{ext,1}} < 12.0$ mag) and is not processed by the CU6-DR3 pipeline ($G_{\mathrm{RVS}}^{\mathrm{ext,2}} < 14.0$ mag).

Neither of the two DR2 transits are overlapping any other windows, suggesting that their spectra are not contaminated by other sources.  Residual straylight is the most likely cause of the $G_{\mathrm{RVS}}$ measured from the transit spectra (14.0 and 13.2 mag for the first and second transits respectively) being brighter than S16's external $G_{\mathrm{RVS}}$ (14.8 mag). 

S16's DR2 $T_{\textrm{eff}} = 4938^{+692}_{-514}$ K \citep{rene2018} is consistent with LAMOST's $T_{\textrm{eff}} = 4942\pm273$ K and spectral subclass assignment of K3.  Hence, the RVS transit spectra of S16 are expected to be \ion{Ca}{ii}-dominated so the wavelengths of where these lines are expected to be are marked in \figref[fig:ac_lsf_lamost], both at rest (very similar to S16's LAMOST radial velocity) and Doppler shifted according to the DR2 transit radial velocities.  

\begin{figure*}
\centering
\includegraphics[width=0.75\textwidth]{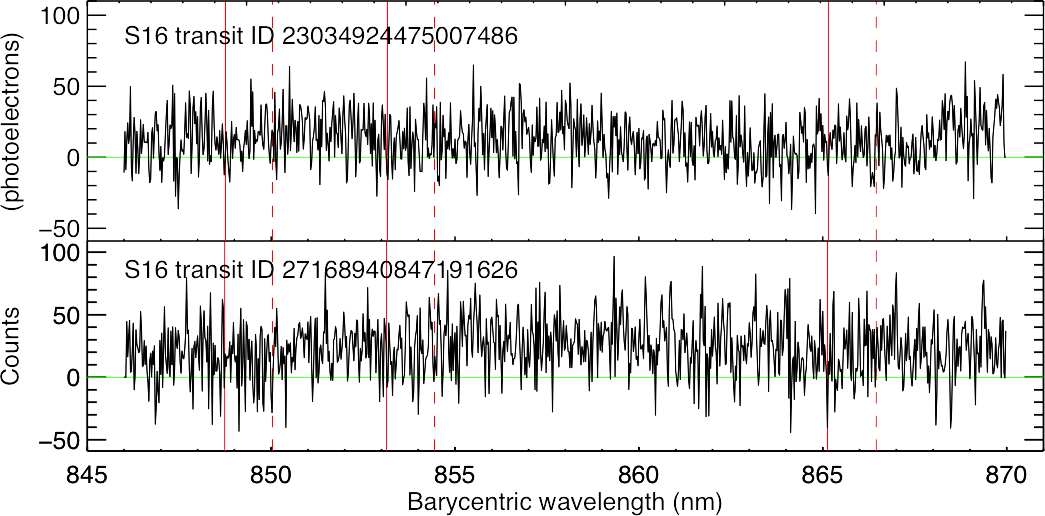}\\
\caption[]{The two DR2 transit (three-CCD averaged) spectra of S16.  The red dashed lines delineate where the \ion{Ca}{ii} absorption lines are expected to be when at rest.  Although not unambiguously identifiable in the spectra, the red solid lines are the same lines, Doppler shifted according to the transit radial velocity.}
\label{fig:ac_lsf_lamost}
\end{figure*}

Even though it is brighter than the first transit, \figref[fig:ac_lsf_lamost] shows the spectrum of the second transit is noisier.  This may be related to it being observed on CCD row 5 and/or the great circle of the scan forming a shallow angle with the Galactic plane.  As explained in Sect. \ref{Sect:5966712023814100736}, CCD row 5 has the worst performance in terms of bias non-uniformity and scans forming a shallow angle with the Galactic plane may cause further bias non-uniformity difficulties.   The first transit was observed on CCD row 4, when the scan did not form a shallow angle with the Galactic plane.  

By coincidence, S16 ($G_{\mathrm{RVS}}^{\mathrm{ext,2}} = 14.8$ mag) has an almost identical brightness to 
S5 ($G_{\mathrm{RVS}}^{\mathrm{ext,2}} = 14.9$ mag), meaning their spectra can be visually compared.  There appears to be fewer features in S16's transit spectra than those of S5 (cf. \figref[fig:ac_lsf5966712023814100736]).  S5's spectra appear to exhibit non-astrophysical features, suggestive of instrumental contamination.  These features are not visible in S16's spectra. 

S16's small DR2 radial velocity uncertainty (4.6 \kms) reveals the two transit radial velocity values are close to each other.  Although instrumental effects have not been ruled out, S16's transit radial velocities being so close to each other could be because of noise in its CCD spectra mimicking the \ion{Ca}{ii} lines at the same spurious radial velocity in both transits.  In the vast majority of cases, noisy spectra would not do this and yield sufficiently different transit radial velocities that the uncertainty on their median value would be larger than the DR2 filter on this uncertainty (20 \kms, \citealt{sartoretti2018}), removing their radial velocity from DR2.  However, the CU6-DR2 pipeline produced 78 million transit radial velocities \citep{sartoretti2018}, suggesting that chance agreement between two transit radial velocities cannot be ruled out.

S16 corresponds to LAMOST designation J200933.29$+$131542.5.  Its single radial velocity measurement of 1.2 $\pm$ 4.6 \kms observed on 2016-06-01 suggests that S16's DR2 radial velocity is spurious.  LAMOST's spectrum of S16 was obtained with a fibre pointing 0.1 arcsec from S16's EDR3 position.  The nearest source to S16 of comparable brightness is 5.3 arcsec away: source ID 1803504050895769472 (hereafter S17, Table \ref{table:symbol}), which is $G_{\mathrm{RVS}}^{\mathrm{ext,2}} = 15.0$ mag.  The LAMOST fibres are 3.3 arcsec in diameter \citep{su2012}.  This suggests that, while S16's LAMOST spectrum may contain some flux from S17, it is unlikely to be enough to contaminate the LAMOST radial velocity, confirming that S16 was correctly excluded from EDR3.

\end{appendix}

\end{document}